\documentclass[11pt,letterpaper]{article}

\pdfoutput=1

\usepackage[utf8x]{inputenc}%
\usepackage{tcolorbox,fancybox}%
\usepackage{ntheorem}

\usepackage[english]{babel}
\usepackage{amsmath,amssymb,amsfonts}
\usepackage{graphicx}
\usepackage{booktabs}
\usepackage{color,xcolor}
\usepackage[numbers,sort&compress]{natbib}
\usepackage{tikz}
\usepackage{physics}
\usepackage{orcidlink}
\usepackage{slashed}
\usepackage{mathrsfs}
\usepackage{enumerate}
\usepackage{mdframed}
\usepackage{url}
\usepackage[makeroom]{cancel}
\usepackage{geometry}

\usepackage{pifont}
\newcommand{\cmark}{\ding{51}}%
\newcommand{\xmark}{\ding{55}}%

\newtheorem*{thm-non}{Theorem}
\newtheorem*{conj}[thm-non]{Conjecture}

\usepackage{hyperref} 
\hypersetup{
    colorlinks=true,       
    linkcolor=red,          
    citecolor=blue,        
    filecolor=magenta,      
    urlcolor=purple           
}
\usepackage[all]{hypcap} 

\def\beqn{\begin{eqnarray}}
\def\eeqn{\end{eqnarray}}
\def\beqs{\begin{subequations}}
\def\eeqs{\end{subequations}}
\def\beq{\begin{equation}}
\def\eeq{\end{equation}}
\def\ba{\begin{array}}
\def\ea{\end{array}}

\def\non{\nonumber\\}

\def\hf{\frac{1}{2}}
\def\[{\left[}
\def\]{\right]}
\def\({\left(}
\def\){\right)}
\newcommand\para{\paragraph{}}

\def\gSU{\rm SU}

\def\gSO{\rm SO}

\newcommand{\rep}[1]{\mathbf{#1}}
\newcommand{\repb}[1]{\mathbf{\overline{#1}}}

\def\Bc{\mathcal{B}}
\def\Cc{\mathcal{C}}
\def\Dc{\mathcal{D}}
\def\Ec{\mathcal{E}}
\def\Fc{\mathcal{F}}
\def\Gc{\mathcal{G}}

\def\Lc{\mathcal{L}}
\def\Mc{\mathcal{M}}
\def\Nc{\mathcal{N}}
\def\Oc{\mathcal{O}}

\def\Sc{\mathcal{S}}
\def\Tc{\mathcal{T}}
\def\Uc{\mathcal{U}}

\def\Xc{\mathcal{X}}
\def\Yc{\mathcal{Y}}

  \def\cG{\mathfrak{c}}
\def\DG{\mathfrak{D}}  
\def\EG{\mathfrak{E}}  \def\eG{\mathfrak{e}}

  \def\nG{\mathfrak{n}}

\def\SG{\mathfrak{S}}  \def\sG{\mathfrak{s}}
  
\def\UG{\mathfrak{U}}

\title{
 {\bf A two-generational ${\rm SU}(7)$ model with extended weak sector: mass hierarchies, mixings, and the flavor non-universality} \\
\author{\large Ning Chen$^{\,\heartsuit}$\,\orcidlink{0000-0002-0032-9012}, Ying-nan Mao$\,^\star$\,\orcidlink{0000-0001-8063-8968}, \\ Zhaolong Teng$^{\,\clubsuit}$\,\orcidlink{0000-0002-7141-2331}, Bin Wang$^{\,\spadesuit}$, Xiangjun Zhao$^{\,\diamondsuit}$}
\date{\small \it
$^\heartsuit\, ^\clubsuit \, ^\spadesuit\, ^\diamondsuit$School of Physics, Nankai University, Tianjin, 300071, China \\
$^\star$ Department of Physics, School of Science, Wuhan University of Technology, \\ Wuhan, 430070, Hubei, China \\
}
}

\begin{document}

\maketitle
\setlength{\parskip}{0.2ex}

\begin{abstract}
\bigskip
We study a possible gauge symmetry breaking pattern in an ${\rm SU}(7)$ grand unified theory, which describes the mass origins of all electrically charged SM fermions of the second and the third generations.
Two intermediate gauge symmetries of $\Gc_{341}\equiv {\rm SU}(3)_c \otimes {\rm SU}(4)_W \otimes {\rm U}(1)_{X_0}$ and $\Gc_{331}\equiv {\rm SU}(3)_c \otimes {\rm SU}(3)_W \otimes {\rm U}(1)_{X_1}$ arise above the electroweak scale.
SM fermion mass hierarchies between two generations can be obtained through a generalized seesaw mechanism.
The mechanism can be achieved with suppressed symmetry breaking VEVs from multiple Higgs fields that are necessary to avoid tadpole terms in the Higgs potential.
Some general features of the ${\rm SU}(7)$ fermion spectrum will be described, which include the existence of vectorlike fermions, the tree-level flavor changing weak currents between the SM fermions and heavy partner fermions, and the flavor non-universality between different SM generations from the extended weak sector.
\end{abstract}

\vspace{6cm}
{\emph{Emails:}\\  
$^{\,\heartsuit}$\url{chenning_symmetry@nankai.edu.cn},\\
 $\,^\star$\url{ynmao@whut.edu.cn},\\
$^{\,\clubsuit}$ \url{tengcl@mail.nankai.edu.cn},\\  
$^{\,\spadesuit}$ \url{wb@mail.nankai.edu.cn},\\ 
 $^{\,\diamondsuit}$\url{zhaoxiangjun@mail.nankai.edu.cn}}

\thispagestyle{empty}  
\newpage  
 
\setcounter{page}{1}  

\vspace{1.0cm}
\tableofcontents

\section{Introduction}
\label{section:intro}
%
%

\para
Grand Unified Theories (GUTs), with their original formulations based on the gauge groups of $\gSU(5)$~\cite{Georgi:1974sy} and $\gSO(10)$~\cite{Fritzsch:1974nn}, were proposed to unify all three fundamental symmetries described by the Standard Model (SM).
Though being successful in achieving the gauge coupling unification, one has to acknowledge that all existing unified models have not completely address some fundamental puzzles in the matter sector of the SM.
In particular, the observed mass hierarchies and their mixings under the charged weak currents of three generational fermions have never been fully probed in the framework of GUT.
A detailed description of the flavor puzzle can be found in recent Refs.~\cite{King:2017guk,Xing:2020ijf}.
All puzzles in the flavor sector originate from the three-generational structure.
In most of the previous studies, one-generational anomaly-free fermion representations were trivially repeated with multiple times for the generational structure.
For example, the minimal set of fermion contents are $3\times \[ \repb{5_F} \oplus \rep{10_F} \]$ in the $\gSU(5)$ GUT, and $3\times \rep{16_F}$ in the $\gSO(10)$ GUTs.

\para
The flavor sector becomes more puzzling with the discovery~\cite{ATLAS:2012yve,CMS:2012qbp} and the measurements~\cite{ATLAS:2016neq} of the $125\,{\rm GeV}$ SM Higgs boson at the Large Hadron Collider (LHC).
All existing measurements of the SM Higgs boson have confirmed its Yukawa couplings to the third generational fermions, both through the $H\to ( b \bar b\,, \tau \tau)$ decays~\cite{CMS:2020zge,ATLAS:2020bhl,ATLAS:2022yrq,CMS:2022kdi} and the $t \bar t H$ associated productions~\cite{CMS:2018uxb,ATLAS:2018mme}.
The ongoing searches are indicating the SM Higgs boson decays into the second-generational fermions of charm quarks~\cite{ATLAS:2022ers,CMS:2022psv} and muons~\cite{ATLAS:2020fzp,CMS:2020xwi}.
{\it These experimental results further strengthen the SM flavor puzzle into a SM flavor paradox.}
In the SM, each generation of fermions transform identically under the $\Gc_{\rm SM} \equiv {\rm SU}(3)_c \otimes {\rm SU}(2)_W \otimes {\rm U}(1)_Y$.
Therefore, a single SM Higgs boson is unlikely to generate hierarchical couplings to different flavors without additional symmetries.
An intrinsic issue with the Higgs mechanism is what can be the natural Yukawa coupling.
The proposal of the anarchical fermion mass scenario~\cite{Hall:1999sn,Haba:2000be} will be adopted in the current discussion. 
From the renormalization group evolutions of the Yukawa coupling, it is unlikely to generate large hierarchies from $\Oc(1)$ to $\Oc(10^{-6})$ when it evolves from the GUT scale to the electroweak (EW) scale. 
Studies of the running SM fermion masses can be found in Refs.~\cite{Fusaoka:1998vc,Xing:2007fb,Xing:2011aa,Antusch:2013jca,Huang:2020hdv,Wang:2021wdq}.
From this point of view, the SM Higgs boson that is responsible for the electroweak symmetry breaking (EWSB) is most likely to give top quark mass, with the natural Yukawa coupling of $y_t\sim \Oc(1)$.
Such a result was recently obtained in the third-generational ${\rm SU}(6)$ model~\cite{Chen:2021zwn}.
Therefore, the question becomes what may be the underlying mechanism to generate all other suppressed SM fermion masses to the EW scale.
Furthermore, three generational SM fermions exhibit mixings under the weak charged currents, which are described by the Cabibbo-Kobayashi-Maskawa (CKM) matrix~\cite{Cabibbo:1963yz,Kobayashi:1973fv} for the quark sector, and the Pontecorvo-Maki-Nakagawa-Sakata (PMNS) matrix~\cite{Pontecorvo:1957qd,Maki:1962mu} for the lepton sector, respectively.
The SM itself does not explain the origins of all measured fermion mixings.
In the context of the minimal ${\rm SU}(5)$ GUT, Georgi and Jarlskog~\cite{Georgi:1979df} realized natural mass relations between the down-type quarks and the charged leptons in three generations with extended Higgs fields of $3\times \rep{5_H} \oplus \rep{45_H}$.

\para
Historically, a unification of the flavor sector was first made by Georgi~\cite{Georgi:1979md}, where he suggested that fermions in different generations may transform differently under the symmetries of the UV theory.
It turns out this can be achieved in the framework of GUT, as long as one extends the gauge groups beyond the ${\rm SU}(5)$.
Furthermore, one must include anti-symmetric irreducible representations (irreps) with rank larger than two to avoid the simple repetition of one generational anomaly-free fermions.
This can be minimally achieved from a unified gauge group of ${\rm SU}(7)$, as was shown by Frampton~\cite{Frampton:1979cw,Frampton:1979tj}.
Some early studies of the ${\rm SU}(7)$ models include Refs.~\cite{Claudson:1980dv,Kim:1980ci,Umemura:1980sa,Cox:1980er,Xue:1981ek,Kim:1981be}.
Collectively, we dub the GUTs with gauge groups larger than ${\rm SU}(5)$ the {\it non-minimal} GUTs.
Besides of some brief constructions of the fermion and Higgs spectrum, the {\it non-minimal} GUT was never studied in details from the physical perspective since its inception.
A separate observation is that the {\it non-minimal} GUT groups are usually broken to the SM gauge group via several intermediate scales.
For example, the third-generational ${\rm SU}(6)$ GUT undergoes two symmetry breaking stages of ${\rm SU}(6) \to {\rm SU}(3)_c \otimes {\rm SU}(3)_W \otimes {\rm U}(1)_X \to \Gc_{\rm SM}$~\cite{Chen:2021haa,Chen:2021zwn}.
During the intermediate symmetry breaking stages, some vectorlike fermions in the spectrum can obtain masses and also mix with the SM fermions.
The existence of some Higgs mixing operators is likely to generate EWSB vacuum expectation values (VEVs) that contribute to light SM fermion masses, such as bottom quark and tau lepton in the third generation~\cite{Chen:2021zwn}.
Based on the previous studies to the ${\rm SU}(6)$ toy model, we propose that
\begin{conj}

The realistic GUT can give arise to the observed mass hierarchies as well as the weak mixings of the SM fermions through its realistic symmetry breaking pattern, with the natural Yukawa couplings of $\Oc(1)$.

\end{conj}

\para
A separate feature of the {\it non-minimal} GUT is the automatic emergence of the global symmetries, as long as one relaxes the Georgi's third law~\cite{Georgi:1979md} in formulating the anomaly-free fermion irreps.
As was first observed by Dimopoulos, Raby and Susskind (DRS)~\cite{Dimopoulos:1980hn}, an anomaly-free ${\rm SU}(N+4)$ gauge theory with $N$ chiral fermions in the anti-fundamental irrep and one chiral fermion in the rank-$2$ anti-symmetric irrep enjoys a global DRS symmetry of 
\beqn\label{eq:DRS_global}
\Gc_{\rm DRS}&=& {\rm U}(N) = {\rm SU}(N) \otimes {\rm U}(1) \,,~ N \geq 2 \,.
\eeqn
Notice that the global DRS symmetries in Eq.~\eqref{eq:DRS_global} come from the anomaly-free condition, and can be generally true with anomaly-free rank-$k$ ($k\geq 3$) fermion irreps.
Though the original study of Ref.~\cite{Dimopoulos:1980hn} deals with a strong interacting theory with the gauge symmetry broken through the bi-linear fermionic condensates, the emergence of the DRS global symmetries is valid when the underlying gauge symmetry is spontaneously broken by the Higgs mechanism.
This result can be generalized to the {\it non-minimal} GUTs beyond the ${\rm SU}(5)$.
Indeed, this was soon pointed out by Georgi, Hall, and Wise in Ref.~\cite{Georgi:1981pu}, where they conjectured the Abelian component of the global DRS symmetry to be the automatic Peccei-Quinn (PQ) symmetry~\cite{Peccei:1977hh}.
The emergent PQ symmetry from the {\it non-minimal} GUT may be an appealing source of another long-standing PQ quality problem~\cite{Georgi:1981pu,Dine:1986bg,Barr:1992qq,Kamionkowski:1992mf,Holman:1992us,Ghigna:1992iv,Kallosh:1995hi} in the axion models, e.g., a minimal construction may be realized in the supersymmetric ${\rm SU}(6)$ model with additional discrete $\mathbb{Z}_4$ symmetry~\cite{Chen:2021haa}.

\para
Motivated by our main conjecture, we study a two-generational ${\rm SU}(7)$ toy model.
There are two major purposes of the current study, which are
\begin{enumerate}

    
    \item to probe the origin of the SM fermion mass hierarchies and their mixings through the realistic symmetry breaking patterns;
    
    \item to describe the gauge couplings of fermion sector in the {\it non-minimal} GUTs.
    
\end{enumerate}
The rest of the paper is organized as follows.
In Sec.~\ref{section:model}, we describe the two-generational fermion content in the ${\rm SU}(7)$ model, and describe the gauge symmetry breaking pattern followed by ${\rm SU}(7)\to {\rm SU}(3)_{c} \otimes {\rm SU}(4)_W \otimes {\rm U}(1)_{X_0}$ at the GUT scale.
In Sec.~\ref{section:SU6}, we review some necessary ingredients from the third-generational ${\rm SU}(6)$ GUT.
In particular, we focus on a gauge-invariant $\nu$-term and its contribution to the Higgs potential.
In Sec.~\ref{section:Higgs}, we discuss the Higgs sector of the ${\rm SU}(7)$ in details, with focus on the contributions from the gauge-invariant and DRS-invariant operators in the Higgs potential.
These terms can be viewed as the generalized $\nu$-terms in the ${\rm SU}(6)$ model.
We found that some of such terms automatically generate a chain of additional VEVs through the tadpole-free condition in the Higgs potential.
The gauge sector from the symmetry breaking pattern is described in Sec.~\ref{section:gauge}.
Sec.~\ref{section:breaking} is the core of this work, where we analyze the details of the realistic gauge symmetry breaking pattern.
Based on the analyses, we obtain the fermion mass terms and mixings together with the Higgs VEVs described in Sec.~\ref{section:Higgs}.
In particular, the analyses in this section will justify our identification of SM fermions in Tabs.~\ref{tab:SU7_2gen_7barferm}, \ref{tab:SU7_2gen_21ferm}, and \ref{tab:SU7_2gen_35ferm}. 
In Sec.~\ref{section:fermions}, we summarize the two-generational fermion masses, the quark mixings under the charged EW currents.
Particularly, we display the flavor non-universality between different generational fermions through their gauge couplings to flavor-conserving neutral gauge boson $Z_\mu^{\prime \prime}$.
We summarize our results and make outlook in Sec.~\ref{section:conclusion}.
In App.~\ref{section:Br}, we give the decomposition rules and the charge quantization according to the desirable symmetry breaking pattern in the ${\rm SU}(7)$ GUT.
All related Lie group calculations are carried out by {\tt LieART}~\cite{Feger:2012bs,Feger:2019tvk}.
In App.~\ref{section:name}, we define the particle names and indices used for different gauge groups in the ${\rm SU}(7)$ model.
In App.~\ref{section:SU7_HiggsOp}, we give explicit derivations of the Higgs VEV generations through the complete set of Higgs mixing operators in the context of the ${\rm SU}(7)$.

\para
We wish to make some disclaimers before presenting the main results.
Throughout the current discussions, we focus on the mass origins and their mixings of two-generational SM fermions, while we do not address some general issues in the GUTs, such as (i) the gauge coupling unification, (ii) the proton lifetime, and (iii) the supersymmetric extension.
We believe all these issues would better be studied in the realistic GUTs with $n_g=3$.

\section{Motivation and the fermion sector in ${\rm SU}(7)$}
\label{section:model}

\subsection{The flavor unification in GUTs}

\para
In a seminal paper~\cite{Georgi:1979md} of the GUT model building including the flavor structure of the SM fermions, Georgi proposed three laws for any extension of the GUT group with the following chiral fermion content
\beqn\label{eq:SUN_fermions}
\{ f_L \}_{ {\rm SU}(N)}^{n_g}&=& \sum_{ \{ k\} }  m_k \, \[ N\,, k \]_{ \rep{F}} \,.
\eeqn
Here, $\[ N\,,k \]_{\rep{F}}$ represents a rank-$k$ anti-symmetric irreps of the ${\rm SU}(N)$ for the chiral fermions.
It was also argued that only the $\[ N\,,k \]_{\rep{F}}$ should be considered in order to avoid the exotic fermions beyond the SM irreps in the spectrum.
In the third law, he required no repetition of a particular irrep of $\[ N\,, k \]_{ \rep{F}}$.
In other words, one can only have $m_k=0$ or $m_k=1$ in Eq.~\eqref{eq:SUN_fermions}.
Correspondingly, he found that the minimal possible GUT group is ${\rm SU}(11)$ with $561$ chiral fermions in total.

\para
The essence of Georgi's third law~\cite{Georgi:1979md} is to prohibit the simple repetition of one set of anomaly-free chiral fermions in the flavor sector, such as $3\times \[ \repb{5_F} \oplus \rep{10_F} \]$ in the ${\rm SU}(5)$ model or $3\times \rep{16_F}$ in the ${\rm SO}(10)$ model.
However, this was soon reconsidered by allowing the repetition of some fermion representations as long as the gauge anomaly is cancelled~\cite{Frampton:1979cw,Frampton:1979tj}.
It is reasonable to conjecture that the third law for the realistic GUT building with multiple fermion generations can be modified by partitioning the ${\rm SU}(N)$ chiral fermions into several irreducible anomaly-free sets as follows
\beqn\label{eq:SUN_fermions_realistic}
\{ f_L \}_{ {\rm SU}(N)}^{n_g}&=& \bigoplus_{ \{ k \}^\prime } \Big\{  {\rm Anom} ( \[ N\,,k \]_{\rep{F}}) \times \overline{ \[ N\,, 1 \]_{ \rep{F}} } \oplus   \[ N\,, k \]_{ \rep{F}}  \Big\} \,,~ {\rm for}~  2 \leq k\leq \[ \frac{N}{2} \]\,.
\eeqn
The anomaly factor for a generic rank-$k$ anti-symmetric chiral fermion is given by~\cite{Banks:1976yg,Okubo:1977sc,Georgi:1979md}
\beqn\label{eq:rk_anomaly}
{\rm Anom} ( \[ N\,,k \]_{\rep{F}}) &=& \frac{ (N-2k)\, ( N-3)! }{ (N-k-1)!\, (k-1)!} \,.
\eeqn
Each set of anomaly-free fermions are composed of a rank-$k$ (with $2 \leq k\leq \[ \frac{N}{2} \] $) anti-symmetric chiral fermion together with ${\rm Anom} ( \[ N\,,k \]_{\rep{F}})$ copies of anti-fundamental chiral fermions.
We propose that the third law of the flavor unification in GUT should be
\begin{conj}

Simple repetition of any irreducible anomaly-free fermion set in Eq.~\eqref{eq:SUN_fermions_realistic} is not allowed.

\end{conj}
The notion of $\{ k \}^\prime$ in Eq.~\eqref{eq:SUN_fermions_realistic} represents the choices of the rank-$k$ anti-symmetric fermion irreps without repeating itself. 
The conjectured third law thus leads to global DRS symmetries of
\beqn\label{eq:DRS_general}
\Gc_{\rm DRS}&=& \bigotimes_{ \{ k \}^\prime } \Big[  {\rm SU}\Big( {\rm Anom} ( \[ N\,,k \]_{\rep{F}})  \Big)_k \otimes {\rm U}(1)_{k} \Big]  \,.
\eeqn
This is certainly a generalization of the global DRS symmetry in the rank-$2$ anti-symmetric ${\rm SU}(N+4)$ theory~\cite{Dimopoulos:1980hn}.
In the current ${\rm SU}(7)$ toy model, the specific choices of $k$ in Eq.~\eqref{eq:SUN_fermions_realistic} will be required to reproduce $n_g=2$ according to the rules described below.
In the realistic GUT, one should reproduce the observed $n_g=3$ for all SM fermions.

\para
Georgi also gave the rules of counting the SM fermion generations, which read as follows
\begin{itemize}

\item The ${\rm SU}(N)$ fundamental irrep is decomposed under the ${\rm SU}(5)$ as $\[ N\,,1 \]_{\rep{F}}= (N-5) \times \rep{1_F} \oplus \rep{5_F}$. 
The decompositions of other higher-rank irreps can be obtained by tensor products.

\item For an ${\rm SU}(N)$ GUT, one can eventually decompose the set of anomaly-free fermion irreps into the ${\rm SU}(5)$ irreps of $(\rep{1_F}\,, \rep{5_F}\,, \repb{5_F}\,, \rep{10_F}\,, \repb{10_F} )$.

\item Count the multiplicity of each ${\rm SU}(5)$ irrep as $\nu_{\rep{5_F}}$ and so on, and the anomaly-free condition must lead to $\nu_{\rep{5_F}} + \nu_{\rep{10_F}} = \nu_{\repb{5_F}} + \nu_{\repb{10_F}}$.

\item The SM fermion generation is determined by $n_g= \nu_{\repb{5_F}} - \nu_{\rep{5_F}} = \nu_{\rep{10_F}} - \nu_{\repb{10_F}}$.

\end{itemize}
It turns out that the multiplicity difference between $\rep{10_F}$ and $\repb{10_F}$ from a given irrep of $\[ N\,,k \]_{\rep{F}}$ can be expressed as
\beqn\label{eq:gen_rkirrep}
\nu_{\rep{10_F}} \[ N\,,k \]_{\rep{F}} - \nu_{\repb{10_F}} \[ N\,,k \]_{\rep{F}}&=&  \frac{( N-2k ) (N-5)! }{ (k-2)!\, (N -k -2)! }\,, ~ (k\geq 2\,, k\neq N-1) \,.
\eeqn
Obviously for $k=2$, one always has $\nu_{\rep{10_F}} \[N\,,2 \]_{\rep{F}} - \nu_{\repb{10_F}} \[ N\,,2 \]_{\rep{F}} =1$.
To avoid naive replication of generations, it is therefore necessary to consider anti-symmetric irreps with rank greater than or equal to $3$.
For the ${\rm SU}(6)$ case, the irrep of $\[ 6\,, 3 \]_{\rep{F}}=\rep{20_F}$ is self-conjugate~\footnote{More generally, any self-conjugate irrep of the ${\rm SU}(2N)$ GUT cannot contribute to a SM fermion generation at the EW scale.} and leads to $\nu_{\rep{10_F}} \[6\,,3\]_{\rep{F}} - \nu_{\repb{10_F}} \[6\,,3 \]_{\rep{F}}=0$ according to Eq.~\eqref{eq:gen_rkirrep}.
In this regard, the ${\rm SU}(7)$ group is expected to be the leading GUT group that generates multiple generations non-trivially.

\para 
The ${\rm SU}(7)$ group was first considered by Frampton in Ref.~\cite{Frampton:1979cw}, where he suggested the ${\rm SU}(7)$ group could generate multiple SM fermion generations. 
However, we wish to point out this was not true.
To see this, we list two following sets of fermions from Ref.~\cite{Frampton:1979cw}
\beqs\label{eqs:Frampton_SU7} 
\beqn 
\{ f_L \}_{ {\rm SU}(7)}&=& \[ 8 \times \repb{7_F} \] \bigoplus \[ 2 \times \rep{21_F} \] \bigoplus \rep{35_F} \,, ~ {\rm dim}_{ \rep{F}}= 133 \,,\\[1mm]
\{ f_L \}_{ {\rm SU}(7)}^\prime &=& \[ 7 \times \repb{7_F} \] \bigoplus  \rep{21_F} \bigoplus \[ 2\times \rep{35_F} \] \,, ~ {\rm dim}_{ \rep{F}}= 140\,,
\eeqn 
\eeqs 
that can both lead to $n_g=3$ according to Georgi's rule.
However, these fermions become
\beqs\label{eqs:Frampton_SU7_partition} 
\beqn 
\{ f_L \}_{ {\rm SU}(7)}&=& \left\{ 2\times \[ 3 \times \repb{7_F} \oplus  \rep{21_F} \] \right\} \bigoplus \[2 \times \repb{7_F} \oplus \rep{35_F} \]   \,, \\[1mm]
\{ f_L \}_{ {\rm SU}(7)}^\prime &=& \[ 3 \times \repb{7_F} \oplus  \rep{21_F} \] \bigoplus \left\{ 2 \times \[2 \times \repb{7_F} \oplus \rep{35_F} \] \right\} \,,
\eeqn 
\eeqs 
if one partitions the fermions in terms of irreducible anomaly-free sets.
Hence, there are trivial repetitions of one set of anomaly-free fermions for both cases suggested by Frampton.
Through detailed analyses of the ${\rm SU}(7)$ symmetry breaking pattern below, we shall show that the ${\rm SU}(7)$ can never generate realistic SM fermion mass hierarchies for three generations.

\subsection{Possible symmetry breaking patterns}

\para
Gauge symmetry breaking patterns are determined by Higgs representations, which was first studied in Ref.~\cite{Li:1973mq}.
For this purpose, we tabulate the patterns of the symmetry breaking for the ${\rm SU}(N)$ groups with various Higgs representations in table~\ref{tab:HiggsPatterns}.

\begin{table}[htp]
\begin{center}
\begin{tabular}{c|cc}
\hline
 Higgs irrep&  dim & pattern   \\
\hline
fundamental  &  $N$  &  $ {\rm SU}(N-1)$  \\
rank-$2$ symmetric  &  $\frac{1}{2}N(N+1)$  & ${\rm SO}(N)$   \\
  &  & or ${\rm SU} (N-1)$ \\
rank-$2$ anti-symmetric  &  $\frac{1}{2}N(N-1)$  & ${\rm SO}(2k+1)\,, k=[ \frac{N}{2}]$  \\
  &  & or ${\rm SU} (N-2)$ \\
adjoint  &  $N^2 -1$  &  ${\rm SU}(N-k) \otimes {\rm SU}(k)\otimes {\rm U}(1)\,, k=[ \frac{N}{2}]$ \\
 &   &  or $ {\rm SU}(N-1)$  \\
\hline
\end{tabular}
\end{center}
\caption{The Higgs representations and the corresponding symmetry breaking patterns in the ${\rm SU}(N)$ group.}
\label{tab:HiggsPatterns}
\end{table}%

\para
One expects four different symmetry breaking stages for the ${\rm SU}(7)$ model.
The zeroth-stage symmetry breaking occurs at the GUT scale, and one expects a maximally broken pattern of ${\rm SU}(7) \to {\rm SU}(3) \otimes {\rm SU}(4) \otimes {\rm U}(1)$ due to the Higgs VEVs of an adjoint Higgs field of $\rep{48_H}$.
Since this stage leads to massive gauge bosons that can mediate the proton decays in the context of the minimal ${\rm SU}(5)$ GUT, it would be mostly proper to occur at the zeroth stage.
However, one cannot determine whether the ${\rm SU}(4)$ or the ${\rm SU}(3)$ subgroup will describe the strong or weak sector.
In the current discussion, we consider the following symmetry breaking pattern of the ${\rm SU}(7)$
\beqn\label{eq:SU7_pattern}
&& {\rm SU}(7)\xrightarrow{ \Lambda_{\rm GUT} } \Gc_{341} \xrightarrow{ v_{341} } \Gc_{331} \xrightarrow{ v_{331} }  \Gc_{\rm SM} \xrightarrow{ v_{\rm EW} }  \Gc_{\rm SM}^\prime  \,, \non
&& \Gc_{341} \equiv {\rm SU}(3)_c \otimes {\rm SU}(4)_W \otimes {\rm U}(1)_{X_0} \,,~   \Gc_{331} \equiv  {\rm SU}(3)_c \otimes {\rm SU}(3)_W \otimes {\rm U}(1)_{X_1} \,, \non
&&  \Gc_{\rm SM} \equiv {\rm SU}(3)_c \otimes {\rm SU}(2)_W \otimes {\rm U}(1)_{Y} \,, ~  \Gc_{\rm SM}^\prime \equiv {\rm SU}(3)_c \otimes  {\rm U}(1)_{\rm EM} \,, \non 
&& \Lambda_{\rm GUT} \gg v_{341} \gg v_{331} \gg v_{\rm EW} \,.
\eeqn
This was previously considered in Refs.~\cite{Claudson:1980dv,Cox:1980er}.
An alternative zeroth-stage symmetry breaking pattern of ${\rm SU}(7)\to {\rm SU}(4)_s \otimes {\rm SU}(3)_W \otimes {\rm U}(1)_{X_0}$ was previously discussed in Refs.~\cite{Kim:1980ci,Umemura:1980sa,Gao:1980up,Xue:1981ek,Kim:1981be}.
We also wish to mention that such an ambiguity no longer arises when one considers the GUT groups of ${\rm SU}(8)$~\cite{Barr:1979xt,Ma:1981pr,Barr:2008pn} and ${\rm SU}(9)$~\cite{Frampton:1979tj}, where non-trivial embedding of three-generational SM fermions can be achieved.
Given the ${\rm SU}(7)$ symmetry breaking pattern in Eq.~\eqref{eq:SU7_pattern}, we define the decomposition rules and charge quantizations in App.~\ref{section:Br}.
Above the EWSB scale, the effective theory is described by a 331 model, which was previously studied in various aspects~\cite{Lee:1977qs,Lee:1977tx,Singer:1980sw,Pisano:1992bxx,Frampton:1992wt,Foot:1992rh,Montero:1992jk,Ng:1992st,Liu:1993gy,Pal:1994ba,Long:1995ctv,Tonasse:1996cx,Ponce:2002sg,Dias:2003zt,Dias:2004dc,Ferreira:2011hm,Dong:2012bf,Buras:2012dp,Machado:2013jca,Buras:2013dea,Boucenna:2014ela,Boucenna:2014dia,Boucenna:2015zwa,Deppisch:2016jzl,Cao:2016uur,Li:2019qxy,CarcamoHernandez:2021tlv,Chen:2021haa,Buras:2021rdg,Hernandez:2021zje,Chen:2021zwn,Alves:2022hcp,Cherchiglia:2022zfy}.

\subsection{The fermion content}

\para
The gauge anomaly factors of several leading ${\rm SU}(7)$ irreps are listed below
\beqn
&& {\rm Anom}(  \rep{7_F}  ) =+1 \,,~  {\rm Anom}( \rep{ 21_F} ) =+3 \,, ~  {\rm Anom}( \rep{35_F } ) =+2 \,,
\eeqn
according to Eq.~\eqref{eq:rk_anomaly}.
Any anti-symmetric irreps with higher ranks of the ${\rm SU}(7)$ are their conjugates.
By decomposing the ${\rm SU}(7)$ fermions in terms of the ${\rm SU}(5)$ irreps, we have
\beqn
\rep{ 7_F}&=& 2\times \[ 5\,,0 \]_{ \rep{F}} \oplus \[ 5\,, 1 \]_{ \rep{F}}  \,,\non
\rep{ 21_F}&=&   \[ 5\,,0 \]_{ \rep{F}} \oplus 2\times \[ 5\,, 1 \]_{ \rep{F}}  \oplus  \[ 5\,, 2 \]_{ \rep{F}}  \,,\non
\rep{ 35_F}&=&   \[ 5\,, 1 \]_{ \rep{F}}  \oplus  2\times \[5\,, 2 \]_{ \rep{F}} \oplus \[ 5\,, 3 \]_{ \rep{F}} \,.
\eeqn
According to Ref.~\cite{Georgi:1979md}, one identifies one generational left-handed quarks from the $\rep{ 21_F}$, and $2-1$ generational left-handed quarks from the $\rep{35_F}$.
The two-generational ${\rm SU}(7)$ model contains the following fermions
\beqn\label{eq:SU7_2gen_fermions}
\{ f_L \}_{ {\rm SU}(7)}^{n_g=2}&=& \Big[ 3\times \repb{7_F} \oplus \rep{21_F} \Big] \bigoplus \Big[ 2\times \repb{7_F} \oplus \rep{35_F} \Big] \,,~ {\rm dim}_{ \mathbf{F}}= 91\,.
\eeqn
In other words, the fermions can be viewed by joining a rank-$2$ ${\rm SU}(7)$ model with a rank-$3$ ${\rm SU}(7)$ model.
The corresponding global DRS symmetries are given by
\beqn\label{eq:DRS_SU7_2gen}
\Gc_{\rm DRS} \[{\rm SU}(7)\,, n_g=2 \]&=&\Big[  {\rm SU}(3)_2 \otimes {\rm U}(1)_2 \Big]  \bigotimes \Big[ {\rm SU}(2)_3 \otimes {\rm U}(1)_3  \Big] \,,
\eeqn
according to Eq.~\eqref{eq:DRS_general}.

\para
Given the global DRS symmetries in Eq.~\eqref{eq:DRS_SU7_2gen}, we label the flavor indices as follows
\beqn\label{eq:2gen_indices}
&&\Lambda\equiv ( \lambda\,, \dot \lambda )\,,~  \lambda \equiv ( {\rm I}\,,{\rm II}\,,3 ) \,,~ \dot \lambda \equiv ( \dot {\rm I}\,, \dot 2)\,,
\eeqn
where the undotted and dotted indices are used to distinguish the ${\rm SU}(3)_2$ flavors and the ${\rm SU}(2)_3$ flavors.
Throughout the context, the Roman numbers and the Arabic numbers are used for the heavy fermion flavors and the SM fermion flavors, respectively.
Fields that are contracted by ${\rm SU}(3)_2$-invariant and/or ${\rm SU}(2)_3$-invariant $\epsilon$-tensors, as well as their possible combinations are dubbed the DRS-singlets.
Fields or their combinations carrying the ${\rm U}(1)_2$ and/or ${\rm U}(1)_3$ charges are dubbed the DRS-charged states.
Note that the DRS-singlets may not be gauge-invariant in general.
The DRS-invariant terms are both DRS-singlets and DRS-neutral states.

\begin{table}[htp]
\begin{center}
\begin{tabular}{c|c|c|c}
\hline \hline
   $\gSU(7)$   &  $\Gc_{341}$  & $\Gc_{331}$  &  $\Gc_{\rm SM}$  \\
\hline \hline
 $\repb{ 7_F}^\Lambda$   & $( \repb{3} \,, \rep{1}\,,  +\frac{1}{3})_{ \mathbf{F} }^\Lambda$  &  $(\repb{3} \,, \rep{1} \,, +\frac{1}{3} )_{ \mathbf{F} }^\Lambda$  &  $( \repb{3} \,, \rep{ 1}  \,, +\frac{1}{3} )_{ \mathbf{F} }^\Lambda~:~ (\Dc_R^\Lambda)^c$  \\[1mm]
 & $(\rep{1}\,, \repb{4}  \,,  -\frac{1}{4})_{ \mathbf{F} }^\Lambda$  &  $( \rep{1} \,, \repb{3} \,,  -\frac{1}{3})_{ \mathbf{F} }^{\Lambda}$  &  $( \rep{1} \,, \repb{2} \,,  -\frac{1}{2})_{ \mathbf{F} }^\Lambda~:~ ( \Ec_L^\Lambda \,, - \Nc_L^\Lambda )^T$   \\
 &   &    &  $( \rep{1} \,, \rep{1} \,,  0)_{ \mathbf{F} }^\Lambda~:~ \check \Nc_L^\Lambda$  \\
  &   &  $( \rep{1} \,, \rep{1} \,, 0)_{ \mathbf{F} }^{\Lambda^\prime }$ &   $( \rep{1} \,, \rep{1} \,, 0)_{ \mathbf{F} }^{\Lambda^\prime}~:~ \check \Nc_L^{\Lambda^\prime } $   \\ 
\hline\hline
\end{tabular}
\end{center}
\caption{
The $\gSU(7)$ fermion representation of $\repb{7_F}^\Lambda$ under the $\Gc_{341}\,,\Gc_{331}\,, \Gc_{\rm SM}$ subgroups for the ${\rm SU}(7)$ model, with $\Lambda=(\lambda\,, \dot \lambda)$, $\lambda=({\rm I}\,,{\rm II}\,,3)$, and $\dot \lambda = ( \dot {\rm I}\,, \dot 2)$.
Here, we denote $\underline{ (\Dc_R^\Lambda)^c=(d_R^\Lambda)^c}$ for the SM right-handed down-type quarks, and $(\Dc_R^\Lambda)^c=(\DG_R^\Lambda)^c$ for the heavy right-handed down-type partner quarks.
Similarly, we denote $\underline{ ( \Ec_L^\Lambda \,, - \Nc_L^\Lambda )^T = ( e_L^\Lambda \,, - \nu_L^\Lambda)^T}$ for the SM left-handed lepton doublets, and $( \Ec_L^\Lambda \,, - \Nc_L^\Lambda )^T =( \eG_L^\Lambda \,, - \nG_L^\Lambda )^T$ for the heavy left-handed lepton doublets.
}
\label{tab:SU7_2gen_7barferm}
\end{table}%

\begin{table}[htp]
\begin{center}
\begin{tabular}{c|c|c|c}
\hline \hline
   $\gSU(7)$   &  $\Gc_{341}$  & $\Gc_{331}$  &  $\Gc_{\rm SM}$  \\
\hline \hline
 $\rep{21_F}$   & $( \repb{3}\,, \rep{ 1} \,, - \frac{2}{3})_{ \mathbf{F}}$ & $( \repb{3} \,, \rep{1}  \,, - \frac{2}{3} )_{ \mathbf{F} }$   &  $\underline{ ( \repb{3} \,, \rep{1}  \,, - \frac{2}{3} )_{ \mathbf{F} }~:~{t_R}^c}$ \\[1mm]
  & $( \rep{ 3} \,, \rep{4} \,, - \frac{1}{12})_{ \mathbf{F}}$  &  $( \rep{ 3} \,, \rep{3} \,, 0)_{ \mathbf{F}}$ &  $\underline{ ( \rep{ 3} \,, \rep{2} \,, +\frac{1}{6} )_{ \mathbf{F}}~:~ (t_L \,, b_L)^T }$ \\
  &   &    &   $( \rep{ 3} \,, \rep{1} \,, -\frac{1}{3} )_{ \mathbf{F}}~:~\DG_L$   \\
  &    & $(\rep{3} \,, \rep{1} \,, -\frac{1}{3})_{ \mathbf{F}}^\prime$  &  $(\rep{3} \,, \rep{1} \,, -\frac{1}{3})_{ \mathbf{F}}^\prime~:~\DG_L^\prime$    \\[1mm]
  & $( \rep{ 1} \,, \rep{6} \,, +\frac{1}{2})_{ \mathbf{F} }$  &  $(\rep{1} \,, \repb{3} \,, +\frac{2}{3})_{ \mathbf{F} }$   &  $(\rep{ 1} \,, \repb{2} \,, +\frac{1}{2} )_{ \mathbf{F} }~:~ ( {\nG_R}^c\,, -{\eG_R}^c)^T$   \\
  &    &    &  $\underline{ (\rep{ 1} \,, \rep{1} \,, +1 )_{ \mathbf{F} }~:~ {\tau_R}^c}$ \\
  &   &  $(\rep{ 1} \,, \rep{3} \,, +\frac{1}{3} )_{ \mathbf{F} }$  &  $(\rep{ 1} \,, \rep{2} \,, +\frac{1}{2} )_{ \mathbf{F} }^\prime~:~ ( {\eG_R^\prime}^c\,, {\nG_R^\prime}^c)^T$ \\
  &  &  & $(\rep{ 1} \,, \rep{1} \,, 0 )_{ \mathbf{F} }~:~\check \nG_R^c$  \\ 
\hline\hline
\end{tabular}
\end{center}
\caption{
The $\gSU(7)$ fermion representation of $\rep{21_F}$ under the $\Gc_{341}\,,\Gc_{331}\,, \Gc_{\rm SM}$ subgroups for the ${\rm SU}(7)$ model.
All SM fermion names and irreps are identified with underlines.}
\label{tab:SU7_2gen_21ferm}
\end{table}%

\begin{table}[htp]
\begin{center}
\begin{tabular}{c|c|c|c}
\hline \hline
   $\gSU(7)$   &  $\Gc_{341}$  & $\Gc_{331}$  &  $\Gc_{\rm SM}$  \\
\hline \hline
 $\rep{35_F}$  &  $( \rep{1}\,, \rep{1} \,, -1)_{ \mathbf{F}}$  & $( \rep{1}\,, \rep{1} \,, -1)_{ \mathbf{F}}$  & $( \rep{1}\,, \rep{1} \,, -1)_{ \mathbf{F}}~:~\EG_L$  \\[1mm]
                            &  $( \rep{ 1}\,, \repb{4} \,, +\frac{3}{4})_{ \mathbf{F}}$  & $( \rep{1}\,, \repb{3} \,, +\frac{2}{3})_{ \mathbf{F}}^{\prime\prime}$  & $(  \rep{1} \,, \repb{2} \,, +\frac{1}{2} )_{ \mathbf{F}}^{\prime \prime}~:~ ( {\nG_R^{\prime\prime} }^c\,, -{\eG_R^{\prime\prime} }^c)^T$  \\ 
                            &   &    &  $\underline{ (  \rep{1} \,, \rep{1} \,, +1)_{ \mathbf{F}}^{\prime\prime}~:~{\mu_R}^c }$   \\
                            &  & $( \rep{1}\,, \rep{1} \,, +1)_{ \mathbf{F}}^\prime$  & $( \rep{1}\,, \rep{1} \,, +1)_{ \mathbf{F}}^{\prime}~:~{\EG_R}^c$  \\[1mm]
                            &  $( \repb{ 3}\,, \rep{ 4} \,, -\frac{5}{12})_{ \mathbf{F}}$  & $( \repb{3}\,, \rep{3} \,, -\frac{1}{3} )_{ \mathbf{F}}$  & $( \repb{3}\,, \rep{2} \,, -\frac{1}{6} )_{ \mathbf{F}}~:~ ( {\sG_R}^c \,, {\cG_R}^c )^T$  \\
                            &  &  & $( \repb{3}\,, \rep{1} \,, -\frac{2}{3} )_{ \mathbf{F}}^\prime~:~{\UG_R}^c$  \\
                            &  &  $( \repb{3}\,, \rep{1} \,, -\frac{2}{3} )_{ \mathbf{F}}^{\prime\prime}$ &  $\underline{ ( \repb{3}\,, \rep{1} \,, -\frac{2}{3} )_{ \mathbf{F}}^{\prime\prime}~:~{c_R}^c}$ \\[1mm]
                            &  $( \rep{ 3}\,, \rep{6} \,, +\frac{1}{6})_{ \mathbf{F}}$  &  $( \rep{3}\,, \rep{3} \,, 0)_{ \mathbf{F}}^\prime$  &  $\underline{ ( \rep{3} \,,\rep{2} \,,+\frac{1}{6} )_{ \mathbf{F}}^{\prime}~:~(c_L \,, s_L)^T}$   \\ 
                            & & &  $( \rep{3} \,,\rep{1} \,,-\frac{1}{3} )_{ \mathbf{F}}^{\prime\prime}~:~\SG_L$   \\ 
                            &  & $( \rep{3}\,, \repb{3} \,, +\frac{1}{3})_{ \mathbf{F}}$ & $( \rep{3} \,, \repb{2} \,, +\frac{1}{6} )_{ \mathbf{F}}^{\prime\prime}~:~ (\sG_L \,, - \cG_L )^T$ \\ 
                            & & & $( \rep{3} \,, \rep{1} \,, +\frac{2}{3})_{ \mathbf{F}}~:~\UG_L$   \\
\hline\hline
\end{tabular}
\end{center}
\caption{
The $\gSU(7)$ fermion representation of $\rep{35_F}$ under the $\Gc_{341}\,,\Gc_{331}\,, \Gc_{\rm SM}$ subgroups for the ${\rm SU}(7)$ model.
All SM fermion names and irreps are identified with underlines.
}
\label{tab:SU7_2gen_35ferm}
\end{table}%

\para
We tabulate the ${\rm SU}(7)$ fermion spectrum in Tabs.~\ref{tab:SU7_2gen_7barferm}, \ref{tab:SU7_2gen_21ferm}, and \ref{tab:SU7_2gen_35ferm}. 
The ${\rm U}(1)_{X_0\,, X_1\,,Y}$ charges are obtained according to the assignments given in Eqs.~\eqref{eq:X1charge} and~\eqref{eq:Ycharge}.
From the decomposition, we find one quark doublet of $(\rep{3}\,, \rep{2}\,, +\frac{1}{6} )_{ \mathbf{F}}$ from $\rep{21_F}$, two quark doublets of $(\rep{3}\,, \rep{2}\,, +\frac{1}{6} )_{ \mathbf{F}}^\prime$ and $(\rep{3}\,, \repb{2}\,, +\frac{1}{6} )_{ \mathbf{F}}^{\prime\prime}$,  plus one mirror quark doublet of $(\repb{3}\,, \rep{2}\,, -\frac{1}{6} )_{ \mathbf{F}}$ from $\rep{35_F}$.
The existence of the mirror quark doublet~\cite{Maalampi:1988va} is a distinctive feature from the previous one-generational ${\rm SU}(6)$ toy model~\cite{Chen:2021zwn}, and they can emerge from {\it non-minimal} GUTs in general.
According to the counting rule by Georgi~\cite{Georgi:1979md}, it is straightforward to find $n_g=2$ in the current setup.
All chiral fermions are named by their SM irreps.
For the right-handed quarks of $(\Dc_R^\Lambda)^c$, they are named as follows
\beqn\label{eq:rightD_names}
&& (\Dc_R^{{\rm I}} )^c \equiv {\DG_R}^c \,, ~  (\Dc_R^{{\rm II}} )^c \equiv {\DG_R^\prime}^c \,,~ (\Dc_R^{3} )^c \equiv {b_R}^c \,, \non
&& (\Dc_R^{ \dot {\rm I}} )^c \equiv {\SG_R}^c \,, ~ (\Dc_R^{  \dot 2} )^c \equiv {s_R}^c \,.
\eeqn
For the left-handed lepton doublets of $(\Ec_L^\Lambda \,, -\Nc_L^\Lambda )$, they are named as follows
\beqn\label{eq:leftlep_names}
&& ( \Ec_L^{\rm I} \,, \Nc_L^{\rm I} )  \equiv (\eG_L \,, -\nG_L) \,, ~ 
( \Ec_L^{ {\rm II}} \,, \Nc_L^{ {\rm II}} )  \equiv (\eG_L^\prime \,, -\nG_L^\prime ) \,, ~
( \Ec_L^{ \dot {\rm I}} \,, \Nc_L^{\dot {\rm I}} )  \equiv (\eG_L^{\prime\prime} \,, -\nG_L^{\prime\prime} ) \,,\non
&&  ( \Ec_L^{ \dot 2} \,,  - \Nc_L^{  \dot 2}) \equiv ( \mu_L  \,,  -\nu_{\mu\,L} )\,, ~( \Ec_L^{ 3 } \,,  - \Nc_L^3)  \equiv ( \tau_L \,, - \nu_{\tau\, L} )  \,.
\eeqn
All fermions are categorized according to their electrical charges in App.~\ref{section:name}.
Names of all SM fermions will become transparent with the analyses of their mass origins from the symmetry breaking pattern in Sec.~\ref{section:breaking}.

\section{Review of the third-generational ${\rm SU}(6)$ toy model}
\label{section:SU6}

\subsection{The ${\rm SU}(6)$ model setup and the DRS symmetry}

\para
Before presenting the fermion masses in the two-generational ${\rm SU}(7)$ model, we review some necessary ingredients from the third-generational ${\rm SU}(6)$ toy model~\cite{Chen:2021zwn}.
This was recently shown to split the bottom quark and tau lepton masses from the top quark mass in the third generation with natural Yukawa couplings of $\sim \Oc(1)$.
The minimal set of anomaly-free fermion content in the ${\rm SU}(6)$ model is given by
\beqn\label{eq:SU6_fermions}
\{ f_L \}_{ {\rm SU}(6)}^{n_g=1}&=& 2\times \repb{6_F} \oplus \rep{15_F}   \,,~ {\rm dim}_{ \mathbf{F}}= 27\,.
\eeqn
This model enjoys a global DRS symmetry of
\beqn\label{eq:DRS_SU6}
\Gc_{\rm DRS} \[{\rm SU}(6) \,, n_g=1 \]&=&  {\rm SU}(2)_F \otimes {\rm U}(1)  \,,
\eeqn
according to Eq.~\eqref{eq:DRS_general}.
The model setup can be summarized in Tab.~\ref{tab:SU6_setup}.
The symmetry breaking pattern of the ${\rm SU}(6)$ model simply reads $ {\rm SU}(6)\xrightarrow{ \Lambda_{\rm GUT} }  \Gc_{331} \xrightarrow{ v_{331} }  \Gc_{\rm SM} \xrightarrow{ v_{\rm EW} }  \Gc_{\rm SM}^\prime$.

\begin{table}[htp]
\begin{center}
\begin{tabular}{c|cc}
\hline \hline
${\rm SU}(6)$   &  ${\rm SU}(2)_F$  &  ${\rm U}(1)$    \\
\hline
$\repb{6_F}^{\lambda}$    & $\rep{2}$   &  $p$   \\ 
$\rep{15_F}$     & $\rep{1}$  & $q$  \\
\hline
$\repb{6_H}_{\,,\lambda}$   &  $\repb{2}$   & $-p -q$   \\
$\rep{15_H}$   & $\rep{1}$     &  $-2 q$  \\
$\rep{35_H}$   & $\rep{1}$     &  $0$  \\
\hline\hline
\end{tabular}
\end{center}
\caption{
The ${\rm SU}(6)$ fermions and Higgs representations under the global DRS symmetries in Eq.~\eqref{eq:DRS_SU6}.
}
\label{tab:SU6_setup}
\end{table}%

\para
In the Higgs sector of the ${\rm SU}(6)$ model, the Higgs components that can develop VEVs for the sequential stages of symmetry breaking after the GUT-scale symmetry breaking are
\beqs\label{eqs:SU6_Higgs_Br}
\beqn
\repb{6_H}_{\,,\lambda}&\supset&   \boxed{ \overbrace{(  \rep{1}\,, \repb{3}\,, -\frac{1}{3} )_{\rep{H}\,,\lambda}}^{  \Phi_{\repb{3}\,,\lambda}} } \supset  \boxed{ (  \rep{1}\,, \repb{2}\,, -\frac{1}{2} )_{\rep{H}\,,\lambda} }   \,, \label{eq:SU6_Higgs_Br01} \\[1mm]
\rep{15_H}&\supset& \underline{ \overbrace{(  \rep{1}\,, \repb{3}\,, +\frac{2}{3} )_{\rep{H}}}^{ \Phi_{\repb{3}}^\prime } } \supset \boxed{ (  \rep{1}\,, \repb{2}\,, +\frac{1}{2} )_{\rep{H}} }\,, \label{eq:SU6_Higgs_Br02}
\eeqn
\eeqs
with $\lambda = 1\,,2$ being the ${\rm SU}(2)_F$ indices according to Eq.~\eqref{eq:DRS_SU6}.
All Higgs components that can develop VEVs for the symmetry breaking are framed with boxes.
The Yukawa couplings contain the terms of
\beqn\label{eq:SU6Yukawa_SU2}
&& {(Y_\Dc)_\lambda}^\kappa  \repb{6_F}^\lambda \rep{15_F} \repb{6_H}_{\,, \kappa }  + H.c. \non
&\supset&{(Y_\Dc)_\lambda}^\kappa \Big[ ( \mathbf{3}\,, \mathbf{ 3}\,, 0 )_{ \mathbf{F}}  \otimes  ( \mathbf{\bar 3}\,, \mathbf{ 1}\,, +\frac{1}{3} )_{ \mathbf{F}}^{ \lambda}  \oplus (  \mathbf{ 1}\,, \mathbf{\bar 3}\,, +\frac{2}{3} )_{ \mathbf{F}}   \otimes (  \mathbf{ 1}\,, \mathbf{\bar 3}\,, -\frac{1}{3} )_{ \mathbf{F}}^{\lambda} \Big] \non
&\otimes&  ( \mathbf{1}\,, \mathbf{ \bar 3}\,,  -\frac{1}{3} )_{ \mathbf{H}\,, \kappa }  + H.c.   \,.
\eeqn
These Yukawa couplings lead to massive vectorlike fermions from one copy of $\repb{6_F}^\lambda$, and will be integrated out after the $\Gc_{331}$ breaking.
Accordingly, at least one of the $ ( \mathbf{1}\,, \mathbf{ \bar 3}\,,  -\frac{1}{3} )_{ \mathbf{H}\,, \lambda } \subset \repb{6_H}_{\,, \lambda}$ should develop the VEV.
We dub this the ``fermion-Higgs matching pattern'' for the symmetry breaking.
When both of the $ ( \mathbf{1}\,, \mathbf{ \bar 3}\,,  -\frac{1}{3} )_{ \mathbf{H}\,, \lambda } \subset \repb{6_H}_{\,, \lambda}$ develop VEVs for the $\Gc_{331}$ breaking, there is still one copy of $\repb{6_F}^\lambda$ become massive at this stage.
We dub this the ``fermion-Higgs mismatching pattern'' for the symmetry breaking~\cite{Chen:2021zwn}.
Correspondingly, one naturally denote the VEVs of two $( \mathbf{1}\,, \mathbf{ \bar 3}\,,  -\frac{1}{3} )_{ \mathbf{H}\,, \lambda }$ as
\beqn\label{eq:SU6_3barHiggs_VEV01}
\langle  \Phi_{ \repb{3}\,, \lambda }  \rangle &=& \frac{1}{\sqrt{2}} \left( \ba{c}   0   \\  0  \\ V_{\repb{3}\,, \lambda } \ea  \right)   \,.
\eeqn

\subsection{The Higgs potential and the $\nu$-term}

\para
For the ${\rm SU}(6)$ toy model, one can write down the following Higgs potential
\beqs\label{eqs:2HTM_potential}
\beqn
V_{\Gc_{331}}&=& V( \Phi_{ \repb{3}\,, \lambda } ) + V( \Phi_{ \repb{3}}^\prime ) + V( \Phi_{ \repb{3}\,, \lambda }\,,\Phi_{ \repb{3}}^\prime )   \,,\\[1mm]
V( \Phi_{ \repb{3}\,, \lambda } )&=& \mu_{11}^2 | \Phi_{ \repb{3}\,, 1 } |^2 + \mu_{22}^2 | \Phi_{ \repb{3}\,, 2 } |^2 - \Big( \mu_{12}^2 \Phi_{ \repb{3}\,, 1 }^\dag \Phi_{ \repb{3}\,, 2 } + H.c.  \Big) \non
&+& \frac{ \lambda_1}{2} | \Phi_{ \repb{3}\,, 1 }  |^4 + \frac{ \lambda_2}{2} | \Phi_{ \repb{3}\,, 2 }  |^4 + \lambda_3 | \Phi_{ \repb{3}\,, 1}|^2 |\Phi_{ \repb{3}\,, 2 }|^2  \,,\label{eq:2HTM_potential01} \\[1mm]
V( \Phi_{ \repb{3}}^\prime )&=&  \mu_t^2 | \Phi_{ \repb{3}}^\prime |^2 + \lambda_t | \Phi_{ \repb{3}}^\prime |^4  \,,\label{eq:2HTM_potential02} \\[1mm]
V( \Phi_{ \repb{3}\,, \lambda}\,,\Phi_{ \repb{3}}^\prime ) &=&  \kappa_\lambda |\Phi_{ \repb{3}\,,\lambda} |^2  |  \Phi_{ \repb{3}}^\prime |^2  + \Big(  \nu \epsilon^{\tilde i  \tilde j  \tilde k } (\Phi_{ \repb{3}\,, 1})_{\tilde i} ( \Phi_{ \repb{3}\,, 2} )_{\tilde j}  (\Phi_{ \repb{3}}^\prime)_{\tilde k}  + H.c. \Big)  \,. \label{eq:2HTM_potential03}
\eeqn
\eeqs
after the symmetry breaking of ${\rm SU}(6)\to \Gc_{331}$.
Obviously, the $V( \Phi_{ \repb{3}\,, \lambda } )$ mainly describes the $\Gc_{331}$ symmetry breaking, and the $V( \Phi_{ \repb{3}}^\prime )$ mainly describes the EWSB.
The global DRS symmetry in Eq.~\eqref{eq:DRS_SU6} can be restored when $\mu_{11}^2 = \mu_{22}^2$, $\mu_{12}^2=0$, $\lambda_1 = \lambda_2 = \lambda_3$, and $\kappa_1 = \kappa_2$.

\para
It turns out the ${\rm SU}(6)$ Higgs sector can naturally include a gauge-invariant mixing $\nu$-term of
\beqn\label{eq:SU6_nuterm}
V_{\Gc_{331}}&\supset& \nu \epsilon^{\lambda \kappa}\,  \repb{6_H}_{\,,\lambda} \repb{6_H}_{\,,\kappa} \rep{15_H} + H.c. \non 
&\supset& \nu  \epsilon^{\tilde i \tilde j \tilde k } \, (  \rep{1}\,, \repb{3}\,, -\frac{1}{3} )_{\rep{H}\,,\tilde i \, \lambda} \otimes (  \rep{1}\,, \repb{3}\,, -\frac{1}{3} )_{\rep{H}\,,\tilde j\, \kappa} \otimes (  \rep{1}\,, \repb{3}\,, +\frac{2}{3} )_{\rep{H}\,, \tilde k}  + H.c. \,.
\eeqn
According to Tab.~\ref{tab:notations}, $(\tilde i \,, \tilde j\,, \tilde k)$ represent the fundamental/anti-fundamental indices of the ${\rm SU}(3)_W$ group.
This $\nu$-term can also be DRS-invariant as long as $p=-2q$ according to Tab.~\ref{tab:SU6_setup}.
The $\Gc_{\rm SM}$-singlet terms that can develop the $\Gc_{331}$ breaking VEVs correspond to the $\tilde i =3$ components of two $(  \rep{1}\,, \repb{3}\,, -\frac{1}{3} )_{\rep{H}\,,\lambda}$.
Meanwhile, the $(  \rep{1}\,, \repb{3}\,, +\frac{2}{3} )_{\rep{H}}$ only contains the EWSB component with $\tilde k=1$.
Therefore, the above $\nu$-term can lead to a following tadpole term in the Higgs potential
\beqn\label{eq:SU6_tadpole}
&\sim& -  \nu \epsilon^{\lambda \kappa} \, V_{\repb{3}\,, \lambda} v_t  (  \rep{1}\,, \repb{3}\,, -\frac{1}{3} )_{\rep{H}\,, 2\, \kappa}   + H.c. \,,
\eeqn
where we have denoted $\langle (  \rep{1}\,, \repb{3}\,, -\frac{1}{3} )_{\rep{H}\,, 3\, \lambda} \rangle \equiv \frac{1}{ \sqrt{2}} (0 \,, 0 \,, V_{\repb{3}\,, \lambda})^T$ and $\langle (  \rep{1}\,, \repb{3}\,, +\frac{2}{3} )_{\rep{H}\,, 1}  \rangle \equiv \frac{1}{ \sqrt{2}} ( v_t \,, 0 \,,0)^T$.
To remove this tadpole term, it is necessary to have the $\tilde j=2$ component in the $(  \rep{1}\,, \repb{3}\,, -\frac{1}{3} )_{\rep{H}\,,\tilde j\, \kappa}$ develop a VEV for the EWSB as well.
Thus, two VEVs of $\Phi_{ \repb{3}\,, \lambda }$ in Eq.~\eqref{eq:SU6_3barHiggs_VEV01} should be modified into
\beqn\label{eq:SU6_3barHiggs_VEV02}
\langle  \Phi_{ \repb{3}\,, \lambda }  \rangle &=& \frac{1}{\sqrt{2}} \left( \ba{c}   0   \\  u_{\repb{2}\,, \lambda }  \\ V_{\repb{3}\,, \lambda } \ea  \right)   \,.
\eeqn
We parametrize the $\Gc_{331}$-breaking and the EWSB VEVs as follows~\footnote{Throughout the context, we always use the short-handed notations of $(c_{ \tilde \beta} \,, s_{\tilde \beta} ) \equiv ( \cos\tilde \beta\,, \sin\tilde \beta)$.}
\beqn\label{eq:SU6VEVs_relation}
&& ( V_{\repb{3}\,, 1 } \,, V_{\repb{3}\,, 2})=( c_{\tilde \beta}\,, s_{\tilde \beta}) v_{331} \,, ~ ( u_{\repb{2}\,, 1 } \,, u_{\repb{2}\,, 2})=( c_{ \beta^\prime}\,, s_{ \beta^\prime}) u_\phi \,.
\eeqn 
The EWSB VEV of $u_\phi$ will give masses to the bottom quark and the tau lepton.
Hence, a hierarchy of $u_\phi \ll v_{331}$ is expected.
One can further demand no mass mixing term between the $W^\pm$ and $W^{\prime\, \pm}$.
To see this, we write down the charged gauge boson masses from the ${\rm SU}(3)_W \otimes {\rm U}(1)_{X_1}$ covariant derivative in Eq.~\eqref{eq:331_connection_gauge} by using the VEVs in Eq.~\eqref{eq:SU6_3barHiggs_VEV02}, and they read
\beqn\label{eq:331_chargedWmass}
&& \frac{1}{4} g_{3W}^2 ( W_\mu^-\,, W_\mu^{\prime\, -}) \cdot \left( \ba{cc}
 v_{\rm EW}^2 &  u_{ \repb{2}\,, \lambda} V_{ \repb{3}\,,\lambda} \\
 u_{ \repb{2}\,, \lambda} V_{ \repb{3}\,,\lambda}  &   v_{331}^2 + v_t^2  \\  \ea \right)  \cdot \left( \ba{c}  W^{+\,\mu}  \\  W^{\prime\,+\,\mu}  \\     \ea \right) \,.
\eeqn
It is straightforward to obtain the following orthogonal relation of
\beqn
&& \sum_{\lambda=1\,,2} \, u_{\repb{2}\,, \lambda }  V_{\repb{3}\,, \lambda } =0\,,
\eeqn
through the gauge transformations to $\Phi_{ \repb{3}\,, \lambda }$, which assures the absence of mass mixing between the $W^\pm$ and $W^{\prime\, \pm}$.
The VEV ratios in Eq.~\eqref{eq:SU6VEVs_relation} are related as $\beta^\prime = \tilde \beta - \pi/2$.
Thus, one can perform the following orthogonal transformations into the Higgs basis
\beqn
\left(   \ba{c} \Phi_{\repb{3}\,,1}^\prime  \\   \Phi_{\repb{3}\,,2}^\prime  \ea  \right) &=& \left( \ba{cc}
 c_{\tilde \beta} &  s_{ \tilde \beta}  \\
 -s_{ \tilde \beta} & c_{\tilde \beta}  \\  \ea \right) \cdot \left(   \ba{c} \Phi_{\repb{3}\,,1} \\   \Phi_{\repb{3}\,,2} \ea  \right)  \,,
\eeqn
such that
\beqn\label{eq:SU6_3barHiggs_VEV02b}
&& \langle  \Phi_{ \repb{3}\,, 1 }^\prime  \rangle = \frac{1}{\sqrt{2}} \left( \ba{c}   0   \\  0  \\ v_{331 } \ea  \right) \,, ~ \langle  \Phi_{ \repb{3}\,, 2 }^\prime  \rangle = \frac{1}{\sqrt{2}} \left( \ba{c}   0   \\  - u_{\phi }  \\  0 \ea  \right)   \,.
\eeqn

\para
The $\nu$-term in Eq.~\eqref{eq:SU6_tadpole} will contribute to a VEV term of $\frac{1}{\sqrt{2}} \nu ( u_2 V_1 - u_1 V_2) v_t$ to the Higgs potential.
The minimization of the Higgs potential in Eqs.~\eqref{eqs:2HTM_potential} leads to
\beqs\label{eqs:2HTM_min_mod}
\beqn
\frac{\partial V}{\partial V_{\repb{3}\,, 1 }} =0 &\Rightarrow& \mu_{11}^2 = \mu_{12}^2 t_{\tilde \beta} - \frac{\lambda_1}{2}(  V_{\repb{3}\,, 1 }^2 +  u_{\repb{2}\,, 1 }^2 )  - \frac{ \lambda_{3}}{2} ( V_{\repb{3}\,, 2}^2 +  u_{\repb{2}\,, 2}^2 ) -  \frac{  \kappa_{1}}{2}   v_t^2 + \frac{ \nu u_{\repb{2}\,, 2 } v_t}{  \sqrt{2} V_{\repb{3}\,, 1 }}  \,,\label{eq:2HTM_min_mod01} \\[1mm]
\frac{\partial V}{\partial V_{\repb{3}\,, 2 }} =0 &\Rightarrow& \mu_{22}^2 = \frac{ \mu_{12}^2}{ t_{\tilde\beta} } - \frac{ \lambda_2}{2}( V_{\repb{3}\,, 2 }^2 + u_{\repb{2}\,, 2}^2 ) -\frac{ \lambda_{3}}{2} ( V_{\repb{3}\,, 1 }^2 +  u_{\repb{2}\,, 1 }^2 )  -  \frac{ \kappa_{2}}{2} v_t^2 - \frac{ \nu u_{\repb{2}\,, 1 } v_t}{ \sqrt{2} V_{\repb{3}\,, 2 }} \,, \label{eq:2HTM_min_mod02}  \\[1mm]
\frac{\partial V}{\partial v_u} =0 &\Rightarrow& -  \mu_t^2 = \lambda_t v_t^2 +  \frac{ \kappa_{1} }{2}( V_{\repb{3}\,, 1 }^2 + u_{\repb{2}\,, 1 }^2 ) + \frac{ \kappa_{2}}{2} ( V_{\repb{3}\,, 2}^2 + u_{\repb{2}\,, 2 }^2 ) \non
&+& \frac{ \nu }{ \sqrt{2}  v_t} ( u_{\repb{2}\,, 1 } V_{\repb{3}\,, 2}  - u_{\repb{2}\,, 2 } V_{\repb{3}\,, 1 } )  \,, \label{eq:2HTM_min_mod03} \\[1mm]
\frac{\partial V}{\partial u_{\repb{2}\,, 1 }} =0 &\Rightarrow&  \mu_{11}^2 =  - \frac{  \mu_{12}^2}{ t_{\tilde \beta} }  - \frac{\lambda_1 }{2} (  V_{\repb{3}\,, 1 }^2 +  u_{\repb{2}\,, 1 }^2 ) - \frac{ \lambda_{3}}{2} ( V_{\repb{3}\,, 2}^2 + u_{\repb{2}\,, 2 }^2 )-  \frac{  \kappa_{1}}{2}  v_t^2 - \frac{ \nu  V_{\repb{3}\,, 2} v_t}{ \sqrt{2}  u_{\repb{2}\,, 1 }} \,,\label{eq:2HTM_min_mod04} \\[1mm]
\frac{\partial V}{\partial u_{\repb{2}\,, 2 }} =0 &\Rightarrow&  \mu_{22}^2 =  - \mu_{12}^2 t_{ \tilde \beta }  - \frac{ \lambda_2}{2}(  V_{\repb{3}\,, 2 }^2 + u_{\repb{2}\,, 2 }^2 ) -\frac{ \lambda_{3}}{2} ( V_{\repb{3}\,, 1 }^2 + u_{\repb{2}\,, 1 }^2 ) - \frac{ \kappa_{2}}{2}  v_t^2 + \frac{ \nu  V_{\repb{3}\,, 1 } v_t}{ \sqrt{2} u_{\repb{2}\,, 2 }}  \,.\label{eq:2HTM_min_mod05}
\eeqn
\eeqs
By equating Eqs.~\eqref{eq:2HTM_min_mod01} with \eqref{eq:2HTM_min_mod04}, and Eqs.~\eqref{eq:2HTM_min_mod02} with \eqref{eq:2HTM_min_mod05}, we have a relation of
\beqn\label{eq:331_constraint}
&&  (\frac{u_\phi}{  v_{331}} )^2 - \frac{ \sqrt{2} \mu_{12}^2 }{\nu v_t s_{ \tilde \beta} c_{ \tilde \beta} }\, \frac{u_\phi}{v_{331}} -1=0 \,.
\eeqn
The solution of Eq.~\eqref{eq:331_constraint} naturally leads to $u_\phi \sim \frac{\nu v_{331}}{ \mu_{12}^2 } v_t$.
For example, if one takes parameters of $\mu_{12}  \sim \Oc(v_{331}) \sim \Oc(10)\,{\rm TeV}$ and $ \nu \sim \Oc(100)\,{\rm GeV}$, one can have suppressed VEVs of $u_{\repb{2}\,, \lambda }\sim \Oc(1)\,{\rm GeV}$ for the third-generational $b$ quark and tau lepton masses in Eq.~\eqref{eq:331_constraint}.
This means some fine-tuning of the parameter $\nu$ in the ${\rm SU}(6)$ model is necessary~\cite{Chen:2021zwn}.

\para
The origin of the fine-tuning in the toy ${\rm SU}(6)$ model is simply because the $\nu$-term is renormalizable.
Such a fine-tuning problem is analogous to the $\mu$-problem in the minimal supersymmetric Standard Model (MSSM)~\cite{Martin:1997ns}.
A widely accepted solution is the Kim-Nilles mechanism~\cite{Kim:1983dt}, where the $\mu$-term in the MSSM superpotential is generated by non-renormalizable operator.
Such operator can be possible with additional symmetries, for instance, the PQ symmetry.
Analogously, we shall probe whether non-renormalizable operators that contribute to the Higgs VEV terms can emerge in the {\it non-minimal} GUTs with extended gauge symmetries beyond the third-generational toy ${\rm SU}(6)$.
Obviously, the global DRS symmetries also differ between the two-generational ${\rm SU}(7)$ in Eq.~\eqref{eq:DRS_SU7_2gen} and the third-generational toy ${\rm SU}(6)$ in Eq.~\eqref{eq:DRS_SU6}.

\section{The Higgs sector and the VEV generations}
\label{section:Higgs}

\subsection{The Higgs fields}

\para
The minimal set of Higgs fields can be obtained from the following ${\rm SU}(7)$ gauge-invariant Yukawa couplings~\footnote{Throughout the context, we always sum over one superscript flavor index with one subscript flavor index.}
\beqn\label{eq:SU7_Yukawa}
-\Lc_Y &=&{(Y_{\Bc} )_{\lambda}}^\kappa  \repb{ 7_F}^\lambda \rep{ 21_F} \repb{ 7_H}_{\,,\kappa} + {(Y_{\Sc} )_{\dot \lambda }}^{\dot \kappa}   \repb{ 7_F}^{\dot\lambda } \rep{ 35_F} \repb{ 21_{H}}_{\,,\dot \kappa } \non
&+& Y_{\Tc} \rep{ 21_F} \rep{ 21_F} \rep{ 35_H} + Y_{\Tc\Cc} \rep{ 21_F} \rep{ 35_F} \rep{ 21_H} + Y_{\Cc}\rep{ 35_F} \rep{ 35_F} \rep{ 7_H}  + H.c. \,.
\eeqn
Names for Yukawa couplings will become manifest in the fermion mass generation.
In the DRS limit, the Yukawa couplings become
\beqn\label{eq:SU7_Yukawa_DRS}
&&  {(Y_{\Bc} )_{\lambda}}^\kappa = Y_{\Bc} \, {\delta_{\lambda}}^\kappa \,,~  {(Y_{\Sc} )_{\dot \lambda }}^{\dot \kappa}  = Y_{\Sc}\,  {\delta_{\dot \lambda }}^{\dot \kappa}  \,,
\eeqn
while other Yukawa couplings of $(Y_\Tc \,, Y_{\Tc \Cc} \,, Y_\Cc)$ involve the DRS singlet fermions.
Altogether, we collect the two-generational ${\rm SU}(7)$ Higgs fields as follows
\beqn\label{eq:SU7_2gen_Higgs}
\{ H \}_{ {\rm SU}(7)}^{n_g=2}&=& \[ 3\times \repb{7_H} \] \oplus  \[ 2\times \repb{21_H} \] \oplus \rep{7_H} \oplus \rep{21_H} \oplus \rep{35_H} \oplus \rep{48_H} \,.
\eeqn
According to Ref.~\cite{Li:1973mq}, the adjoint Higgs field of $\rep{48_H}$ will be responsible for the GUT symmetry breaking of ${\rm SU}(7)\to \Gc_{341}$ through its VEV of
\beqn\label{eq:48H_VEV}
\langle  \rep{48_H}\rangle &=&\frac{1}{ 2 \sqrt{42}} {\rm diag}(-4\,, -4\,, -4\,, +3\,, +3\,, +3\,, +3) v_U \,.
\eeqn
One can consider a generic Higgs potential for the adjoint Higgs field of $\rep{48_H}$ as follows
\beqn
V( \rep{48_H} )&=&\mu^2 {\rm Tr}( \rep{48_H} )^2 + \lambda_1 {\rm Tr} [ ( \rep{48_H} )^2 ]^2 + \lambda_2 {\rm Tr}( \rep{48_H} )^4 \,.
\eeqn
It turns out the desirable symmetry breaking of ${\rm SU}(7)\to \Gc_{341}$ can be achieved with $\lambda_2>0$~\cite{Li:1973mq}.
The $\rep{48_H} $ can be decomposed into two $3\times 3$ and $4\times 4$ diagonal blocks, plus two $3\times 4$ off-diagonal blocks.
Obviously, the scalar components in two $3\times 4$ off-diagonal blocks are the Nambu-Goldstone bosons for the massive gauge bosons (vectorial leptoquarks) at the GUT scale symmetry breaking.
By combining the fermions in Eq.~\eqref{eq:SU7_2gen_fermions} and Higgs fields in Eq.~\eqref{eq:SU7_2gen_Higgs}, we tabulate their transformations and the most general charge assignments under the global DRS symmetries in Tab.~\ref{tab:SU7_setup}.

\begin{table}[htp]
\begin{center}
\begin{tabular}{c|cccc}
\hline \hline
${\rm SU}(7)$ &    ${\rm SU}(3)_2$ &  ${\rm SU}(2)_3$  &  ${\rm U}(1)_2$  &  ${\rm U}(1)_3$  \\
\hline
$\repb{7_F}^{\lambda}$    & $\rep{3}$  & $\rep{1}$  &  $p_1$ &  $0$  \\ 
$\repb{7_F}^{\dot \lambda}$   & $\rep{1}$  & $\rep{2}$  &  $0$  &  $p_2$  \\ 
$\rep{21_F}$    &  $\rep{1}$ & $\rep{1}$  &  $q_1$ & $0$  \\ 
$\rep{35_F}$   & $\rep{1}$  & $\rep{1}$  &  $0$ & $q_2$  \\ 
\hline
$\repb{7_H}_{\,,\lambda}$   &  $\repb{3}$  & $\rep{1}$  & $-p_1 -q_1$  &  $0$  \\
$\repb{21_H}_{\,,\dot\lambda }$   & $\rep{1}$  & $\repb{2}$  & $0$  &  $-p_2 -q_2$  \\
$\rep{7_H}$   & $\rep{1}$  & $\rep{1}$  &  $0$ & $-2q_2$ \\
$\rep{21_H}$   & $\rep{1}$  &  $\rep{1}$  &  $-q_1$ &  $-q_2$  \\
$\rep{35_H}$   & $\rep{1}$  & $\rep{1}$  &  $-2q_1$ & $0$ \\
$\rep{48_H}$   & $\rep{1}$  & $\rep{1}$  &  $0$ & $0$ \\
\hline\hline
\end{tabular}
\end{center}
\caption{
The ${\rm SU}(7)$ fermions and Higgs representations under the global DRS symmetry in Eq.~\eqref{eq:DRS_SU7_2gen}.
}
\label{tab:SU7_setup}
\end{table}%

\begin{table}[htp]
\begin{center}
\begin{tabular}{c|ccc}
\hline \hline
Higgs & $\Gc_{341}$ &  $\Gc_{331}$ &  $\Gc_{\rm SM}$  \\
\hline
$\repb{7_H}_{\,,\lambda}$ & \cmark$\,(\lambda={\rm II})$  & \cmark$\,(\lambda= {\rm I})$  &  \cmark$\,(\lambda= 3)$ \\
$\repb{21_H}_{\,,\dot\lambda }$ & \xmark  & \cmark$\,(\dot \lambda=\dot {\rm I})$  &  \cmark$\,(\dot \lambda= \dot 2)$ \\
$\rep{7_H}$ & \cmark  & \cmark  &  \cmark  \\
$\rep{21_H}$ & \xmark  & \cmark  &  $\triangle$  \\
$\rep{35_H}$ & \xmark  & \xmark  &  \cmark  \\
\hline\hline
\end{tabular}
\end{center}
\caption{
The Higgs fields and their symmetry breaking directions in the $\gSU(7)$ model.
The \cmark and \xmark represent possible and impossible symmetry breaking directions for a given Higgs field.
The $\triangle$ represents the symmetry breaking direction that is mathematically possible, while can bring potential phenomenological constraints.
The flavor indices for the Higgs fields that develop VEVs at different stages are also specified in the parentheses.
}
\label{tab:SU7Higgs_directions}
\end{table}%

\para
Before we analyze the details of the symmetry breaking, it will be useful to decompose all Higgs fields in Eq.~\eqref{eq:SU7_2gen_Higgs} and to look for the singlet directions for the sequential symmetry breaking stages.
The zeroth-stage symmetry breaking occurs at the GUT scale, which follows the pattern determined by the adjoint Higgs field and the charge quantization in Eq.~\eqref{eq:X0charge}.
After the zeroth-stage GUT symmetry breaking, they read
\beqs\label{eqs:SU7_Higgs_Br}
\beqn
 \repb{7_{H}}_{\,,\lambda}&=& ( \repb{3}\,, \rep{ 1}\,, +\frac{1}{3})_{ \mathbf{H}\,,\lambda } \oplus \boxed{ \overbrace{( \rep{1}\,, \repb{ 4}\,, - \frac{1}{4})_{ \mathbf{H}\,,\lambda}}^{ \rep{\Phi}_{\repb{4}\,, \lambda } }  }  \supset \boxed{ ( \rep{1}\,, \repb{ 3}\,, - \frac{1}{3})_{ \mathbf{H}\,,\lambda} }  \supset \boxed{ ( \rep{1}\,, \repb{ 2}\,, - \frac{1}{2})_{ \mathbf{H}\,,\lambda} }  \,, \\[1mm]
\repb{21_{H}}_{\,,\dot \lambda} &=&  ( \rep{3} \,, \rep{ 1}\,,+\frac{2}{3} )_{\mathbf{H}\,, \dot \lambda}  \oplus ( \repb{3} \,, \repb{ 4}\,,+\frac{1}{12} )_{\mathbf{H}\,, \dot \lambda} \oplus \underline{ \overbrace{ ( \rep{1} \,, \rep{ 6}\,,-\frac{1}{2} )_{\mathbf{H}\,, \dot \lambda } }^{ \rep{\Phi}_{ \rep{6}\,, \dot \lambda}} } \non
&\supset&  \boxed{ ( \rep{1} \,, \repb{ 3}\,,-\frac{1}{3} )_{\mathbf{H}\,, \dot \lambda } } \oplus  \underline{ ( \rep{1} \,, \rep{ 3}\,,-\frac{2}{3} )_{\mathbf{H}\,, \dot \lambda }}  \supset  \boxed{ ( \rep{1} \,, \repb{ 2}\,,-\frac{1}{2} )_{\mathbf{H}\,, \dot \lambda } } \oplus  \boxed{ ( \rep{1} \,, \rep{ 2}\,,-\frac{1}{2} )_{\mathbf{H}\,, \dot \lambda }  }   \,, \\[1mm]
\rep{7_{H} } &=&  ( \rep{3}\,, \rep{ 1}\,, -\frac{1}{3})_{ \mathbf{H}} \oplus \boxed{ \overbrace{ ( \rep{1}\,, \rep{ 4}\,, +\frac{1}{4})_{ \mathbf{H}} }^{ \rep{\Phi}_{\rep{4}} }  } \supset \boxed{ ( \rep{1}\,, \rep{ 3}\,, +\frac{1}{3})_{ \mathbf{H}}} \supset \boxed{ ( \rep{1}\,, \rep{ 2}\,, +\frac{1}{2})_{ \mathbf{H}}}  \,, \\[1mm]
\mathbf{21_H} &=&  ( \repb{3} \,, \rep{ 1}\,, -\frac{2}{3} )_{\mathbf{H}}  \oplus ( \rep{3} \,, \rep{ 4}\,,-\frac{1}{12} )_{\mathbf{H}}  \oplus \underline{ \overbrace{ ( \rep{1} \,, \rep{ 6}\,,+\frac{1}{2} )_{\mathbf{H}}  }^{ \rep{\Phi}_{ \rep{6}}  }  } \non
&\supset&  \boxed{ ( \rep{1} \,, \rep{ 3}\,, +\frac{1}{3} )_{\mathbf{H}}^\prime } \oplus \underline{ ( \rep{1} \,, \repb{ 3}\,,+\frac{2}{3} )_{\mathbf{H} }}  \supset  \boxed{ ( \rep{1} \,, \rep{ 2}\,, +\frac{1}{2} )_{\mathbf{H}}^\prime } \oplus  \boxed{ ( \rep{1} \,, \repb{ 2}\,,+\frac{1}{2} )_{\mathbf{H}}  }   \,,  \\[1mm]
\rep{35_H} &=&    ( \rep{1} \,, \rep{1}\,, -1)_{\mathbf{H}}  \oplus \underline{ \overbrace{ ( \rep{1} \,, \repb{4}\,,+\frac{3}{4} )_{\mathbf{H}} }^{ \rep{\Phi}_{\repb{4}} }  } \oplus ( \repb{3} \,, \rep{4}\,, -\frac{5}{12} )_{\mathbf{H}} \oplus ( \rep{3} \,, \rep{6}\,, +\frac{1}{6} )_{\mathbf{H}} \non
&\supset& \underline{ ( \rep{1} \,, \repb{ 3}\,, +\frac{2}{3} )_{\mathbf{H}}^\prime } \supset \boxed{ ( \rep{1} \,, \repb{ 2}\,, +\frac{1}{2} )_{\mathbf{H}}^\prime }   \,.
\eeqn
\eeqs
with all possible Higgs VEV components responsible for the sequential symmetry breaking framed with boxes.
All Higgs components without underlines or boxes are prohibited to develop VEVs so that the ${\rm SU}(3)_c \otimes {\rm U}(1)_{\rm EM}$ remain exact.
For our later convenience, we also give names to all Higgs fields that are responsible for the extended weak symmetry breakings.
These components are determined by whether they contain the singlet directions according to the desirable symmetry breaking pattern in Eq.~\eqref{eq:SU7_pattern}.
Note that the $\rep{35_H}$ only contains the EWSB component of $( \rep{1} \,, \repb{2} \,, +\frac{1}{2} )_{\mathbf{H}}^{\prime  }$.
Thus, we expect the $\rep{35_H}$ plays the similar roles as the $\rep{15_H}$ in the ${\rm SU}(6)$ model, and it gives top quark mass with natural Yukawa coupling of $Y_\Tc \sim \Oc(1)$.
Based on the decompositions in Eqs.~\eqref{eqs:SU7_Higgs_Br}, we summarize the results in Tab.~\ref{tab:SU7Higgs_directions}.
For DRS-transforming Higgs fields of $\repb{7_H}_{\,,\lambda}$ and $\repb{21_H}_{\,, \dot \lambda}$, we assign their VEVs according to the so-called fermion-Higgs matching pattern.
As will be described in Sec.~\ref{section:breaking}, the number of Higgs VEVs will exactly match the copies of anti-fundamental fermions that acquire their vectorlike masses at each symmetry breaking stage.
For each individual Higgs field, we assign VEVs for the highest symmetry breaking scale that is allowed by the symmetry breaking pattern given in Eq.~\eqref{eq:SU7_pattern}.
According to Tab.~\ref{tab:SU7Higgs_directions}, we expect that the $\rep{7_H}$ is mainly responsible for the $\Gc_{341}$ symmetry breaking, the $\rep{21_H}$ is mainly responsible for the $\Gc_{331}$ symmetry breaking, and so on.
Altogether, we arrive at the minimal set of Higgs VEVs for the sequential symmetry breaking stages as follows
\beqs\label{eqs:SU7_Higgs_VEVs_simple}
\beqn
\Gc_{341} \to \Gc_{331} ~&:&~ \langle ( \rep{1} \,, \repb{4} \,, -\frac{1}{4} )_{\mathbf{H}\,, {\rm II} } \rangle \equiv  w_{\repb{4}\,, {\rm II}}\,,~    \langle ( \rep{1} \,, \rep{4} \,, +\frac{1}{4} )_{\mathbf{H} }  \rangle \equiv w_{\rep{4}} \,,\\[1mm]
\Gc_{331} \to \Gc_{\rm SM} ~&:&~  \langle ( \rep{1} \,, \repb{3} \,, -\frac{1}{3} )_{\mathbf{H}\,, {\rm I}  } \rangle \equiv V_{ \repb{3}\,, {\rm I}} \,,~ \langle ( \rep{1} \,, \repb{3} \,, -\frac{1}{3} )_{\mathbf{H}\,, \dot {\rm I}} \rangle \equiv V_{ \repb{3}\,, \dot {\rm I}} \,,\non
&&\langle ( \rep{1} \,, \rep{3} \,, +\frac{1}{3} )_{\mathbf{H} }^\prime \rangle \equiv V_{ \rep{3}}^\prime  \,, \\[1mm]
 {\rm EWSB} ~&:&~  \langle ( \rep{1} \,, \repb{2} \,, +\frac{1}{2} )_{\mathbf{H} }^{ \prime } \rangle \equiv v_t   \,.
\eeqn
\eeqs
It is natural to expect all these minimal set of VEVs at the particular symmetry breaking stage are of the same order, i.e., we have $w_{\repb{4}\,, {\rm II}} \sim w_{\rep{4}}$ and $V_{\repb{3}\,, {\rm I}} \sim V_{\repb{3}\,, \dot {\rm I}} \sim V_{\rep{3}}^\prime$.
Meanwhile, such an expectation are no longer valid for additional Higgs VEVs to be generated through the Higgs mixing operators below.
Here, the flavor indices for all DRS-transforming Higgs fields will be chosen in accordance to the symmetry breaking pattern in Sec.~\ref{section:breaking}.

\subsection{The Higgs potential}

\begin{table}[htp]
\begin{center}
\begin{tabular}{c|cc|c}
\hline \hline
  $\Oc_{\rm mix}$  &   ${\rm U}(1)_2$  &  ${\rm U}(1)_3$  & VEV terms \\
\hline
 $\Oc_{\mathscr A}^{d=3} \equiv (\rep{21_H} )^2  \cdot \rep{35_H}$   &    $ -4q_1 \,,(0)$  &  $-2 q_2\,,(p_2)$  & \xmark  \\
   $\Oc_{\mathscr B}^{d=3} \equiv \rep{7_H} \cdot ( \rep{35_H} )^2 $    &   $-4q_1\,,(0)$  &  $-2q_2\,,(p_2)$  & \xmark   \\[1mm]
   \hline
  $\Oc_{\mathscr A}^{d=4} \equiv  \epsilon\cdot (\repb{7_H}_{\,,\lambda})^3  \cdot \rep{35_H}$   &    $-( 3p_1 + 5 q_1)\,,(0)$ & $0$  & \cmark  \\
    $\Oc_{\mathscr B}^{d=4} \equiv \epsilon\cdot ( \repb{21_H}_{\,,\dot \lambda})^2 \cdot \rep{7_H} \cdot \rep{35_H}$   &   $-2 q_1\,,( 0)$ & $-2( p_2 + 2 q_2) \,,(0)$  &   \cmark   \\
    $\Oc_{\mathscr C}^{d=4} \equiv\epsilon\cdot ( \repb{21_H}_{\,,\dot \lambda})^2 \cdot  ( \rep{21_H} )^2$    &   $-2 q_1\,,( 0)$ & $-2( p_2 + 2 q_2)\,,(0)$  &  \cmark  \\
    $\Oc_{\mathscr D}^{d=4} \equiv \rep{7_H} \cdot (\rep{21_H} )^3 $   &   $-3q_1\,,( 0)$ & $-5 q_2\,,(\frac{5}{2} p_2)$ & \xmark    \\
\hline\hline
\end{tabular}
\end{center}
\caption{
The renormalizable ${\rm SU}(7)$ Higgs mixing operators and their ${\rm U}(1)_{2\,,3}$ charges.
The charge assignments with $p_1=q_1=0$ and $q_2=- \hf p_2 \neq 0$ are marked in parentheses.
}
\label{tab:SU7_HiggsMix_renOp}
\end{table}%

\begin{table}[htp]
\begin{center}
\begin{tabular}{c|cc|c}
\hline \hline
  $\Oc_{\rm mix}$  &   ${\rm U}(1)_2$  &  ${\rm U}(1)_3$  & VEV terms \\
\hline
  $\Oc_{\mathscr A}^{d=5} \equiv \epsilon\cdot (\repb{7_H}_{\,,\lambda})^3 \cdot  \epsilon\cdot ( \repb{21_H}_{\,,\dot \lambda})^2$   &    $-3(p_1 + q_1)\,,(0)$ & $-2(p_2 + q_2)\,,(-p_2)$  &  \xmark  \\
        $\Oc_{\mathscr B}^{d=5} \equiv ( \rep{7_H} )^4 \cdot \rep{35_H}$   &    $-2 q_1\,,(0)$ & $-8 q_2\,,(4p_2)$  & \xmark  \\
      $\Oc_{\mathscr C}^{d=5} \equiv  \epsilon\cdot (\repb{7_H}_{\,,\lambda})^3 \cdot \rep{7_H} \cdot \rep{21_H}$   &    $-(3p_1 + 4 q_1)\,,(0)$ & $-3 q_2\,,(\frac{3}{2} p_2)$  &  \cmark  \\
      $\Oc_{\mathscr D}^{d=5} \equiv \Big[ \epsilon\cdot ( \repb{21_H}_{\,,\dot \lambda})^2 \Big]^2 \cdot \rep{7_H}$   &  $0$ & $-2(2p_2 + 3 q_2),(-p_2)$  &  \xmark  \\[1mm]
\hline
  $\Oc_{\mathscr A}^{d=6} \equiv \epsilon\cdot (\repb{7_H}_{\,,\lambda})^3 \cdot ( \rep{7_H}  )^3$  &  $-3(p_1 + q_1)\,,(0)$  &  $-6 q_2\,,(3p_2)$  & \cmark  \\
        $\Oc_{\mathscr B}^{d=6} \equiv \Big[ \epsilon\cdot ( \repb{21_H}_{\,,\dot \lambda})^2 \Big] \cdot ( \rep{7_H} )^4$  &  $0$  &  $-2(p_2 + 5 q_2 )\,,(3p_2)$ &  \xmark \\
         $\Oc_{\mathscr C}^{d=6} \equiv ( \rep{21_H}  )^4 \cdot  ( \rep{35_H}  )^2$  &  $-8q_1\,,(0)$  &   $-4q_2\,,(2p_2)$   & \xmark    \\
       $\Oc_{\mathscr D}^{d=6} \equiv\rep{7_H} \cdot ( \rep{21_H}  )^2 \cdot  ( \rep{35_H}  )^3$  &  $-8q_1\,,(0)$  &   $-4q_2\,,(2p_2)$  &  \xmark   \\[1mm]
  \hline
$\Oc^{d=8} \equiv \Big[ \epsilon\cdot ( \repb{21_H}_{\,,\dot \lambda})^2 \Big]^3 \cdot \rep{21_H}\cdot \rep{35_H}$  &  $-3 q_1\,,(0)$  &  $-( 6 p_2 + 7 q_2)\,,(-\frac{5}{2}p_2)$ & \xmark   \\[1mm]
\hline
$\Oc_{\mathscr A}^{d=9} \equiv \Big[ \epsilon\cdot (\repb{7_H}_{\,,\lambda})^3  \Big]^2 \cdot (\rep{21_H})^3$  &  $-3( 2 p_1 + 3 q_1)\,,(0)$  &  $-3 q_2\,,(\frac{3}{2} p_2)$ & \cmark   \\
$\Oc_{\mathscr B}^{d=9} \equiv  \Big[ \epsilon\cdot ( \repb{21_H}_{\,,\dot \lambda})^2 \Big]^3\cdot \rep{7_H} \cdot (\rep{21_H})^2$  &  $-2 q_1\,,(0)$  &  $-2 ( 3 p_2 + 5 q_2)\,,(-p_2)$ & \xmark   \\
$\Oc_{\mathscr C}^{d=9} \equiv  \Big[ \epsilon\cdot ( \repb{21_H}_{\,,\dot \lambda})^2 \Big]^4\cdot \rep{21_H}$  &  $- q_1\,,(0)$  &  $-( 8 p_2 + 9 q_2)\,,(-\frac{7}{2}p_2)$ & \xmark   \\
\hline\hline
\end{tabular}
\end{center}
\caption{
The non-renormalizable ${\rm SU}(7)$ Higgs mixing operators and their ${\rm U}(1)_{2\,,3}$ charges.
The \cmark and \xmark represent whether the specific operator can contribute to VEV terms in the Higgs potential or not.
The charge assignments with $p_1=q_1=0$ and $q_2=- \hf p_2 \neq 0$ are marked in parentheses.
}
\label{tab:SU7_HiggsMix_nonrenOp}
\end{table}%

\para
The most generic Higgs potential for the $\{ \rep{\Phi}_k \}=\{  \repb{7_H}_{\,,\lambda} \,, \repb{21_H}_{\,, \dot \lambda}  \,, \rep{7_H}\,, \rep{21_H}\,, \rep{35_H} \}$ include the following terms
\beqn\label{eq:SU7_potential}
V&=& \sum_{\rep{\Phi}_k   } V( |\rep{\Phi}_k|^2 )   + \Big( V_{\rm mix}^{d\leq 4}  + V_{\rm mix}^{d\geq 5} + H.c. \Big) \,.
\eeqn
with the renormalizable moduli terms expressed as
\beqn\label{eq:SU7_potential_moduli}
V( | \rep{\Phi}_k|^2 ) &=& \mu_k^2 | \rep{\Phi}_k |^2 + \lambda_k  | \rep{\Phi}_k|^4 + \sum_{j\neq k} \kappa_{jk} |\rep{\Phi}_j|^2 \cdot |\rep{\Phi}_k|^2 \,.
\eeqn
Obviously, the moduli terms are both gauge-invariant and DRS-invariant.
For all mixing terms of $\Oc_{\rm mix}$ between Higgs fields, the DRS-invariance becomes non-trivial given that the possible operators may be DRS-charged. 
Here, we require that all renormalizable operators of $\Oc_{\rm mix}^{d\leq 4}$ listed in Tab.~\ref{tab:SU7_HiggsMix_renOp} to be DRS-neutral.
For this reason, the only possible ${\rm U}(1)_{2\,,3}$ charge assignments are 
\beqn\label{eq:DRS_assign}
&& p_1=q_1=0 \,,\quad q_2 = - \frac{1}{2} p_2\neq 0 \,.
\eeqn
In particular, three $d=4$ operators of $(\Oc_{\mathscr A}^{d=4} \,, \Oc_{\mathscr B}^{d=4} \,, \Oc_{\mathscr C}^{d=4} )$ will play crucial roles in generating additional Higgs VEVs for the SM fermion masses.
With the ${\rm U}(1)_{1\,,2}$ charge assignments in Eq.~\eqref{eq:DRS_assign}, all non-renormalizable mixing operators of $\Oc_{\rm mix}^{d\geq 5}$ in Tab.~\ref{tab:SU7_HiggsMix_nonrenOp} are inevitably DRS-charged.
However, they can be allowed in the Higgs potential, in the sense that they can only be violated by the gravitational effects~\cite{Dine:1986bg,Barr:1992qq,Kamionkowski:1992mf,Holman:1992us,Ghigna:1992iv,Kallosh:1995hi,Harlow:2018tng}.
In Tabs.~\ref{tab:SU7_HiggsMix_renOp} and \ref{tab:SU7_HiggsMix_nonrenOp}, we use the short-handed notations of
\beqs
\beqn
\epsilon\cdot (\repb{7_H}_{\,,\lambda})^3&\equiv& \epsilon^{\lambda_1 \lambda_2  \lambda_3} \, \repb{7_H}_{\,,\lambda_1} \repb{7_H}_{\,,\lambda_2} \repb{7_H}_{\,,\lambda_3}  \,,\\[1mm]
\epsilon\cdot ( \repb{21_H}_{\,,\dot \lambda})^2&\equiv& \epsilon^{ \dot \lambda_1  \dot \lambda_2}\, \repb{21_H}_{\,,\dot \lambda_1} \repb{21_H}_{\,, \dot \lambda_2} \,.
\eeqn
\eeqs
Thus, we express the mixing terms in the Higgs potential as follows
\beqn\label{eq:SU7_potential_mix}
&& V_{\rm mix}^{d\leq 4} = g_{d} \Oc_{\rm mix}^{ d \leq 4}\,, \quad V_{\rm mix}^{d\geq 5} = \frac{g_{d}}{ M_{\rm pl}^{d-4} } \Oc_{\rm mix}^{ d \geq 5} \,.
\eeqn
Obviously, only the Higgs VEV components should be taken into account in Eq.~\eqref{eq:SU7_potential} when one minimizes the potential.
This was previously pointed out in analyzing the explicit PQ-breaking terms by Ref.~\cite{Dobrescu:1996jp}.
In App.~\ref{section:SU7_HiggsOp}, we derive and determine whether a specific operator in Tabs.~\ref{tab:SU7_HiggsMix_renOp} and~\ref{tab:SU7_HiggsMix_nonrenOp} can lead to VEV terms to the Higgs potential.

\para
For three renormalizable $d=4$ operators that can lead to Higgs VEV terms, a suppression of induced EWSB VEVs are possible.
Let us consider the generation of the suppressed VEV of $u_{\repb{2}\,,3}$. 
After the GUT scale symmetry breaking, the Higgs fields that lead to a VEV term in the operator of $\Oc_{\mathscr A}^{d=4}$ are $(\Phi_{\repb{4}\,,\lambda} \,, \Phi_{\repb{4}})$.
The Higgs potential must contain the following terms
\beqn
V_{341}&\supset& \mu_\lambda^2 | \Phi_{\repb{4}\,,\lambda}  |^2 + g_{4 {\mathscr A}}  \epsilon^{ \lambda \kappa \delta }  \Phi_{\repb{4}\,,\lambda} \Phi_{\repb{4}\,,\kappa } \Phi_{\repb{4}\,,\delta } \Phi_{ \repb{4}}  \,.
\eeqn
By minimizing the potential along the direction of the generated VEV $u_{\repb{2}\,,3}$ according to Eq.~\eqref{eq:Oc4A_VEVgen}, one finds a relation of $\mu_3^2 u_{\repb{2}\,,3} \sim g_{4 {\mathscr A}}  w_{\repb{4}} V_{\repb{3}} v_t$.
By further assuming the natural relations of $\mu_3\sim w_{\repb{4}} $ and $g_{4 {\mathscr A}} \sim \Oc(1)$, one finds that $u_{\repb{2}\,,3} \sim (\frac{ V_{\repb{3}} }{ w_{\repb{4}} } ) v_t \ll v_t $.
In addition to renormalizable $d=4$ operators, we also find three non-renormalizable operators of $( \Oc_{\mathscr C}^{d=5}\,,\Oc_{\mathscr A}^{d=6} \,, \Oc_{\mathscr A}^{d=9}  )$ that can generate additional Higgs VEV terms.

\begin{figure}
\centering
\includegraphics[width=10cm]{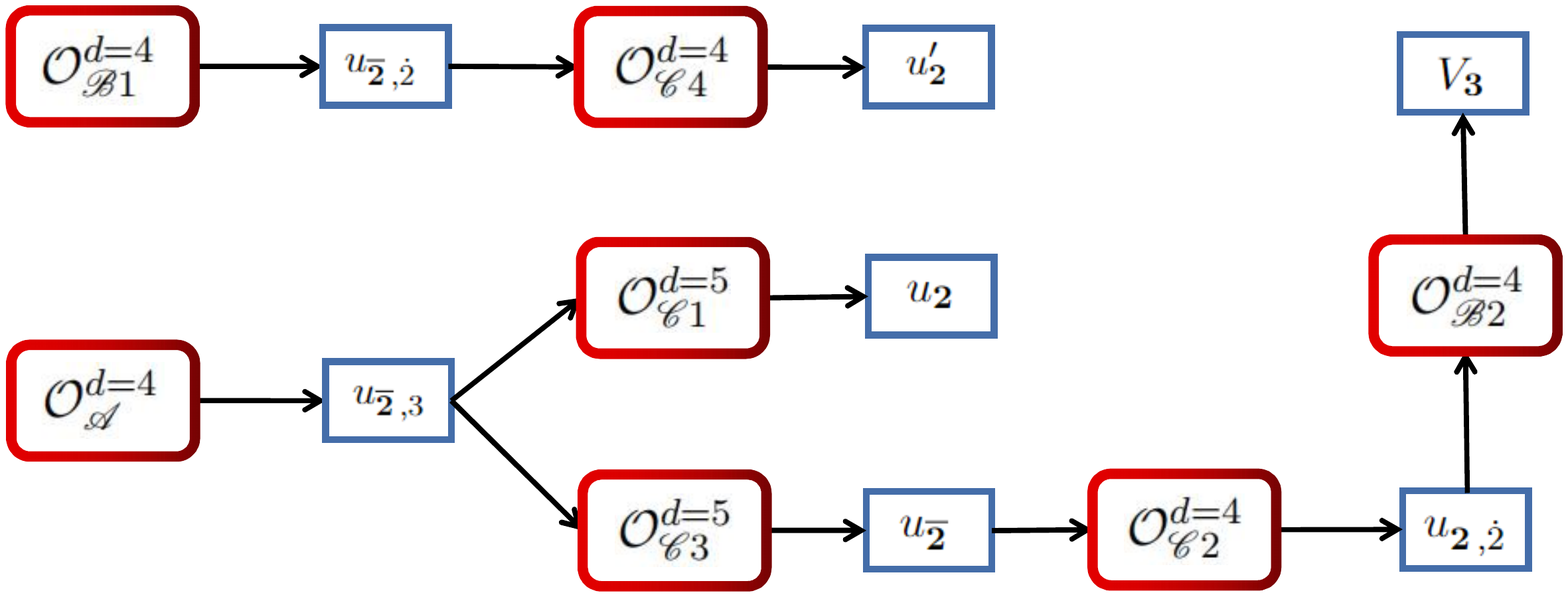}
 \caption{
 The generation chains of all additional Higgs VEVs through the gauge-invariant operators composed by ${\rm SU}(7)$ Higgs fields.
}
\label{fig:VEV_chain} 
\end{figure}

\para
Altogether, we end up with the following Higgs VEVs at each stage of symmetry breaking
\beqs\label{eqs:SU7_Higgs_VEVs_full}
\beqn
\Gc_{341} \to \Gc_{331} ~&:&~ \langle ( \rep{1} \,, \repb{4} \,, -\frac{1}{4} )_{\mathbf{H}\,, {\rm II} } \rangle \equiv w_{ \repb{4}\,, {\rm II}}\,, ~   \langle ( \rep{1} \,, \rep{4} \,, +\frac{1}{4} )_{\mathbf{H} } \rangle \equiv w_{\rep{4}}  \,, \label{eq:SU7_Higgs_VEVs_full01} \\[1mm]
\Gc_{331} \to \Gc_{\rm SM} ~&:&~  \langle ( \rep{1} \,, \repb{3} \,, -\frac{1}{3} )_{\mathbf{H}\,, {\rm I} } \rangle \equiv V_{ \repb{3}\,, {\rm I}} \,,~ \langle ( \rep{1} \,, \repb{3} \,, -\frac{1}{3} )_{\mathbf{H}\,, \dot {\rm I} } \rangle \equiv V_{ \repb{3}\,, \dot {\rm I} }  \,,\non
&&\boxed{ \langle ( \rep{1} \,, \rep{3} \,, +\frac{1}{3} )_{\mathbf{H} } \rangle \equiv V_{ \rep{3}} } \,,~  \langle ( \rep{1} \,, \rep{3} \,, +\frac{1}{3} )_{\mathbf{H} }^\prime \rangle \equiv V_{ \rep{3}}^\prime  \,, \label{eq:SU7_Higgs_VEVs_full02} \\[1mm]
 {\rm EWSB} ~&:&~ \boxed{ \langle ( \rep{1} \,, \repb{2} \,, -\frac{1}{2} )_{\mathbf{H}\,, 3 } \rangle \equiv u_{ \repb{2}\,, 3 } } \,,  \non
 && \boxed{ \langle ( \rep{1} \,, \repb{2} \,, -\frac{1}{2} )_{\mathbf{H}\,, \dot 2 } \rangle \equiv u_{ \repb{2}\,, \dot 2 }}\,,  ~ \boxed{ \langle ( \rep{1} \,, \rep{2} \,, -\frac{1}{2} )_{\mathbf{H}\,, \dot 2 } \rangle \equiv u_{ \rep{2}\,, \dot 2 }}\,, \non
&&  \boxed{ \langle ( \rep{1} \,, \rep{2} \,, +\frac{1}{2} )_{\mathbf{H} } \rangle \equiv u_{ \rep{2}} } \,, ~ \boxed{ \langle ( \rep{1} \,, \rep{2} \,, +\frac{1}{2} )_{\mathbf{H} }^{ \prime} \rangle \equiv u_{ \rep{2} }^\prime } \,, ~ \boxed{ \langle ( \rep{1} \,, \repb{2} \,, +\frac{1}{2} )_{\mathbf{H} } \rangle \equiv u_{ \repb{2} } } \,, \non
 &&  \langle ( \rep{1} \,, \repb{2} \,, +\frac{1}{2} )_{\mathbf{H} }^{ \prime} \rangle \equiv v_t \,. \label{eq:SU7_Higgs_VEVs_full03}
\eeqn
\eeqs
The additional Higgs VEVs generated by the Higgs mixing operators in App.~\ref{section:SU7_HiggsOp} are marked with boxes.
We display the VEV generation chains through the series of operators in Fig.~\ref{fig:VEV_chain}, with the minimal set of Higgs VEVs given in Eq.~\eqref{eqs:SU7_Higgs_VEVs_simple} as their input parameters.
The EWSB VEV from the $\rep{21_H}$ may lead to flavor-changing decay mode of $t\to h_{\rm SM} c$, which should be further studied with the ongoing LHC searches for this rare decay mode.
For our later usage, we denote the Higgs VEVs at different scales collectively as follows
\beqs
\beqn
&& w_{ \repb{4}\,,{\rm II} }^2 + w_{ \rep{4}}^2 \equiv v_{341}^2\,,\label{eq:SU7_Higgs_VEVs01} \\[1mm]
&& \sum_{\Lambda ={\rm I}\,, \dot {\rm I} } V_{ \repb{3}\,, \Lambda }^2 + V_{ \rep{3}}^2 + (V_{ \rep{3}}^\prime )^2  \equiv v_{331}^2  \,,\label{eq:SU7_Higgs_VEVs02} \\[1mm]
&& \sum_{ \Lambda =\dot 2\,, 3 } u_{ \repb{2}\,, \Lambda}^2 + ( u_{ \rep{2}\,, \dot 2} )^2 + u_{ \rep{2}}^2  + (u_{ \rep{2}}^\prime )^2  + u_{ \repb{2}}^2 + v_t^2 \equiv v_{\rm EW}^2 \,. \label{eq:SU7_Higgs_VEVs03}
\eeqn
\eeqs

\section{The gauge sector}
\label{section:gauge}

\para
After the GUT scale symmetry breaking, the effective theory is described by an extended electroweak symmetry of ${\rm SU}(3)_c \otimes {\rm SU}(4)_W  \otimes {\rm U}(1)_{X_0}$.
The sequential symmetry breaking of ${\rm SU}(4)_W  \otimes {\rm U}(1)_{X_0} \to {\rm SU}(3)_W  \otimes {\rm U}(1)_{X_1}$ and ${\rm SU}(3)_W  \otimes {\rm U}(1)_{X_1} \to {\rm SU}(2)_W  \otimes {\rm U}(1)_{Y}$ lead to seven and five massive gauge bosons, respectively.
In this section, we describe the massive gauge bosons during two stages of symmetry breaking.
Results obtained in this section will be used to describe the gauge couplings of fermions in Sec.~\ref{section:fermions}.
All group indices in this section follow the conventions defined in Tab.~\ref{tab:notations}.

\subsection{The ${\rm SU}(4)_W \otimes {\rm U}(1)_{X_0}$ gauge bosons}
\label{section:gauge341}

\para
We express the ${\rm SU}(4)_W \otimes {\rm U}(1)_{X_0}$ covariant derivatives as follows
\beqn\label{eq:341_cov_fund}
D_\mu &\equiv& \partial_\mu -i g_{4W} A_\mu^{\bar I} T_{ {\rm SU}(4)}^{\bar I} -i g_{X_0} \Xc_0 \mathbb{I}_4 X_{0\,\mu}\,,
\eeqn
for the ${\rm SU}(4)_W $ fundamental representation.
It becomes
\beqn\label{eq:341_cov_antifund}
D_\mu &\equiv&  \partial_\mu +i g_{4W} A_\mu^{\bar I} ( T_{ {\rm SU}(4)}^{\bar I} )^T - i g_{X_0} \Xc_0 \mathbb{I}_4 X_{0\,\mu} \,,
\eeqn
for the ${\rm SU}(4)_W $ anti-fundamental representation.
The ${\rm SU}(4)_W $ generators of $T_{ {\rm SU}(4)}^{\bar I}$ are normalized such that $ \Tr\Big( T_{ {\rm SU}(4)}^{\bar I} T_{ {\rm SU}(4)}^{\bar J}  \Big)= \hf \delta^{\bar I \bar J}$.
The explicit form for the gauge fields of $g_{4W} A_\mu^{\bar I}  T_{ {\rm SU}(4)}^{\bar I} +g_{X_0} \Xc_0 \mathbb{I}_4 X_{0\,\mu}$ can be expressed in terms of a $4\times 4$ matrix as follows
\beqn\label{eq:341_connection_fund}
&& g_{4W} A^{\bar I}_\mu T_{ {\rm SU}(4)}^{\bar I} + g_{X_0} \Xc_0 \mathbb{I}_4 X_{0\,\mu}=  \non
&& \frac{ g_{4W}}{\sqrt{2} }  \,  \left( 
\ba{cccc}  
     0  &   W_\mu^+  &   W_\mu^{\prime\, +}  &  W_\mu^{\prime\prime \, +}   \\ 
 W_\mu^-  &    &   &    \\ 
 W_\mu^{\prime\, -}   &     \multicolumn{3}{c}{  0_{3\times 3} }  \\
 W_\mu^{\prime\prime \, -}  &  &  &   \\   \ea  \right) + \frac{ g_{4W}}{\sqrt{2} } \,  \left( 
\ba{cccc}  
     0  & 0  &  0  &  0  \\ 
 0  &  0   &  N_\mu &  N_\mu^\prime  \\ 
0  &  \bar N_\mu   &  0 &  N_\mu^{\prime\prime}  \\
 0  &   \bar N_\mu^{\prime}  &  \bar N_\mu^{\prime\prime}  &  0  \\   \ea  \right)  \non
 &+& \frac{ g_{4W}  }{2} {\rm diag}  \Big(   A_\mu^3 + \frac{1}{\sqrt{3} } A_\mu^8 \,, - A_\mu^3 + \frac{1}{\sqrt{3} } A_\mu^8  \,,  - \frac{ 2 }{\sqrt{3} }  A_\mu^8  \,,  0 \Big) \non
&+& \frac{ g_{4W} }{2 \sqrt{6}} {\rm diag}  \Big(  \[   A_\mu^{15}  + 12 t_{\theta_G } \Xc_0 X_{0\,\mu} \] \mathbb{I}_{3\times 3} \,,   - 3  A_\mu^{15} +  12 t_{\theta_G }  \Xc_0 X_{0\,\mu}  \Big) \,,
\eeqn
where we have defined a mixing angle of
\beqn\label{eq:Glashow_angle}
 t_{\theta_G} &\equiv& \tan\theta_{G} = \frac{g_{X_0} }{ \sqrt{6} g_{4W } }\,.
\eeqn
The notions of massive gauge bosons are determined according to their electric charges from the relation of $[ \hat Q_e( \rep{15_W}) \,, A^{\bar I}_\mu T_{ {\rm SU}(4)}^{\bar I}  ] = Q_e\, A^{\bar I}_\mu T_{ {\rm SU}(4)}^{\bar I}$, with $\hat Q_e( \rep{15_W})$ defined in Eq.~\eqref{eq:Qcharge_4adj}.
Explicitly, we find the electrically charged gauge bosons of
\beqn
&& Q_e ( A_\mu^1 \mp i A_\mu^2 ) =  Q_e ( A_\mu^4 \mp i A_\mu^5 ) =  Q_e ( A_\mu^9 \mp i A_\mu^{10} ) = \pm 1 \,,
\eeqn
while all other gauge bosons are electrically neutral.

\para
Through the analyses in Sec.~\ref{subsection:breaking1}, this stage of symmetry breaking can be achieved by ${\rm SU}(4)_W$ anti-fundamental Higgs fields of $\rep{\Phi}_{ \repb{4} \,, {\rm II} } \equiv ( \rep{1}\,, \repb{4}\,, -\frac{1}{4} )_{\mathbf{H} \,, {\rm II} } \subset \repb{7_H}_{\,,{\rm II} }$ and one ${\rm SU}(4)_W$ fundamental Higgs field of $\rep{\Phi}_{ \rep{4} } \equiv ( \rep{1}\,, \rep{4}\,, +\frac{1}{4} )_{\mathbf{H} }  \subset \rep{7_H}$.
With the ${\rm SU}(4)_W$-breaking VEVs in Eqs.~\eqref{eqs:SU7_Higgs_VEVs_full}, the Higgs kinematic terms lead to the $(W_\mu^{\prime\prime\, \pm} \,, N_\mu^\prime\,, N_\mu^{\prime\prime} )$ gauge boson masses of 
\beqn
&& \frac{1}{4} g_{4W}^2 v_{341}^2   \Big( W_\mu^{\prime \prime +}  W^{\prime \prime -\, \mu} +N_\mu^\prime \bar N^{\prime \, \mu} +N_\mu^{\prime \prime } \bar N^{\prime \prime \, \mu}  \Big)  \,,
\eeqn
with the Higgs VEV of $v_{341}$ given in Eq.~\eqref{eq:SU7_Higgs_VEVs01}.
For the flavor-conserving neutral gauge bosons of $(A_\mu^{15}\,, X_{0\, \mu} )$, the mass squared matrix reads
\beqn
&& \frac{3 }{16  } g_{4W}^2 v_{341}^2 (   A^{15}_\mu \,, X_{0\,\mu}  )  \cdot \left( 
\ba{cc}  
 1   &  - t_{\theta_G}  \\ 
- t_{\theta_G}  & t_{\theta_G}^2   \\   \ea \right)  \cdot  \left( \ba{c}  A^{15\,\mu}  \\ X_{0}^\mu \\   \ea \right)\,.
\eeqn
Obviously it contains a zero eigenvalue which corresponds to the massless gauge boson of $X_{1\, \mu}$ after the ${\rm SU}(4)_W$ symmetry breaking.
The mass eigenstates can be diagonalized in terms of the mixing angle in Eq.~\eqref{eq:Glashow_angle} as follows
\beqn
 \left( \ba{c}  Z^{\prime\prime}_\mu  \\ X_{1\,\mu}  \\    \ea \right) &=& \left( \ba{cc}  
   c_{\theta_{G}}  &   - s_{\theta_{G}} \\ 
  s_{\theta_{G}}  &  c_{\theta_{G}}  \\   \ea \right) \cdot  \left( \ba{c}  A^{15}_\mu  \\ X_{0\,\mu} \\     \ea \right)\,.
\eeqn
The ${\rm SU}(4)_W \otimes {\rm U}(1)_{X_0}$ gauge couplings of $(\alpha_{4W}\,, \alpha_{X_0} )$ match with the ${\rm SU}(3)_W \otimes {\rm U}(1)_{X_1}$ gauge couplings as follows
\beqn\label{eq:341_coupMatch}
&&  \alpha_{3W }^{-1} (v_{341}) =  \alpha_{4W }^{-1} (v_{341}) \,,~ \alpha_{X_1}^{-1} (v_{341})  = \frac{1}{6} \alpha_{4W }^{-1} (v_{341})  +  \alpha_{X_0}^{-1} (v_{341})  \,, \non
&& \frac{1}{6} \alpha_{4W }^{-1} = \alpha_{X_1}^{-1} s_{\theta_{G}}^2 \,,~ \alpha_{X_0}^{-1} = \alpha_{X_1}^{-1} c_{\theta_{G}}^2  \,.
\eeqn
From the definitions of two mixing angles in Eqs.~\eqref{eq:Glashow_angle} and~\eqref{eq:Salam_angle}, we find a relation of
\beqn\label{eq:GS_relation}
&& \boxed{ \sin\theta_G = \frac{1}{ \sqrt{2} } \tan\theta_S } \,.
\eeqn
The tree-level masses for seven gauge bosons at this stage read
\beqs
\beqn
&& m_{W_\mu^{\prime\prime\, \pm} }^2 =  m_{ N_\mu^\prime\,, \bar N_\mu^\prime }^2   = m_{ N_\mu^{\prime\prime}\,, \bar N_\mu^{\prime\prime}  }^2   =  \frac{g_{X_1}^2}{ 24 s_{\theta_G}^2 }   v_{341}^2 \,,\\[1mm]
&& m_{Z_\mu^{\prime\prime} }^2 = \frac{ g_{X_1}^2}{ 16 s_{\theta_G}^2 c_{\theta_G}^2  }   v_{341}^2 \,.
\eeqn
\eeqs
After the first-stage symmetry breaking, the remaining massless gauge bosons are $(W_\mu^\pm\,, W_\mu^{\prime\, \pm})$, $(N_\mu\,,\bar N_\mu )$, and $(A_\mu^3\,, A_\mu^8\,, X_{1\, \mu})$.

\para
In terms of mass eigenstates, the gauge bosons from the covariant derivative in Eq.~\eqref{eq:341_connection_fund} are expressed as
\beqn\label{eq:341_connection_fundmass}
&& g_{4W} A^{\bar I}_\mu T_{ {\rm SU}(4)}^{\bar I} + g_{X_0} \Xc_0 \mathbb{I}_4 X_{0\,\mu}=  \non
&& \frac{ g_{4W}}{\sqrt{2} }  \,  \left( 
\ba{cccc}  
     0  &   W_\mu^+  &   W_\mu^{\prime\, +}  &  W_\mu^{\prime\prime \, +}   \\ 
 W_\mu^-  &    &   &    \\ 
 W_\mu^{\prime\, -}   &     \multicolumn{3}{c}{  0_{3\times 3} }  \\
 W_\mu^{\prime\prime \, -}  &  &  &   \\   \ea  \right) + \frac{ g_{4W}}{\sqrt{2} } \,  \left( 
\ba{cccc}  
     0  & 0  &  0  &  0  \\ 
 0  &  0   &  N_\mu &  N_\mu^\prime  \\ 
0  &  \bar N_\mu   &  0 &  N_\mu^{\prime\prime}  \\
 0  &   \bar N_\mu^{\prime}  &  \bar N_\mu^{\prime\prime}  &  0  \\   \ea  \right)  \non
 &+& \frac{ g_{4W}  }{2} {\rm diag}  \Big(   A_\mu^3 + \frac{1}{\sqrt{3} } A_\mu^8 \,, - A_\mu^3 + \frac{1}{\sqrt{3} } A_\mu^8  \,,  - \frac{ 2 }{\sqrt{3} }  A_\mu^8  \,,  0 \Big) \non
&+&  g_{X_1} {\rm diag}  \Big(   ( \frac{1}{12} +  \Xc_0 ) \mathbb{I}_{3\times 3}  \,,  - \frac{1}{4} +  \Xc_0  \Big) X_{1\,\mu}  \non
&+& \frac{ g_{X_1}}{  s_{\theta_G} c_{\theta_G}  } {\rm diag}  \Big(  \[   \frac{1}{ 12} - (  \frac{1}{12} +  \Xc_0) s_{\theta_G}^2 \] \mathbb{I}_{3\times 3} \,,    - \frac{1}{4} +  ( \frac{1}{4} -  \Xc_0) s_{\theta_G}^2  \Big) Z_\mu^{\prime\prime }  \,.
\eeqn
As a consistent check, the $(44)$-component in Eq.~\eqref{eq:341_connection_fundmass} is reduced to $-\frac{ g_{X_1} }{ 4 s_{\theta_G} c_{\theta_G} }  Z_\mu^{\prime\prime }$ when setting $\Xc_0=+\frac{1}{4}$ for the fundamental representation.
Likewise, we find the explicit form the gauge fields of $- g_{4W} A_\mu^{\bar I}  (T_{ {\rm SU}(4)}^{\bar I} )^T +g_{X_0} \Xc_0 \mathbb{I}_4 X_{0\,\mu}$ for the anti-fundamental representation as follows
\beqn\label{eq:341_connection_antifundmass}
&& - g_{4W} A^{\bar I}_\mu ( T_{ {\rm SU}(4)}^{\bar I} )^T + g_{X_0} \Xc_0 \mathbb{I}_4 X_{0\,\mu}=  \non
&& - \frac{ g_{4W}}{\sqrt{2} }  \,  \left( 
\ba{cccc}  
     0  &   W_\mu^-  &   W_\mu^{\prime\, -}  &  W_\mu^{\prime\prime \, -}   \\ 
 W_\mu^+  &    &   &    \\ 
 W_\mu^{\prime\, +}   &     \multicolumn{3}{c}{  0_{3\times 3} }  \\
 W_\mu^{\prime\prime \, +}  &  &  &   \\   \ea  \right) - \frac{ g_{4W}}{\sqrt{2} } \,  \left( 
\ba{cccc}  
     0  & 0  &  0  &  0  \\ 
 0  &  0   &  \bar N_\mu &  \bar N_\mu^\prime  \\ 
0  &  N_\mu   &  0 &  \bar N_\mu^{\prime\prime}  \\
 0  &  N_\mu^{\prime}  & N_\mu^{\prime\prime}  &  0  \\   \ea  \right) \non
 &-& \frac{g_{4W}}{2} {\rm diag}\,  \Big(    A_\mu^3 + \frac{1}{\sqrt{3} } A_\mu^8  \,,   - A_\mu^3 + \frac{1}{\sqrt{3} } A_\mu^8  \,, - \frac{ 2 }{\sqrt{3} }  A_\mu^8 \,, 0 \Big) \non
&+& g_{X_1} {\rm diag} \Big( ( - \frac{1}{12} + \Xc_0 ) \mathbb{I}_{3\times 3}  \,,   \frac{1}{4} +  \Xc_0  \Big) X_{1\,\mu}  \non
&+& \frac{ g_{X_1}}{ s_{\theta_G} c_{\theta_G}  } {\rm diag}  \Big(  \[  - \frac{1}{12} + ( \frac{1}{12} - \Xc_0) s_{\theta_G}^2 \] \mathbb{I}_{3\times 3} \,, \frac{1}{4} -  (\frac{1}{4} +  \Xc_0) s_{\theta_G}^2  \Big) Z_\mu^{\prime\prime }  \,.
\eeqn

\subsection{The ${\rm SU}(3)_W \otimes {\rm U}(1)_{X_1}$ gauge bosons}
\label{section:gauge331}

\para
We express the ${\rm SU}(3)_W \otimes {\rm U}(1)_{X_1}$ covariant derivatives for the ${\rm SU}(3)_W $ fundamental and anti-fundamental representations as follows
\beqs\label{eqs:331_covariant}
\beqn
D_\mu &\equiv&  \partial_\mu -i  g_{3W} A_\mu^{\tilde I} \frac{\lambda_{ {\rm SU}(3)}^{\tilde I} }{2} -i  g_{X_1} \Xc_1 \mathbb{I}_3  X_{1\,\mu}  \,,\label{eq:331_covariant_fund} \\[1mm]
D_\mu &\equiv&  \partial_\mu +i g_{3W} A_\mu^{\tilde I} ( \frac{\lambda_{ {\rm SU}(3)}^{\tilde I} }{2} )^T -i   g_{X_1} \Xc_1 \mathbb{I}_3  X_{1\,\mu}  \,. \label{eq:331_covariant_antifund}
\eeqn
\eeqs
The $\lambda_{ {\rm SU}(3)}^{\tilde I}$ are Gell-Mann matrices, which are normalized such that $ \Tr\Big( \lambda_{ {\rm SU}(3)}^{\tilde I} \lambda_{ {\rm SU}(3)}^{\tilde J}  \Big)= 2 \delta^{\tilde I  \tilde J}$.
The gauge fields from Eq.~\eqref{eq:331_covariant_fund} can be expressed in terms of a $3\times 3$ matrix 
\beqn\label{eq:331_connection_gauge}
&& g_{3W} A^{\tilde I}_\mu \frac{\lambda_{ {\rm SU}(3)}^{\tilde I}}{2} + g_{X_1} \Xc_1 \mathbb{I}_3  X_{1\,\mu}  \non
&=&  \frac{g_{3W}}{ \sqrt{2} }\left( 
\ba{ccc}  
     0  &   W_\mu^+  &   W_\mu^{\prime\, +}  \\ 
 W_\mu^- &  0   &  0  \\ 
 W_\mu^{\prime\, -}   &  0 &  0   \\  \ea  \right)  + \frac{g_{3W}}{ \sqrt{2} }\left( 
\ba{ccc}  
     0  &  0  &  0  \\ 
 0 &  0   &  N_\mu  \\ 
 0   &  \bar N_\mu  &  0   \\  \ea  \right)  \non
 &+& \frac{ g_{3W} }{2} {\rm diag}  \Big(    A_\mu^3 + \frac{1}{\sqrt{3} } A_\mu^8  + 2 \sqrt{3} t_{\theta_S} \Xc_1 X_{1\,\mu} \,,  \non
&&  - A_\mu^3 + \frac{1}{\sqrt{3} } A_\mu^8  + 2 \sqrt{3} t_{\theta_S} \Xc_1 X_{1\,\mu} \,, - \frac{ 2 }{\sqrt{3} }  A_\mu^8  + 2 \sqrt{3} t_{\theta_S} \Xc_1 X_{1\,\mu} \Big)  \,,
\eeqn
with
\beqn\label{eq:Salam_angle}
t_{\theta_S} &\equiv& \tan\theta_{S} = \frac{g_{X_1} }{ \sqrt{3} g_{3W } }\,.
\eeqn

\para
This stage of symmetry breaking is achieved by Higgs VEVs in Eq.~\eqref{eq:SU7_Higgs_VEVs_full02}.
The gauge boson masses from the Higgs kinematic terms are
\beqn
&& \frac{1}{4} g_{3W}^2 v_{331}^2 \Big[   \Big(  W_\mu^{\prime\,+} W^{\prime\,-\,\mu} + N_\mu \bar N^\mu \Big) +  \frac{2}{3}  \Big( A_\mu^8 - t_{\theta_S } X_{1\,\mu} \Big)^2 \Big] \,.
\eeqn
It is straightforward to obtain the mass eigenstates in terms of the mixing angle in Eq.~\eqref{eq:Salam_angle} for this case
\beqn
 \left( \ba{c}  Z^{\prime}_\mu  \\ B_{\mu}  \\  \ea \right) &=& \left( \ba{cc}  
   c_{\theta_{S}}  &  - s_{\theta_{S}}  \\ 
  s_{\theta_{S}}   &  c_{\theta_{S}}    \\   \ea \right) \cdot  \left( \ba{c}  A^{8}_\mu  \\ X_{1\,\mu} \\     \ea \right)\,.
\eeqn
The ${\rm SU}(3)_W \otimes {\rm U}(1)_{X_1}$ gauge couplings of $(\alpha_{3W}\,, \alpha_{X_1} )$ match with the EW gauge couplings as follows
\beqn\label{eq:331_coupMatch}
&& \alpha_{2W }^{-1} (v_{331}) = \alpha_{3W }^{-1} (v_{331}) \,, ~ \alpha_Y^{-1} (v_{331})  = \frac{1}{3} \alpha_{3W }^{-1} (v_{331})  +  \alpha_{X_1}^{-1} (v_{331}) \,,  \non
&& \frac{1}{3} \alpha_{3W}^{-1} = \alpha_Y^{-1} s_{\theta_{S} }^2 \,,~ \alpha_{X_1}^{-1} = \alpha_Y^{-1} c_{\theta_{S} }^2 \,.
\eeqn
From the definitions of two mixing angles in Eqs.~\eqref{eq:Salam_angle} and \eqref{eq:Weinberg_angle}, we find a relation of
\beqn\label{eq:SW_relation}
&& \boxed{ \sin\theta_S = \frac{1}{ \sqrt{3} } \tan\theta_W } \,.
\eeqn
The tree-level masses for five gauge bosons at this stage read
\beqs\label{eqs:331_GBmasses}
\beqn
&& m_{W_\mu^{\prime\, \pm} }^2    = m_{ N_\mu\,, \bar N_\mu  }^2   =  \frac{ g_Y^2 }{ 12 s_{\theta_S}^2 }  v_{331}^2 \,,\label{eq:331_GBmass01}\\
&& m_{Z_\mu^{\prime} }^2 =  \frac{g_{Y}^2}{ 9 s_{\theta_S}^2 c_{\theta_S}^2 }  v_{331}^2 \,.\label{eq:331_GBmass02}
\eeqn
\eeqs

\para
In terms of mass eigenstates, the gauge bosons from the covariant derivative in Eq.~\eqref{eq:331_connection_gauge} become
\beqn\label{eq:331_connection_fund}
&& \frac{g_{3W}}{ \sqrt{2} } \left( 
\ba{ccc}  
 0  & W_\mu^+ & 0  \\ 
W_\mu^-  &  0 & 0 \\ 
0  & 0 &  0  \\  \ea  \right)   + \frac{g_{3W}}{ \sqrt{2} } \left( 
\ba{ccc}  
 0  & 0  & W_\mu^{\prime +}  \\ 
0  &  0 &  N_\mu  \\ 
W_\mu^{\prime\, -}   &  \bar N_\mu  &  0  \\  \ea  \right) \non
&+&  g_{3W} {\rm diag}  \Big( \hf \,, - \hf \,,  0 \Big) A_\mu^3 + g_Y {\rm diag}  \Big( ( \frac{1}{6} + \Xc_1) \mathbb{I}_{2\times 2}  \,, -\frac{1}{3} +  \Xc_1 \Big) B_\mu \non
&+& \frac{g_Y}{ s_{\theta_S}  c_{\theta_S}  }   {\rm diag}  \Big(  \[ \frac{1}{6} -  ( \frac{1}{6} +  \Xc_1 ) s_{\theta_S}^2 \] \mathbb{I}_{2\times 2}  \,, - \frac{1}{3} + (\frac{1}{3} -  \Xc_1 ) s_{ \theta_S }^2 \Big) Z_\mu^\prime \,.
\eeqn
Likewise, the covariant derivative for the ${\rm SU}(3)_W$ anti-fundamental representation gives
\beqn\label{eq:331_connection_antifund}
&& - \frac{g_{3W}}{ \sqrt{2} } \left( 
\ba{ccc}  
 0  & W_\mu^- & 0  \\ 
W_\mu^+  &  0 & 0 \\ 
0 &  0  &  0  \\  \ea  \right)  -  \frac{g_{3W}}{ \sqrt{2} } \left( 
\ba{ccc}  
 0  & 0  & W_\mu^{\prime\,-}  \\ 
0  &  0 & \bar N_\mu  \\ 
W_\mu^{\prime\, +}  & N_\mu  &  0  \\  \ea  \right) \non 
&-&   g_{3W}  {\rm diag}\, \Big( \hf \,, - \hf \,,  0 \Big) A_\mu^3  + g_Y {\rm diag}\,\Big( (-\frac{1}{6} + \Xc_1 ) \mathbb{I}_{2\times 2} \,,  \frac{1}{3} + \Xc_1 \Big) B_\mu \non
&+& \frac{g_Y}{ s_{\theta_S}  c_{\theta_S}  } {\rm diag} \Big( \[ - \frac{1}{6} +  ( \frac{1}{6} - \Xc_1 ) s_{\theta_S}^2 \] \mathbb{I}_{2 \times 2}  \,, \frac{1}{3} - (\frac{1}{3} +  \Xc_1 ) s_{ \theta_S }^2 \Big) Z_\mu^\prime \,.
\eeqn

\subsection{The ${\rm SU}(2)_W \otimes {\rm U}(1)_{ Y}$ gauge bosons}
\label{section:gaugeEW}

\para
We express the ${\rm SU}(2)_W \otimes {\rm U}(1)_{Y}$ covariant derivatives in the EW sector for the fundamental and anti-fundamental representations as follows
\beqs\label{eqs:EW_covariant}
\beqn
 D_\mu  &\equiv&   \partial_\mu -i  g_{2W} A_\mu^{ I} T_{ {\rm SU}(2 )}^{ I}  -i  g_{Y} \Yc \mathbb{I}_2  B_{\mu}  \,,\label{eq:EW_covariant_fund} \\[1mm]
 D_\mu &\equiv&  \partial_\mu + i  g_{2W} A_\mu^{ I} ( T_{ {\rm SU}(2 )}^{ I} )^T - i g_{Y} \Yc \mathbb{I}_2  B_{\mu} \,, \label{eq:EW_covariant_antifund}
\eeqn
\eeqs
with $T_{ {\rm SU}(2 )}^{ I} \equiv \hf \sigma^I$.
The gauge fields from Eqs.~\eqref{eqs:EW_covariant} can be expressed in terms of a $2\times 2$ matrix
\beqs
\beqn
&& g_{2W} A^{ I}_\mu T_{ {\rm SU}(2 )}^{ I}  + g_{Y} \Yc \mathbb{I}_2  B_{\mu}  \non
&=&  \frac{g_{2W}}{ \sqrt{2} }\left(
\ba{cc}  
     0  &   W_\mu^+    \\
 W_\mu^- &  0   \\   \ea  \right)  + \frac{ g_{2W} }{2} {\rm diag}  \Big(  A_\mu^3 +  2  t_{\theta_W} \Yc B_{ \mu} \,,   - A_\mu^3  + 2  t_{\theta_W} \Yc B_{ \mu}  \Big)  \,, \\[1mm]
 && - g_{2W} A^{ I}_\mu ( T_{ {\rm SU}(2 )}^{ I} )^T + g_{Y} \Yc \mathbb{I}_2  B_{\mu}  \non
&=&  - \frac{g_{2W}}{ \sqrt{2} }\left(
\ba{cc}  
     0  &   W_\mu^-    \\
 W_\mu^+ &  0   \\   \ea  \right)  +  \frac{ g_{2W} }{2} {\rm diag}  \Big( - A_\mu^3 +  2  t_{\theta_W} \Yc B_{ \mu} \,,   A_\mu^3  + 2  t_{\theta_W} \Yc B_{ \mu}   \Big) \,,
\eeqn
\eeqs
with the Weinberg angle defined by
\beqn\label{eq:Weinberg_angle}
t_{\theta_W} &\equiv& \tan\theta_{W} = \frac{g_{Y} }{ g_{2W } }\,.
\eeqn
%
%

\section{The symmetry breaking patterns and Yukawa couplings in the ${\rm SU}(7)$}
\label{section:breaking}

\para
In this section, we analyze the Yukawa couplings in the ${\rm SU}(7)$ symmetry breaking pattern.
All fermions obtain their masses through the Yukawa couplings to the minimal set of Higgs fields given in Eq.~\eqref{eq:SU7_Yukawa} at each stage of symmetry breaking.

\subsection{The $\Gc_{341} \to \Gc_{331}$ symmetry breaking}
\label{subsection:breaking1}

\para
At the first stage, we consider the Higgs fields of $( \rep{1}\,, \repb{4} \,, -\frac{1}{4} )_{\mathbf{H} \,, \lambda } \subset \repb{ 7_H}_{\,,\lambda}$ and $( \rep{1}\,, \rep{4} \,, +\frac{1}{4} )_{\mathbf{H}}  \subset \rep{7_H}$ for the symmetry breaking, according to Tab.~\ref{tab:SU7Higgs_directions} and Eqs.~\eqref{eqs:SU7_Higgs_Br}.
The Yukawa coupling term between the $\repb{7_F}^\lambda$ and $\rep{21_F}$ is explicitly given by
\beqn\label{eq:Yukawa_341_01}
&& {( Y_\Bc )_{\lambda }}^\kappa  \repb{ 7_F}^{\lambda } \rep{ 21_F} \repb{ 7_H}_{\,,\kappa} + H.c. \non
&\supset& {( Y_\Bc )_{\lambda} }^\kappa   \Big[  ( \rep{1}\,, \repb{4} \,, -\frac{1}{4} )_{\mathbf{F}}^{\lambda } \otimes ( \rep{1}\,, \rep{6} \,, +\frac{1}{2} )_{\mathbf{F}}  \oplus  ( \repb{3}\,, \rep{1} \,, +\frac{1}{3} )_{\mathbf{F}}^{\lambda} \otimes ( \rep{3}\,, \rep{4} \,, -\frac{1}{12} )_{\mathbf{F}} \Big] \non
&\otimes& ( \rep{1}\,, \repb{4} \,, -\frac{1}{4} )_{\mathbf{H}\,,\kappa} + H.c. \non
&\supset& {( Y_\Bc )_{\lambda }}^\kappa \Big[  ( \rep{1}\,, \repb{3} \,, -\frac{1}{3} )_{\mathbf{F}}^{\lambda } \otimes ( \rep{1}\,, \rep{3} \,, +\frac{1}{3} )_{\mathbf{F}} \oplus ( \repb{3}\,, \rep{1} \,, +\frac{1}{3} )_{\mathbf{F}}^{\lambda } \otimes ( \rep{3}\,, \rep{1} \,, -\frac{1}{3} )_{\mathbf{F}}^\prime  \Big] \non
&\otimes& ( \rep{1}\,, \rep{1} \,, 0 )_{\mathbf{H}\,,\kappa } + H.c. \,,
\eeqn
along the  $\Gc_{331}$-singlet direction.
The corresponding vectorlike fermion masses are 
\beqn\label{eq:masses_341_01}
&& \frac{1}{ \sqrt{2}} Y_\Bc  \Big(  \overline{ \eG_L^\prime} \eG_R^\prime - \overline{ \nG_L^\prime} \nG_R^\prime + \overline{ \check \nG_L } \check \nG_R +  \overline{ \DG_L^\prime } \DG_R^\prime \Big)  w_{ \repb{4}\,,{\rm II} } + H.c. \,.
\eeqn
with the VEVs in Eqs.~\eqref{eqs:SU7_Higgs_VEVs_full} and the DRS limit of Yukawa couplings in Eq.~\eqref{eq:SU7_Yukawa_DRS}.
According to our convention of flavor indices, we identify that $ ( \rep{1}\,, \repb{2} \,, -\frac{1}{2} )_{\mathbf{F}}^{\lambda ={\rm II }} \equiv ( \eG_L^\prime \,, - \nG_L^\prime )^T$, $\check \Nc_L^{ \lambda ={\rm II} }= \check \nG_L$, and $( \repb{3}\,, \rep{1} \,, +\frac{1}{3} )_{ \rep{F}}^{\lambda={\rm II}} \equiv {\DG_R^\prime}^c $.

\para
The Yukawa coupling term between two $\rep{35_F}$'s is explicitly given by
\beqn\label{eq:Yukawa_341_02}
&& Y_\Cc  \rep{ 35_F} \rep{ 35_F} \rep{ 7_H} + H.c. \non
&\supset& Y_\Cc \Big[  ( \rep{1}\,, \rep{1} \,, -1 )_{\mathbf{F}} \otimes ( \rep{1}\,, \repb{4} \,, +\frac{3}{4} )_{\mathbf{F}} \oplus  ( \repb{3}\,, \rep{4} \,, -\frac{5}{12} )_{\mathbf{F}} \otimes ( \rep{3}\,, \rep{6} \,, +\frac{1}{6} )_{\mathbf{F}} \Big] \otimes ( \rep{1}\,, \rep{4} \,, +\frac{1}{4} )_{\mathbf{H}} + H.c. \non
&\supset& Y_\Cc   \Big[  ( \rep{1}\,, \rep{1} \,, -1 )_{\mathbf{F}} \otimes ( \rep{1}\,, \rep{1} \,, +1 )_{\mathbf{F}}^{\prime} \oplus ( \repb{3}\,, \rep{3} \,, -\frac{1}{3} )_{\mathbf{F}} \otimes ( \rep{3}\,, \repb{3} \,, +\frac{1}{3} )_{\mathbf{F}} \Big] \non
&\otimes & ( \rep{1}\,, \rep{1} \,, 0 )_{\mathbf{H} } + H.c. \,. 
\eeqn
They give masses to the following vectorlike fermions
\beqn\label{eq:masses_341_02}
&& Y_\Cc  \Big[ ( \repb{3}\,, \rep{2} \,, -\frac{1}{6} )_{\mathbf{F}} \otimes  ( \rep{3}\,, \repb{2} \,, +\frac{1}{6} )_{\mathbf{F}}^{\prime\prime }  \oplus ( \repb{3}\,, \rep{1} \,, -\frac{2}{3} )_{\mathbf{F}}^\prime \otimes (\rep{3}\,, \rep{1} \,, +\frac{2}{3} )_{\mathbf{F}}  \oplus ( \rep{1}\,, \rep{1} \,, -1 )_{\mathbf{F}} \otimes ( \rep{1}\,, \rep{1} \,, +1 )_{\mathbf{F}}^{\prime}   \Big] \non
&\otimes& ( \rep{1}\,, \rep{1} \,, 0 )_{\mathbf{H}} + H.c. \non
&=& \frac{1}{ \sqrt{2} } Y_\Cc \Big(  \overline{ \sG_L} \sG_R - \overline{ \cG_L} \cG_R + \overline{\UG_L} \UG_R + \overline{\EG_L} \EG_R \Big) w_{ \rep{4}} + H.c. \,.
\eeqn
In particular, the left-handed quark doublet of $ ( \rep{3}\,, \repb{2} \,, +\frac{1}{6} )_{\mathbf{F}}^{\prime\prime}$ and its mirror doublet of $( \repb{3}\,, \rep{2} \,, -\frac{1}{6} )_{\mathbf{F}}$ from the $\rep{35_F}$ become massive.
Names of $( \sG\,, \cG\,, \UG\,, \EG)$ become transparent according to their electrical charges, as well as their heavy masses.
One of $( \rep{1}\,, \rep{1} \,, 0)_{\mathbf{F}}^{\lambda^\prime }$ remains massless at this stage.

\para
We find that one of the $( \rep{1}\,, \repb{3} \,, -\frac{1}{3} )_{\mathbf{F}}^{\lambda } \oplus ( \repb{3}\,, \rep{1} \,, +\frac{1}{3} )_{\mathbf{F}}^{\lambda } \subset \repb{7_F}^{\lambda}$ (chosen to be $\lambda={\rm II}$ at this stage) become massive at this stage~\footnote{Loosely speaking and without confusion, we will say one of the anti-fundamental fermion of $\repb{7_F}^\lambda$ becomes massive and is integrated out from the spectrum.}.
After integrating out the massive fermions, the residual massless fermions for the effective $\Gc_{331}$ theory are
\beqs\label{eqs:fermion_331}
\beqn
&& ( \repb{3}\,, \rep{1} \,, +\frac{1}{3} )_{\mathbf{F}}^\Lambda \oplus ( \rep{1}\,, \repb{3} \,, -\frac{1}{3} )_{\mathbf{F}}^\Lambda \oplus ( \rep{1}\,, \rep{1} \,, 0)_{\mathbf{F}}^{\Lambda^\prime }  \subset \repb{7_F}^\Lambda  \,, ~ \Lambda = ({\rm I}\,,3\,, \dot {\rm I}\,, \dot 2 )\,,  \non
&& ( \rep{1}\,, \rep{1} \,, 0)_{\mathbf{F}}^{{\rm II}^\prime } \subset \repb{7_F}^{\rm II} \,, \\[1mm]
&& ( \repb{3}\,, \rep{1} \,, -\frac{2}{3} )_{\mathbf{F}}  \oplus ( \rep{3}\,, \rep{3} \,,0 )_{\mathbf{F}} \oplus ( \rep{1}\,, \repb{3} \,, +\frac{2}{3} )_{\mathbf{F}}   \subset \rep{21_F}  \,,\\[1mm]
&& ( \rep{1}\,, \repb{3} \,, +\frac{2}{3} )_{\mathbf{F}}^{\prime \prime} \oplus  ( \repb{3}\,, \rep{1} \,, -\frac{2}{3} )_{\mathbf{F}}^{\prime\prime }  \oplus  ( \rep{3}\,, \rep{3} \,, 0 )_{\mathbf{F}}^\prime \subset \rep{35_F}\,.
\eeqn
\eeqs
With the massless fermions in Eqs.~\eqref{eqs:fermion_331}, one can explicitly check the anomaly-free conditions of $\[ {\rm SU}(3)_c \]^2\, \[ {\rm U}(1)_{X_1} \]=0$, $\[ {\rm SU}(3)_W \]^2\, \[ {\rm U}(1)_{X_1} \]=0$, and $\[ {\rm U}(1)_{X_1} \]^3=0$ in the effective $\Gc_{331}$ theory.
Thus, the remaining massless fermion numbers after this stage are ${\rm dim}_{ \mathbf{F}}=59$ from Eqs.~\eqref{eqs:fermion_331}.
At least one of the $( \rep{1}\,, \repb{4} \,, -\frac{1}{4} )_{\mathbf{H} \,, \kappa } \subset \repb{7_H}_{\,,\kappa }$ (chosen to be $\kappa={\rm II}$) in Eq.~\eqref{eq:Yukawa_341_01} should develop VEV for the $\Gc_{341}$ symmetry breaking, and we dub this the ``fermion-Higgs matching pattern''.

\subsection{The $\Gc_{331} \to \Gc_{\rm SM}$ symmetry breaking}
\label{subsection:breaking2}

\para
The second-stage symmetry breaking is due to Higgs fields of $ ( \rep{1}\,, \repb{ 3}\,, - \frac{1}{3})_{ \mathbf{H}\,,\lambda} \subset  \repb{ 7_H}_{\,,\lambda}$, $ ( \rep{1}\,, \repb{ 3}\,, - \frac{1}{3})_{ \mathbf{H}\,, \dot \lambda } \subset  \repb{21_H}_{\,, \dot \lambda}$, $( \rep{1}\,, \rep{ 3}\,, +\frac{1}{3})_{ \mathbf{H}} \subset \rep{7_H}$, and $ ( \rep{1}\,, \rep{ 3}\,, +\frac{1}{3})_{ \mathbf{H}}^\prime \subset  \rep{21_H}$, according to Tab.~\ref{tab:SU7Higgs_directions} and Eqs.~\eqref{eqs:SU7_Higgs_Br}.

\para
The Yukawa coupling term between the $\repb{7_F}^\lambda$ and $\rep{21_F}$ is explicitly given by
\beqn\label{eq:Yukawa_331_01}
&& {( Y_\Bc )_{\lambda}}^\kappa \repb{ 7_F}^{\lambda} \rep{ 21_F} \repb{ 7_H}_{\,,\kappa} + H.c. \non
&\supset& {( Y_\Bc )_{\lambda}}^\kappa \Big[  ( \repb{3}\,, \rep{1} \,, +\frac{1}{3} )_{\mathbf{F}}^\lambda \otimes ( \rep{3}\,, \rep{3} \,, 0 )_{\mathbf{F}}  \oplus ( \rep{1}\,, \repb{3} \,, -\frac{1}{3} )_{\mathbf{F}}^\lambda \otimes ( \rep{1}\,, \repb{3} \,, +\frac{2}{3} )_{\mathbf{F}} \Big] \otimes ( \rep{1}\,, \repb{3} \,, -\frac{1}{3})_{\mathbf{H}\,,\kappa } + H.c. \non
&\supset& {( Y_\Bc )_{\lambda} }^\kappa \Big[ ( \repb{3}\,, \rep{1} \,, +\frac{1}{3} )_{\mathbf{F}}^{\lambda } \otimes ( \rep{3}\,, \rep{1} \,, -\frac{1}{3} )_{\mathbf{F}}  \oplus  ( \rep{1}\,, \repb{2} \,, -\frac{1}{2} )_{\mathbf{F}}^{\lambda} \otimes ( \rep{1}\,, \repb{2} \,, +\frac{1}{2})_{\mathbf{F}} \Big]  \non
&\otimes& ( \rep{1}\,, \rep{1} \,, 0 )_{\mathbf{H}\,, \kappa} + H.c. \,,
\eeqn
along the  $\Gc_{\rm SM}$-singlet direction.
They give masses to the following vectorlike fermions
\beqn\label{eq:masses_331_01}
&& Y_\Bc  \Big[ ( \repb{3}\,, \rep{1} \,, +\frac{1}{3} )_{\mathbf{F}}^{\rm I} \otimes ( \rep{3}\,, \rep{1} \,, -\frac{1}{3} )_{\mathbf{F}}   \oplus ( \rep{1}\,, \repb{2} \,, -\frac{1}{2} )_{\mathbf{F}}^{\rm I} \otimes ( \rep{1}\,, \repb{2} \,, +\frac{1}{2})_{\mathbf{F}} \Big]  \otimes ( \rep{1}\,, \rep{1} \,, 0 )_{\mathbf{H}\,, {\rm I} } + H.c. \non
&=& \frac{1}{ \sqrt{2}} Y_\Bc \Big(  \overline{ \DG_L} \DG_R - \overline{ \eG_L} \eG_R + \overline{ \nG_L} \nG_R  \Big) V_{ \repb{3}\,, {\rm I} } + H.c.  \,,
\eeqn
with the Higgs VEVs in Eqs.~\eqref{eqs:SU7_Higgs_VEVs_full} and the DRS limit of Yukawa couplings in Eq.~\eqref{eq:SU7_Yukawa_DRS}.
Correspondingly, we can identify that $( \repb{3}\,, \rep{1} \,, +\frac{1}{3} )_{\mathbf{F}}^{\lambda ={\rm I}} \equiv {\DG_R}^c$, and $( \rep{1}\,, \repb{2} \,, -\frac{1}{2} )_{\mathbf{F}}^{\lambda = {\rm I}} \equiv ( \eG_L\,, - \nG_L)^T$.

\para
The Yukawa coupling term between the $\repb{7_F}^\lambda$ and $\rep{35_F}$ is explicitly given by
\beqn\label{eq:Yukawa_331_02}
&& {( Y_\Sc )_{\dot {\lambda} }}^{\dot \kappa} \repb{ 7_F}^{\dot {\lambda}}\rep{ 35_F} \repb{ 21_H}_{\,,\dot \kappa } + H.c. \non
&\supset& {( Y_\Sc )_{\dot \lambda }}^{\dot \kappa} \Big[  ( \repb{3}\,, \rep{1} \,, +\frac{1}{3} )_{\mathbf{F}}^{\dot \lambda } \otimes ( \rep{3}\,, \rep{3} \,, 0 )_{\mathbf{F}}^\prime \oplus ( \rep{1}\,, \repb{3} \,, -\frac{1}{3} )_{\mathbf{F}}^{\dot \lambda } \otimes ( \rep{1}\,, \repb{3} \,, +\frac{2}{3} )_{\mathbf{F}}^{\prime\prime}   \Big] \otimes  ( \rep{1}\,, \repb{3} \,, -\frac{1}{3})_{\mathbf{H}\,,\dot \kappa}  + H.c. \non
&\supset& {( Y_\Sc )_{\dot {\lambda}  }}^{\dot \kappa} \Big[  ( \repb{3}\,, \rep{1} \,, +\frac{1}{3} )_{\mathbf{F}}^{\dot \lambda } \otimes   ( \rep{3}\,, \rep{1} \,, -\frac{1}{3} )_{\mathbf{F}}^{\prime\prime} \oplus  ( \rep{1}\,, \repb{2} \,, -\frac{1}{2} )_{\mathbf{F}}^{\dot {\lambda} } \otimes   ( \rep{1}\,, \repb{2} \,, +\frac{1}{2} )_{\mathbf{F}}^{\prime\prime} \Big] \non
& \otimes&  ( \rep{1}\,, \rep{1} \,, 0 )_{\mathbf{H}\,,\dot\kappa } + H.c. \,,
\eeqn
along the  $\Gc_{\rm SM}$-singlet direction.
They give masses to the following vectorlike fermions
\beqn\label{eq:masses_331_02}
&& Y_\Sc   \Big[  ( \repb{3}\,, \rep{1} \,, +\frac{1}{3} )_{\mathbf{F}}^{\dot {\rm I}  } \otimes   ( \rep{3}\,, \rep{1} \,, -\frac{1}{3} )_{\mathbf{F}}^{\prime\prime} \oplus  ( \rep{1}\,, \repb{2} \,, -\frac{1}{2} )_{\mathbf{F}}^{\dot {\rm I}} \otimes   ( \rep{1}\,, \repb{2} \,, +\frac{1}{2} )_{\mathbf{F}}^{\prime\prime} \Big]  \otimes  ( \rep{1}\,, \rep{1} \,, 0 )_{\mathbf{H}\,,\dot {\rm I} } + H.c. \non
&=& \frac{1}{ \sqrt{2}}   Y_\Sc  \Big( \overline{\SG_L} \SG_R - \overline{ \eG_L^{\prime\prime} } \eG_R^{\prime\prime} + \overline{ \nG_L^{\prime\prime} } \nG_R^{\prime\prime}   \Big) V_{ \repb{3}\,, \dot {\rm I}}  + H.c. \,.
\eeqn
Correspondingly, we can identify that $( \repb{3}\,, \rep{1} \,, +\frac{1}{3} )_{\mathbf{F}}^{\dot \lambda = \dot {\rm I}} \equiv {\SG_R}^c$, and $( \rep{1}\,, \repb{2} \,, -\frac{1}{2} )_{\mathbf{F}}^{\dot \lambda = \dot {\rm I}} \equiv ( \eG_L^{\prime \prime } \,, - \nG_L^{\prime\prime} )^T$.

\para
The Yukawa coupling term between two $\rep{35_F}$'s is explicitly given by
\beqn\label{eq:Yukawa_331_03} 
&& Y_{\Cc} \rep{ 35_F} \rep{ 35_F} \rep{ 7_H} + H.c. \non
&\supset& Y_\Cc \Big[  ( \rep{1}\,, \rep{1} \,, -1 )_{\mathbf{F}} \otimes ( \rep{1}\,, \repb{4} \,, +\frac{3}{4} )_{\mathbf{F}} \oplus  ( \repb{3}\,, \rep{4} \,, -\frac{5}{12} )_{\mathbf{F}} \otimes ( \rep{3}\,, \rep{6} \,, +\frac{1}{6} )_{\mathbf{F}} \Big] \otimes ( \rep{1}\,, \rep{4} \,, +\frac{1}{4} )_{\mathbf{H}} + H.c. \non
&\supset& Y_{\Cc} \Big[ ( \rep{1}\,, \rep{1} \,, -1 )_{\mathbf{F}} \otimes ( \rep{1}\,, \repb{3} \,, +\frac{2}{3} )_{\mathbf{F}}^{\prime\prime }  \oplus ( \repb{3}\,, \rep{3} \,, -\frac{1}{3} )_{\mathbf{F}} \otimes ( \rep{3}\,, \rep{3} \,, 0 )_{\mathbf{F}}^\prime \non
& \oplus&  ( \repb{3}\,, \rep{1} \,, -\frac{2}{3} )_{\mathbf{F}}^{\prime \prime} \otimes ( \rep{3}\,, \repb{3} \,, +\frac{1}{3} )_{\mathbf{F}}    \Big] \otimes ( \rep{1}\,, \rep{3} \,, +\frac{1}{3} )_{\mathbf{H}} + H.c. \non
&\supset& Y_{\Cc} \Big[ ( \rep{1}\,, \rep{1} \,, -1 )_{\mathbf{F}} \otimes ( \rep{1}\,, \rep{1} \,, +1 )_{\mathbf{F}}^{\prime\prime }  \oplus ( \repb{3}\,, \rep{2} \,, -\frac{1}{6} )_{\mathbf{F}} \otimes ( \rep{3}\,, \rep{2} \,, +\frac{1}{6} )_{\mathbf{F}}^\prime \non
& \oplus&  ( \repb{3}\,, \rep{1} \,, -\frac{2}{3} )_{\mathbf{F}}^{\prime \prime} \otimes ( \rep{3}\,, \rep{1} \,, +\frac{2}{3} )_{\mathbf{F}} \Big]  \otimes ( \rep{1}\,, \rep{1} \,, 0)_{\mathbf{H}} + H.c. \,, 
\eeqn
along the  $\Gc_{\rm SM}$-singlet direction.
They give the following fermion mass mixing terms
\beqn\label{eq:masses_331_03}
&& Y_{\Cc} \Big[ ( \rep{1}\,, \rep{1} \,, -1 )_{\mathbf{F}} \otimes ( \rep{1}\,, \rep{1} \,, +1 )_{\mathbf{F}}^{\prime\prime }  \oplus ( \repb{3}\,, \rep{2} \,, -\frac{1}{6} )_{\mathbf{F}} \otimes ( \rep{3}\,, \rep{2} \,, +\frac{1}{6} )_{\mathbf{F}}^\prime  \oplus  ( \repb{3}\,, \rep{1} \,, -\frac{2}{3} )_{\mathbf{F}}^{\prime \prime} \otimes ( \rep{3}\,, \rep{1} \,, +\frac{2}{3} )_{\mathbf{F}} \Big] \non
&& \otimes ( \rep{1}\,, \rep{1} \,, 0)_{\mathbf{H}} + H.c. \non
&=& \frac{1}{ \sqrt{2}} Y_{\Cc} \Big( \overline{ \EG_L} \mu_R + \overline{ s_L} \sG_R - \overline{ c_L} \cG_R + \overline{ \UG_L} c_R  \Big) V_{ \rep{3}}  + H.c. \,.
\eeqn
Note that the $\rep{7_H}$ has already developed a VEV for the $\Gc_{341}$ symmetry breaking direction, and the VEV of $V_{\rep{3}}$ is generated by Higgs mixing operator according to Fig.~\ref{fig:VEV_chain}.

\para
The Yukawa coupling term that mixes the $\rep{21_F}$ and the $\rep{35_F}$ is explicitly given by
\beqn\label{eq:Yukawa_331_04}
&& Y_{\Tc\Cc} \rep{ 21_F} \rep{ 35_F} \rep{ 21_H} + H.c. \non
&\supset& Y_{\Tc\Cc} \Big[  ( \repb{3}\,, \rep{1} \,, -\frac{2}{3} )_{\mathbf{F}} \otimes ( \rep{3}\,, \repb{3} \,, +\frac{1}{3} )_{\mathbf{F}}  \oplus  ( \rep{3}\,, \rep{3} \,, 0)_{\mathbf{F}} \otimes ( \repb{3}\,, \rep{3} \,, - \frac{1}{3})_{\mathbf{F}}  \non
&\oplus&  ( \rep{1}\,, \repb{3} \,, +\frac{2}{3} )_{\mathbf{F}} \otimes ( \rep{1}\,, \rep{1} \,, -1 )_{\mathbf{F}}   \Big]  \otimes ( \rep{1} \,, \rep{3} \,, +\frac{1}{3} )_{\mathbf{H}}^\prime  + H.c. \non
&\supset& Y_{\Tc\Cc} \Big[  ( \repb{3}\,, \rep{1} \,, -\frac{2}{3} )_{\mathbf{F}} \otimes ( \rep{3}\,, \rep{1} \,, +\frac{2}{3} )_{\mathbf{F}}  \oplus  ( \rep{3}\,, \rep{2} \,, +\frac{1}{6} )_{\mathbf{F}} \otimes ( \repb{3}\,, \rep{2} \,, - \frac{1}{6} )_{\mathbf{F}} \non
&\oplus&  ( \rep{1}\,, \rep{1} \,, +1 )_{\mathbf{F}} \otimes ( \rep{1}\,, \rep{1} \,, -1 )_{\mathbf{F}}  \Big]  \otimes  ( \rep{1}\,, \rep{1} \,, 0)_{\mathbf{H}}^\prime + H.c. \,,
\eeqn
along the  $\Gc_{\rm SM}$-singlet direction.
They give the following fermion mass mixing terms
\beqn\label{eq:masses_331_04}
&& Y_{\Tc\Cc} \Big[  ( \repb{3}\,, \rep{1} \,, -\frac{2}{3} )_{\mathbf{F}} \otimes ( \rep{3}\,, \rep{1} \,, +\frac{2}{3} )_{\mathbf{F}}  \oplus  ( \rep{3}\,, \rep{2} \,, +\frac{1}{6} )_{\mathbf{F}} \otimes ( \repb{3}\,, \rep{2} \,, - \frac{1}{6} )_{\mathbf{F}}  \oplus  ( \rep{1}\,, \rep{1} \,, +1 )_{\mathbf{F}} \otimes ( \rep{1}\,, \rep{1} \,, -1 )_{\mathbf{F}}  \Big] \non
&& \otimes  ( \rep{1}\,, \rep{1} \,, 0)_{\mathbf{H}}^\prime + H.c.  \non
&=& \frac{1}{ \sqrt{2}} Y_{\Tc\Cc} \Big(  \overline{ \UG_L} t_R + \overline{t_L} \cG_R - \overline{b_L} \sG_R  + \overline{ \EG_L } \tau_R \Big)  V_{ \rep{3}}^\prime + H.c. \,.
\eeqn

\para
By analyzing the anomaly-free conditions for the effective $\Gc_{\rm SM}$ theory, we find the residual massless fermions of
\beqs\label{eqs:fermion_321}
\beqn
&& ( \repb{3}\,, \rep{1} \,, +\frac{1}{3} )_{\mathbf{F}}^\Lambda \oplus ( \rep{1}\,, \repb{2} \,, -\frac{1}{2} )_{\mathbf{F}}^\Lambda \oplus ( \rep{1}\,, \rep{1} \,, 0)_{\mathbf{F}}^{\Lambda }  \oplus ( \rep{1}\,, \rep{1} \,, 0)_{\mathbf{F}}^{\Lambda^\prime }  \subset \repb{7_F}^\Lambda\,,~ \Lambda=(\dot 2\,, 3) \,, \label{eq:fermion_321a}
 \\[1mm]
&& ( \repb{3}\,, \rep{1} \,, -\frac{2}{3} )_{\mathbf{F}}  \oplus ( \rep{3}\,, \rep{2} \,, +\frac{1}{6})_{\mathbf{F}} \oplus ( \rep{1}\,, \rep{1} \,, +1 )_{\mathbf{F}}   \subset \rep{21_F}  \,, \label{eq:fermion_321b} \\[1mm]
&& ( \rep{1}\,, \rep{1} \,, +1 )_{\mathbf{F}}^{\prime\prime} \oplus  ( \repb{3}\,, \rep{1} \,, -\frac{2}{3} )_{\mathbf{F}}^{\prime\prime}  \oplus  ( \rep{3}\,, \rep{2} \,, +\frac{1}{6} )_{\mathbf{F}}^{\prime}  \subset \rep{35_F}  \,, \label{eq:fermion_321c}\\[1mm]
&&( \rep{1}\,, \rep{1} \,, 0)_{\mathbf{F}}^{{\rm II}^\prime }  \oplus ( \rep{1}\,, \rep{1} \,, 0)_{\mathbf{F}}^{\rm I } \oplus ( \rep{1}\,, \rep{1} \,, 0)_{\mathbf{F}}^{{\rm I}^\prime}  \oplus ( \rep{1}\,, \rep{1} \,, 0)_{\mathbf{F}}^{\dot {\rm I}} \oplus ( \rep{1}\,, \rep{1} \,, 0)_{\mathbf{F}}^{\dot {\rm I}^\prime}  \,, \label{eq:fermion_321d}
\eeqn
\eeqs
after this stage.
The massive anti-fundamental fermions are chosen to be $\repb{7_F}^{\dot {\rm I}\,, {\rm I}}$ as in Eqs.~\eqref{eq:masses_331_01} and \eqref{eq:masses_331_02}.
Explicitly, the massless fermions in Eqs.~\eqref{eq:fermion_321a}, \eqref{eq:fermion_321b}, and \eqref{eq:fermion_321c} form two-generational SM fermions.

\subsection{The EWSB}
\label{subsection:breaking3}

\para
Through the decomposition of all Higgs fields according to the symmetry breaking pattern in Tab.~\ref{tab:SU7Higgs_directions}, we found that the $\rep{35_H}$ only contains the EWSB direction.
Similar situation also happened in the ${\rm SU}(6)$ toy model, where the Higgs field of $\rep{15_H}$ that gives the top quark Yukawa coupling through the $\rep{15_F} \rep{15_F} \rep{15_H}+H.c.$ contributes mostly to the EWSB~\cite{Chen:2021zwn}.
Based on these facts, we conjecture that the EWSB is mostly achieved by Higgs field of $( \rep{1} \,, \repb{2}\,,+\frac{1}{2} )_{\mathbf{H}}^{\prime} \subset \rep{35_H} $, and the corresponding Yukawa coupling term in Eq.~\eqref{eq:SU7_Yukawa} leads to the top quark mass as follows
\beqn\label{eq:Yukawa_top}
&& Y_\Tc \rep{ 21_F} \rep{ 21_F} \rep{ 35_H} + H.c. \non
&\supset& Y_\Tc ( \repb{3}\,, \rep{1} \,, -\frac{2}{3} )_{\mathbf{F}} \otimes ( \rep{3}\,, \rep{2} \,, +\frac{1}{6} )_{\mathbf{F}}  \otimes  ( \rep{1}\,, \repb{2} \,, +\frac{1}{2} )_{\mathbf{H}}^{\prime}  + H.c. \non
&=&  \frac{1}{\sqrt{2} } Y_\Tc \overline{t_L} t_R  v_t + H.c.  \,.
\eeqn
Below, we list all other fermion mass terms from the additional EWSB VEVs generated from the Higgs mixing operators, as was displayed in Fig.~\ref{fig:VEV_chain}.

\para
The Yukawa coupling term between the $\repb{ 7_F}^\lambda$ and the $\rep{21_F}$ can lead to
\beqn\label{eq:Yukawa_EW_01}
&& {( Y_\Bc )_{\lambda} }^\kappa  \repb{ 7_F}^\lambda \rep{ 21_F} \repb{ 7_H}_{\,,\kappa} + H.c.  \non
&\supset& {( Y_\Bc )_{\lambda} }^\kappa   \Big[ ( \repb{3}\,, \rep{1} \,, +\frac{1}{3} )_{\mathbf{F}}^\lambda \otimes ( \rep{3}\,, \rep{4} \,, -\frac{1}{12} )_{\mathbf{F}}  \oplus ( \rep{1}\,, \repb{4} \,, -\frac{1}{4} )_{\mathbf{F}}^\lambda \otimes ( \rep{1}\,, \rep{6} \,, +\frac{1}{2} )_{\mathbf{F}}  \Big] \otimes ( \rep{1}\,, \repb{4} \,, -\frac{1}{4})_{\mathbf{H}\,, \kappa} +H.c.  \non
&\supset& {( Y_\Bc )_{\lambda} }^\kappa \Big[ ( \repb{3}\,, \rep{1} \,, +\frac{1}{3} )_{\mathbf{F}}^\lambda \otimes ( \rep{3}\,, \rep{3} \,, 0 )_{\mathbf{F}}  \oplus ( \rep{1}\,, \repb{3} \,, -\frac{1}{3} )_{\mathbf{F}}^\lambda  \otimes ( \rep{1}\,, \repb{3} \,, +\frac{2}{3} )_{\mathbf{F}}  \non
&\oplus& ( \rep{1}\,, \rep{1} \,, 0 )_{\mathbf{F}}^{\lambda^\prime } \otimes ( \rep{1}\,, \rep{3} \,, +\frac{1}{3} )_{\mathbf{F}}  \Big] \otimes ( \rep{1}\,, \repb{3} \,, -\frac{1}{3})_{\mathbf{H}\,, \kappa} +H.c.  \non
&\supset& {( Y_\Bc )_{\lambda} }^\kappa   \Big[  ( \repb{3}\,, \rep{1} \,, +\frac{1}{3} )_{\mathbf{F}}^\lambda \otimes ( \rep{3}\,, \rep{2} \,, +\frac{1}{6} )_{\mathbf{F}}  \oplus ( \rep{1}\,, \repb{2} \,, -\frac{1}{2} )_{\mathbf{F}}^\lambda \otimes ( \rep{1}\,, \rep{1} \,, +1 )_{\mathbf{F}} \non
&\oplus& ( \rep{1}\,, \rep{1} \,, 0 )_{\mathbf{F}}^{\lambda } \otimes ( \rep{1}\,, \repb{2} \,, +\frac{1}{2} )_{\mathbf{F}}  \oplus ( \rep{1}\,, \rep{1} \,, 0 )_{\mathbf{F}}^{\lambda^\prime } \otimes ( \rep{1}\,, \rep{2} \,, +\frac{1}{2} )_{\mathbf{F}}^\prime   \Big] \otimes  ( \rep{1}\,, \repb{2} \,, -\frac{1}{2})_{\mathbf{H}\,, \kappa} +H.c.  \,,
\eeqn
along the EWSB direction.
They give fermion masses of
\beqn\label{eq:masses_EW_01}
&& Y_\Bc   \Big[  ( \repb{3}\,, \rep{1} \,, +\frac{1}{3} )_{\mathbf{F}}^3 \otimes ( \rep{3}\,, \rep{2} \,, +\frac{1}{6} )_{\mathbf{F}}  \oplus ( \rep{1}\,, \repb{2} \,, -\frac{1}{2} )_{\mathbf{F}}^3 \otimes ( \rep{1}\,, \rep{1} \,, +1 )_{\mathbf{F}} \non
&& \oplus ( \rep{1}\,, \rep{1} \,, 0 )_{\mathbf{F}}^{ 3 } \otimes ( \rep{1}\,, \repb{2} \,, +\frac{1}{2} )_{\mathbf{F}}  \oplus ( \rep{1}\,, \rep{1} \,, 0 )_{\mathbf{F}}^{ 3^\prime } \otimes ( \rep{1}\,, \rep{2} \,, +\frac{1}{2} )_{\mathbf{F}}^\prime   \Big] \otimes  ( \rep{1}\,, \repb{2} \,, -\frac{1}{2})_{\mathbf{H}\,, 3} +H.c.  \non
&=&  \frac{1}{ \sqrt{2}}  Y_\Bc  \Big( \overline{b_L} b_R + \overline{ \tau_L} \tau_R + \overline{ \check \Nc_L^3} \nG_R  + \overline{ \check \Nc_L^{3^\prime } } \nG_R^\prime  \Big)  u_{ \repb{2}\,, 3} + H.c. \,,
\eeqn
with $\lambda=\kappa=3$ for the DRS limit of Yukawa couplings.
We can identify that $( \repb{3}\,, \rep{1} \,, +\frac{1}{3} )_{\mathbf{F}}^{\lambda = 3} \equiv {b_R}^c$, and $( \rep{1}\,, \repb{2} \,, -\frac{1}{2} )_{\mathbf{F}}^{\lambda = 3} \equiv ( \tau_L\,, - \nu_{\tau\,L})^T$.
Obviously, the Eq.~\eqref{eq:masses_EW_01} gives common tree-level masses to the bottom quark and the tau lepton, which is the same as the prediction in the supersymmetric ${\rm SU}(5)$ Georgi-Glashow model and the third-generational ${\rm SU}(6)$ model~\cite{Chen:2021zwn}.

\para
The Yukawa coupling terms between the $\repb{ 7_F}^{\dot \lambda}$ and the $\rep{35_F}$ can lead to
\beqn\label{eq:Yukawa_EW_02}
&& {( Y_\Sc )_{\dot \lambda} }^{\dot \kappa}  \repb{ 7_F}^{\dot \lambda} \rep{ 35_F} \repb{ 21_H}_{\,,\dot \kappa} + H.c.  \non
&\supset& {( Y_\Sc )_{\dot \lambda} }^{\dot \kappa} \Big[  ( \repb{3}\,, \rep{1} \,, +\frac{1}{3}  )_{ \mathbf{F}}^{\dot \lambda} \otimes ( \rep{3}\,, \rep{6} \,, +\frac{1}{6} )_{ \mathbf{F}}  \oplus  ( \rep{1}\,, \repb{4} \,, -\frac{1}{4} )_{ \mathbf{F}}^{\dot \lambda}  \otimes ( \rep{1}\,, \repb{4} \,, +\frac{3}{4} )_{ \mathbf{F}}  \Big] \otimes ( \rep{1}\,, \rep{6} \,, -\frac{1}{2}  )_{ \mathbf{H}\,, \dot \kappa} \non
&\supset& {( Y_\Sc )_{\dot \lambda} }^{\dot \kappa}  \Big[  ( \repb{3}\,, \rep{1} \,, +\frac{1}{3}  )_{ \mathbf{F}}^{\dot \lambda} \otimes ( \rep{3}\,, \rep{3} \,, 0  )_{ \mathbf{F}}^\prime  \oplus  ( \rep{1} \,, \repb{3}\,, -\frac{1}{3}  )_{ \mathbf{F}}^{\dot \lambda}   \otimes ( \rep{1} \,, \repb{3}\,, +\frac{2}{3}  )_{ \mathbf{F}}^{\prime\prime} \Big] \otimes ( \rep{1}\,, \repb{3} \,, -\frac{1}{3})_{\mathbf{H}\,, \dot \kappa} \non
&+& {( Y_\Sc )_{\dot \lambda} }^{\dot \kappa}  \Big[  ( \repb{3}\,, \rep{1} \,, +\frac{1}{3}  )_{ \mathbf{F}}^{\dot \lambda} \otimes  ( \rep{3}\,, \repb{3} \,, +\frac{1}{3} )_{ \mathbf{F}} \oplus ( \rep{1} \,, \repb{3}\,, -\frac{1}{3}  )_{ \mathbf{F}}^{\dot \lambda}   \otimes  ( \rep{1} \,, \rep{1}\,, +1 )_{ \mathbf{F}}^{\prime  } \non
&\oplus& ( \rep{1}\,, \rep{1} \,, 0  )_{ \mathbf{F}}^{{\dot \lambda}^\prime }  \otimes  ( \rep{1}\,, \repb{3} \,, +\frac{2}{3}  )_{ \mathbf{F}}^{\prime \prime }   \Big] \otimes ( \rep{1}\,, \rep{3} \,, -\frac{2}{3})_{\mathbf{H}\,, \dot \kappa}  + H.c. \non
&\supset& {( Y_\Sc )_{\dot \lambda} }^{\dot \kappa}  \Big[ ( \repb{3}\,, \rep{1} \,, +\frac{1}{3}  )_{ \mathbf{F}}^{\dot \lambda} \otimes ( \rep{3}\,, \rep{2} \,, +\frac{1}{6}  )_{ \mathbf{F}}^\prime   \oplus ( \rep{1}\,, \repb{2} \,, -\frac{1}{2}  )_{ \mathbf{F}}^{\dot \lambda}  \otimes ( \rep{1}\,, \rep{1} \,, +1  )_{ \mathbf{F}}^{\prime \prime } \non
& \oplus& ( \rep{1}\,, \rep{1} \,, 0  )_{ \mathbf{F}}^{\dot \lambda}  \otimes  ( \rep{1}\,, \repb{2} \,, +\frac{1}{2}  )_{ \mathbf{F}}^{\prime \prime }  \Big] \otimes ( \rep{1}\,, \repb{2} \,, -\frac{1}{2})_{\mathbf{H}\,, \dot \kappa} \non
&+& {( Y_\Sc )_{\dot \lambda} }^{\dot \kappa} \Big[ ( \repb{3}\,, \rep{1} \,, +\frac{1}{3}  )_{ \mathbf{F}}^{\dot \lambda} \otimes ( \rep{3}\,, \repb{2} \,, +\frac{1}{6}  )_{ \mathbf{F}}^{\prime \prime}  \oplus ( \rep{1}\,, \repb{2} \,, -\frac{1}{2}  )_{ \mathbf{F}}^{\dot \lambda}  \otimes ( \rep{1}\,, \rep{1} \,, +1  )_{ \mathbf{F}}^{ \prime } \non
& \oplus& ( \rep{1}\,, \rep{1} \,, 0  )_{ \mathbf{F}}^{{\dot \lambda}^\prime }  \otimes  ( \rep{1}\,, \repb{2} \,, +\frac{1}{2}  )_{ \mathbf{F}}^{\prime \prime }  \Big]  \otimes ( \rep{1}\,, \rep{2} \,, -\frac{1}{2})_{\mathbf{H}\,, \dot \kappa}  + H.c. \,,
\eeqn
along the EWSB direction.
They give fermion masses of
\beqs\label{eqs:masses_EW_02}
\beqn
&& Y_\Sc  \Big[ ( \repb{3}\,, \rep{1} \,, +\frac{1}{3}  )_{ \mathbf{F}}^{\dot 2} \otimes ( \rep{3}\,, \rep{2} \,, +\frac{1}{6}  )_{ \mathbf{F}}^\prime   \oplus ( \rep{1}\,, \repb{2} \,, -\frac{1}{2}  )_{ \mathbf{F}}^{\dot 2}  \otimes ( \rep{1}\,, \rep{1} \,, +1  )_{ \mathbf{F}}^{\prime \prime } \oplus ( \rep{1}\,, \rep{1} \,, 0  )_{ \mathbf{F}}^{\dot 2}  \otimes  ( \rep{1}\,, \repb{2} \,, +\frac{1}{2}  )_{ \mathbf{F}}^{\prime \prime }  \Big] \non
&\otimes & ( \rep{1}\,, \repb{2} \,, -\frac{1}{2})_{\mathbf{H}\,, \dot 2}  + H.c. \non
&=& \frac{1}{ \sqrt{2}} Y_\Sc  \Big(  \overline{s_L} s_R  + \overline{\mu_L} \mu_R + \overline{ \check \Nc_L^{ {\dot 2} } } \nG_R^{\prime \prime }   \Big)  u_{ \repb{2}\,,\dot 2} + H.c. \,,\label{eq:masses_EW_02a} \\[1mm]
&& Y_\Sc \Big[ ( \repb{3}\,, \rep{1} \,, +\frac{1}{3}  )_{ \mathbf{F}}^{\dot 2 } \otimes ( \rep{3}\,, \repb{2} \,, +\frac{1}{6}  )_{ \mathbf{F}}^{\prime \prime}  \oplus ( \rep{1}\,, \repb{2} \,, -\frac{1}{2}  )_{ \mathbf{F}}^{\dot 2}  \otimes ( \rep{1}\,, \rep{1} \,, +1  )_{ \mathbf{F}}^{ \prime }  \oplus ( \rep{1}\,, \rep{1} \,, 0  )_{ \mathbf{F}}^{{\dot 2}^\prime }  \otimes  ( \rep{1}\,, \repb{2} \,, +\frac{1}{2}  )_{ \mathbf{F}}^{\prime \prime }  \Big] \non
&\otimes & ( \rep{1}\,, \rep{2} \,, -\frac{1}{2})_{\mathbf{H}\,, \dot 2}  + H.c.   \non
&=& \frac{1}{ \sqrt{2}}   Y_\Sc \Big( \overline{\sG_L } s_R  + \overline{\mu_L} \EG_R + \overline{ \check \Nc_L^{ {\dot 2}^\prime } } \nG_R^{\prime \prime }  \Big) u_{ \rep{2}\,,\dot 2} + H.c. \,.\label{eq:masses_EW_02b}
\eeqn
\eeqs
Eq.~\eqref{eq:masses_EW_02a} gives common tree-level masses to the sea quark and the muon.
The other EWSB VEV from Eq.~\eqref{eq:masses_EW_02b} gives mass mixing terms between the second-generational fermions and heavy partner fermions.

\para
The Yukawa coupling terms between two $\rep{35_F}$'s can lead to
\beqn\label{eq:Yukawa_EW_03}
&& Y_\Cc \rep{35_F} \rep{35_F} \rep{7_H} + H.c. \non
&\supset&   Y_\Cc  \Big[  ( \rep{1}\,, \rep{1} \,, -1  )_{ \mathbf{F}}  \otimes  ( \rep{1}\,, \repb{4} \,, +\frac{3}{4} )_{ \mathbf{F}}   \oplus  ( \repb{3}\,, \rep{4} \,, -\frac{5}{12}  )_{ \mathbf{F}}  \otimes ( \rep{3}\,, \rep{6} \,, +\frac{1}{6}  )_{ \mathbf{F}}  \Big] \otimes ( \rep{1}\,, \rep{4} \,, +\frac{1}{4})_{ \mathbf{H}} +H.c.  \non
&\supset&  Y_\Cc  \Big[  ( \rep{1}\,, \rep{1} \,, -1  )_{ \mathbf{F}}  \otimes  ( \rep{1}\,, \repb{3} \,, +\frac{2}{3} )_{ \mathbf{F}}^{\prime \prime }   \oplus ( \repb{3}\,, \rep{3} \,, - \frac{1}{3}  )_{ \mathbf{F}} \otimes ( \rep{3}\,, \rep{3} \,, 0 )_{ \mathbf{F}}^\prime \non
&\oplus& ( \repb{3}\,, \rep{1} \,, - \frac{2}{3}  )_{ \mathbf{F}}^{\prime \prime} \otimes ( \rep{3}\,, \repb{3} \,, +\frac{1}{3} )_{ \mathbf{F}}  \Big] \otimes ( \rep{1}\,, \rep{3} \,, +\frac{1}{3} )_{ \mathbf{H}} + H.c.  \non
&\supset& Y_\Cc \Big[ ( \rep{1}\,, \rep{1} \,, -1  )_{ \mathbf{F}}  \otimes ( \rep{1}\,, \repb{2} \,, +\frac{1}{2} )_{ \mathbf{F}}^{\prime\prime}   \oplus ( \repb{3}\,, \rep{2} \,, - \frac{1}{6}  )_{ \mathbf{F}} \otimes  ( \rep{3}\,, \rep{1} \,, -\frac{1}{3}  )_{ \mathbf{F}}^{\prime\prime}    \non
&\oplus& ( \repb{3}\,, \rep{1} \,, - \frac{2}{3}  )_{ \mathbf{F}}^\prime \otimes  ( \rep{3}\,, \rep{2} \,, +\frac{1}{6} )_{ \mathbf{F}}^{\prime}   \oplus  ( \repb{3}\,, \rep{1} \,, - \frac{2}{3}  )_{ \mathbf{F}}^{\prime \prime} \otimes  ( \rep{3}\,, \repb{2} \,, +\frac{1}{6} )_{ \mathbf{F}}^{\prime\prime}  \Big]  \otimes ( \rep{1}\,, \rep{2} \,, +\frac{1}{2})_{ \mathbf{H}} +H.c. \,.
\eeqn
along the EWSB direction.
They give fermion mass mixing terms of
\beqn\label{eq:masses_EW_03}
&& Y_\Cc \Big[ ( \rep{1}\,, \rep{1} \,, -1  )_{ \mathbf{F}}  \otimes ( \rep{1}\,, \repb{2} \,, +\frac{1}{2} )_{ \mathbf{F}}^{\prime\prime}   \oplus ( \repb{3}\,, \rep{2} \,, - \frac{1}{6}  )_{ \mathbf{F}} \otimes  ( \rep{3}\,, \rep{1} \,, -\frac{1}{3}  )_{ \mathbf{F}}^{\prime\prime}    \non
&\oplus& ( \repb{3}\,, \rep{1} \,, - \frac{2}{3}  )_{ \mathbf{F}}^\prime \otimes  ( \rep{3}\,, \rep{2} \,, +\frac{1}{6} )_{ \mathbf{F}}^{\prime}   \oplus  ( \repb{3}\,, \rep{1} \,, - \frac{2}{3}  )_{ \mathbf{F}}^{\prime \prime} \otimes  ( \rep{3}\,, \repb{2} \,, +\frac{1}{6} )_{ \mathbf{F}}^{\prime\prime}  \Big] \otimes ( \rep{1}\,, \rep{2} \,, +\frac{1}{2})_{ \mathbf{H}} +H.c. \non
&=&  \frac{1}{ \sqrt{2}} Y_{\Cc} \Big(  - \overline{\EG_L} \eG_R^{\prime\prime }  + \overline{ \SG_L} \sG_R + \overline{ c_L} \UG_R - \overline{ \cG_L } c_R  \Big) u_{ \rep{2}}  + H.c. \,.
\eeqn

\para
The Yukawa coupling term that mixes the $\rep{ 21_F}$ and the $\rep{35_F}$ in Eq.~\eqref{eq:Yukawa_331_04} can further become
\beqn\label{eq:Yukawa_EW_04}
&& Y_{\Tc\Cc} \rep{21_F} \rep{35_F} \rep{21_H} + H.c. \non
&\supset&  Y_{\Tc\Cc} \Big[ ( \repb{3}\,, \rep{1} \,, - \frac{2}{3}  )_{ \mathbf{F}} \otimes ( \rep{3}\,, \rep{6} \,, + \frac{1}{6}  )_{ \mathbf{F}}  \oplus   ( \rep{3}\,, \rep{4} \,, - \frac{1}{12}  )_{ \mathbf{F}}  \otimes ( \repb{3}\,, \rep{4} \,, - \frac{5}{12}  )_{ \mathbf{F}} \non
&\oplus& ( \rep{1}\,, \rep{6} \,, + \frac{1}{2}  )_{ \mathbf{F}}  \otimes ( \rep{1}\,, \rep{1} \,, -1 )_{ \mathbf{F}} \Big] \otimes ( \rep{1}\,, \rep{6} \,, +\frac{1}{2})_{ \mathbf{H}} \non
&\supset&  Y_{\Tc\Cc} \Big[ ( \repb{3}\,, \rep{1} \,, - \frac{2}{3}  )_{ \mathbf{F}} \otimes  ( \rep{3}\,, \repb{3} \,, + \frac{1}{3}  )_{ \mathbf{F}} \oplus  ( \rep{3}\,, \rep{3} \,, 0 )_{ \mathbf{F}} \otimes  ( \repb{3}\,, \rep{3} \,, - \frac{1}{3}  )_{ \mathbf{F}} \non
&\oplus& ( \rep{1}\,, \repb{3} \,, +\frac{2}{3} )_{ \mathbf{F}}  \otimes ( \rep{1}\,, \rep{1} \,, -1 )_{ \mathbf{F}}  \Big]  \otimes  ( \rep{1}\,, \rep{3} \,, +\frac{1}{3})_{ \mathbf{H}}^\prime \non
&+& Y_{\Tc\Cc} \Big[  ( \repb{3}\,, \rep{1} \,, - \frac{2}{3}  )_{ \mathbf{F}} \otimes ( \rep{3}\,, \rep{3} \,, 0  )_{ \mathbf{F}}^\prime  \oplus ( \rep{3}\,, \rep{3} \,, 0  )_{ \mathbf{F}} \otimes ( \repb{3}\,, \rep{1} \,, -\frac{2}{3}  )_{ \mathbf{F}}^{\prime\prime} \non
&\oplus& ( \rep{3}\,, \rep{1} \,, -\frac{1}{3}  )_{ \mathbf{F}}^\prime \otimes ( \repb{3}\,, \rep{3} \,, -\frac{1}{3}  )_{ \mathbf{F}}  \oplus ( \rep{1}\,, \rep{3} \,, +\frac{1}{3} )_{ \mathbf{F}}  \otimes  ( \rep{1}\,, \rep{1} \,, -1 )_{ \mathbf{F}} \Big]  \otimes ( \rep{1}\,, \repb{3} \,, +\frac{2}{3})_{ \mathbf{H}} + H.c. \non
&\supset& Y_{\Tc\Cc} \Big[ ( \repb{3}\,, \rep{1} \,, - \frac{2}{3}  )_{ \mathbf{F}} \otimes ( \rep{3}\,, \repb{2} \,, + \frac{1}{6}  )_{ \mathbf{F}}^{\prime\prime }  \oplus  ( \rep{3}\,, \rep{2} \,, + \frac{1}{6}  )_{ \mathbf{F}} \otimes  ( \repb{3}\,, \rep{1} \,, - \frac{2}{3}  )_{ \mathbf{F}}^\prime \non
&\oplus&( \rep{3}\,, \rep{1} \,, - \frac{1}{3}  )_{ \mathbf{F}}^\prime \otimes ( \repb{3}\,, \rep{2} \,, - \frac{1}{6} )_{ \mathbf{F}}  \oplus ( \rep{1}\,, \repb{2} \,, +\frac{1}{2})_{ \mathbf{F}} \otimes  ( \rep{1}\,, \rep{1} \,, -1)_{ \mathbf{F}} \Big] \otimes  ( \rep{1}\,, \rep{2} \,, +\frac{1}{2})_{ \mathbf{H}}^\prime \non
&+& Y_{\Tc\Cc} \Big[ ( \repb{3}\,, \rep{1} \,, - \frac{2}{3}  )_{ \mathbf{F}} \otimes ( \rep{3}\,, \rep{2} \,, + \frac{1}{6}  )_{ \mathbf{F}}^{\prime }  \oplus  ( \rep{3}\,, \rep{2} \,, + \frac{1}{6}  )_{ \mathbf{F}} \otimes  ( \repb{3}\,, \rep{1} \,, - \frac{2}{3}  )_{ \mathbf{F}}^{\prime \prime} \non
&\oplus&( \rep{3}\,, \rep{1} \,, - \frac{1}{3}  )_{ \mathbf{F}}^\prime \otimes ( \repb{3}\,, \rep{2} \,, - \frac{1}{6} )_{ \mathbf{F}}  \oplus ( \rep{1}\,, \rep{2} \,, +\frac{1}{2})_{ \mathbf{F}}^\prime \otimes  ( \rep{1}\,, \rep{1} \,, -1)_{ \mathbf{F}}  \Big] \otimes  ( \rep{1}\,, \repb{2} \,, +\frac{1}{2})_{ \mathbf{H}} + H.c. \,,
\eeqn
along the EWSB direction.
They give fermion mass mixing terms of
\beqs\label{eqs:masses_EW_04}
\beqn
&&Y_{\Tc\Cc} \Big[ ( \repb{3}\,, \rep{1} \,, - \frac{2}{3}  )_{ \mathbf{F}} \otimes ( \rep{3}\,, \repb{2} \,, + \frac{1}{6}  )_{ \mathbf{F}}^{\prime\prime }  \oplus  ( \rep{3}\,, \rep{2} \,, + \frac{1}{6}  )_{ \mathbf{F}} \otimes  ( \repb{3}\,, \rep{1} \,, - \frac{2}{3}  )_{ \mathbf{F}}^\prime \non
&\oplus&( \rep{3}\,, \rep{1} \,, - \frac{1}{3}  )_{ \mathbf{F}} \otimes ( \repb{3}\,, \rep{2} \,, - \frac{1}{6} )_{ \mathbf{F}}  \oplus ( \rep{1}\,, \repb{2} \,, +\frac{1}{2})_{ \mathbf{F}} \otimes  ( \rep{1}\,, \rep{1} \,, -1)_{ \mathbf{F}} \Big] \otimes  ( \rep{1}\,, \rep{2} \,, +\frac{1}{2})_{ \mathbf{H}}^\prime + H.c. \non
&=& \frac{1}{ \sqrt{2}} Y_{\Tc\Cc} \Big( - \overline{ \cG_L} t_R  + \overline{t_L} \UG_R + \overline{ \DG_L }  \sG_R - \overline{ \EG_L} \eG_R \Big) u_{ \rep{2}}^\prime + H.c. \,,\\[1mm]
&& Y_{\Tc\Cc} \Big[ ( \repb{3}\,, \rep{1} \,, - \frac{2}{3}  )_{ \mathbf{F}} \otimes ( \rep{3}\,, \rep{2} \,, + \frac{1}{6}  )_{ \mathbf{F}}^{\prime }  \oplus  ( \rep{3}\,, \rep{2} \,, + \frac{1}{6}  )_{ \mathbf{F}} \otimes  ( \repb{3}\,, \rep{1} \,, - \frac{2}{3}  )_{ \mathbf{F}}^{\prime \prime} \non
&\oplus&( \rep{3}\,, \rep{1} \,, - \frac{1}{3}  )_{ \mathbf{F}}^\prime \otimes ( \repb{3}\,, \rep{2} \,, - \frac{1}{6} )_{ \mathbf{F}}  \oplus ( \rep{1}\,, \rep{2} \,, +\frac{1}{2})_{ \mathbf{F}}^\prime \otimes  ( \rep{1}\,, \rep{1} \,, -1)_{ \mathbf{F}}    \Big] \otimes  ( \rep{1}\,, \repb{2} \,, +\frac{1}{2})_{ \mathbf{H}} + H.c. \non
&=& \frac{1}{ \sqrt{2}} Y_{\Tc\Cc} \Big(  \overline{c_L } t_R + \overline{t_L } c_R + \overline{ \DG_L^\prime } \sG_R+  \overline{\EG_L } \eG_R^\prime  \Big) u_{ \repb{2}} + H.c.  \,.
\eeqn
\eeqs

\section{The SM fermion masses, mixings, and flavor non-universality}
\label{section:fermions}

\para
In this section, we summarize the fermion mass spectrum according to the symmetry breaking patterns presented in Sec.~\ref{section:breaking}.
With the fermion identifications through the previous analyses, we also obtain their gauge couplings according to the gauge sector described in Sec.~\ref{section:gauge}.

\subsection{The quark mass spectra and their mixings}

\para
We start from the up-type quarks with $Q_e=+\frac{2}{3}$ in Eq.~\eqref{eq:SU7_fermion_names}.
By using the basis of $( c \,, \cG\,, \UG\,,t)$, we find their mass matrix of
\beqn\label{eq:Uquark_mass}
\Mc_\Uc&=& \frac{1}{\sqrt{2} } \left( \ba{cccc}  
 0  & -Y_\Cc V_{ \rep{3}}    &  Y_\Cc u_{ \rep{2}}  &  Y_{ \Tc\Cc} u_{ \repb{2} }   \\
  -Y_\Cc u_{ \rep{2}}   &  - Y_\Cc w_{\rep{4}}   & 0   & -Y_{ \Tc\Cc} u_{ \rep{2}}^\prime  \\
 Y_\Cc V_{ \rep{3}}   &  0   &   Y_\Cc w_{\rep{4}}  & Y_{\Tc\Cc} V_{ \rep{3}}^\prime   \\
   Y_{ \Tc\Cc} u_{ \repb{2} }  &   Y_{\Tc\Cc} V_{ \rep{3}}^\prime  &  Y_{ \Tc\Cc} u_{ \rep{2}}^\prime  &  Y_\Tc v_t  \\
 \ea  \right) \,,
\eeqn
from Eqs.~\eqref{eq:masses_341_02}, \eqref{eq:masses_331_03}, \eqref{eq:masses_331_04}, \eqref{eq:Yukawa_top}, \eqref{eq:masses_EW_03}, and \eqref{eqs:masses_EW_04}.
Several features of Eq.~\eqref{eq:Uquark_mass} should be observed.
First, the charm and top quarks form the $2\times 2$ mass matrix of
\beqn
\Big( \Mc_u \Big)_{2\times 2} &=&  \frac{1}{\sqrt{2} } \left( \ba{cc} 
 0  &  Y_{ \Tc\Cc} u_{ \repb{2} }   \\
 Y_{ \Tc\Cc} u_{ \repb{2} }  &  Y_\Tc v_t  \\  
  \ea  \right) \,,
\eeqn
which resembles the conjectured mass matrix by Georgi and Jarlskog~\cite{Georgi:1979df}.
Second, one finds that 
\beqn\label{eq:Uquark_Detmass}
\det (  \Mc_\Uc \Mc_\Uc^\dag )&=& \frac{ Y_\Cc^4  Y_{ \Tc\Cc}^4  }{16} \Big[  ( u_{ \rep{2}} V_{ \rep{3}}^\prime  - u_{\repb{2} } w_{ \rep{4}}  )^2 - ( u_{\rep{2}}^\prime V_{ \rep{3}} )^2  \Big]^2 = m_c^2 m_t^2 m_{ \cG}^2 m_{ \UG}^2   \,.
\eeqn
%
%
In the limit of the vanishing Yukawa mixing of $Y_{\Tc\Cc}\to 0$, the lightest charm quark must be massless.
The masses of two heaviest vectorlike quarks and the top quark are approximately
\beqn
&& m_{ \cG} \approx m_{ \UG} \approx \frac{Y_\Cc }{\sqrt{2} } w_{ \rep{4}} \,, \quad m_t \approx \frac{Y_\Tc }{\sqrt{2} } v_{ t} \,.
\eeqn
Thus, the charm quark mass can be approximately expressed as
\beqn
m_c &\approx& \frac{ Y_{ \Tc\Cc}^2 }{ 2 m_t  w_{ \rep{4}}^2 }  \Big[  ( u_{ \rep{2}} V_{ \rep{3}}^\prime  - u_{\repb{2} } w_{ \rep{4}}  )^2 - ( u_{\rep{2}}^\prime V_{ \rep{3}} )^2  \Big]  \approx \frac{ Y_{ \Tc\Cc}^2 }{ \sqrt{2} Y_\Tc } \frac{  u_{ \repb{2} }^2 }{ v_t } \,.
\eeqn
Obviously, the charm quark becomes massless when all generated EWSB VEVs of $(u_{\rep{2}}\,, u_{\repb{2}} \,, u_{\rep{2}}^\prime)$ and/or the mixing Yukawa coupling of $Y_{\Tc\Cc}$ vanish.

\para
In general, any fermion mass matrix can be diagonalized in terms of bi-unitary transformation of
\beqn
&& \Fc_L  \Mc_\Fc  \Fc_R^\dag = \Mc_\Fc^{\rm diag} \,, ~ \Mc_\Fc \Mc_\Fc^\dag = \Fc_L^\dag ( \Mc_\Fc^{\rm diag} )^2 \Fc_L  \,, ~ \Mc_\Fc^\dag \Mc_\Fc = \Fc_R^\dag ( \Mc_\Fc^{\rm diag} )^2 \Fc_R\,.
\eeqn
We focus on the $\Mc_\Fc \Mc_\Fc^\dag$ in order to obtain the CKM mixing in the quark sector.
The diagonalization of the up-type quark mass matrix in Eq.~\eqref{eq:Uquark_mass} can be performed in terms of perturbation expansion as follows
\beqs\label{eqs:Uquark_diag}
\beqn
\Mc_\Uc \Mc_\Uc^\dag &\approx& \Big(  \Mc_\Uc \Mc_\Uc^\dag  \Big)^{ (0)} + \Big(  \Mc_\Uc \Mc_\Uc^\dag  \Big)^{ (1)}  \,, \\[1mm]
\Big(  \Mc_\Uc \Mc_\Uc^\dag  \Big)^{ (0)}  &=&  \frac{1}{2 } \left( \ba{cccc}  
 Y_\Cc^2 V_{ \rep{3}}^2  & Y_\Cc^2 V_{ \rep{3}} w_{ \rep{4}}   & 0 & 0 \\
 Y_\Cc^2 V_{ \rep{3}} w_{ \rep{4}}  & Y_\Cc^2 w_{ \rep{4}}^2   & 0  & - Y_\Cc  Y_{\Tc \Cc} V_{ \rep{3}}^\prime w_{ \rep{4}}  \\
 0  & 0   & Y_\Cc^2 w_{ \rep{4}}^2 + Y_{\Tc\Cc}^2 (V_{\rep{3}}^\prime )^2  & Y_\Tc  Y_{\Tc \Cc} V_{ \rep{3}}^\prime v_t  \\
 0   &  - Y_\Cc  Y_{\Tc \Cc} V_{ \rep{3}}^\prime w_{ \rep{4}}   & Y_\Tc  Y_{\Tc \Cc} V_{ \rep{3}}^\prime v_t   & Y_{\Tc\Cc}^2 (V_{\rep{3}}^\prime )^2 + Y_\Tc^2 v_t^2 \\
 \ea  \right) \,, \label{eq:Uquark_diag0} \\[1mm]
 \Big(  \Mc_\Uc \Mc_\Uc^\dag  \Big)^{ (1)}  &=&  \frac{1}{2 }  \left( \ba{cccc}   
0  &  0  & Y_\Cc^2 u_{ \rep{2}} w_{ \rep{4}} + Y_{ \Tc\Cc}^2 u_{ \repb{2}} V_{ \rep{3}}^\prime &  Y_{ \Tc\Cc} Y_\Tc u_{ \repb{2}} v_t   \\ 
0  &  0  & - Y_\Cc^2 u_{ \rep{2}} V_{ \rep{3}} - Y_{\Tc\Cc}^2 u_{ \rep{2}}^\prime V_{ \rep{3}}^\prime   &   - Y_{ \Tc\Cc} Y_\Tc u_{ \rep{2}}^\prime  v_t  \\ 
 Y_\Cc^2 u_{ \rep{2}} w_{ \rep{4}} + Y_{ \Tc\Cc}^2 u_{ \repb{2}} V_{ \rep{3}}^\prime &  - Y_\Cc^2 u_{ \rep{2}} V_{ \rep{3}} - Y_{\Tc\Cc}^2 u_{ \rep{2}}^\prime V_{ \rep{3}}^\prime    & 0  & Y_\Cc Y_{ \Tc \Cc} ( u_{ \repb{2}} V_{ \rep{3}}    \\ 
  &   &   & + u_{ \rep{2}}^\prime w_{ \rep{4}}  )    \\
Y_{ \Tc\Cc} Y_\Tc u_{ \repb{2}} v_t   &  - Y_{ \Tc\Cc} Y_\Tc u_{ \rep{2}}^\prime  v_t  & Y_\Cc Y_{ \Tc \Cc} ( u_{ \repb{2}} V_{ \rep{3}}   &  0   \\ 
 &  & + u_{ \rep{2}}^\prime w_{ \rep{4}}  )   & \\
  \ea  \right) \non
  && \,. \label{eq:Uquark_diag1}
\eeqn
\eeqs
The $\Big(  \Mc_\Uc \Mc_\Uc^\dag  \Big)^{ (0)}$ only contains the mass terms with the ${\rm SU}(4)_W \otimes {\rm U}(1)_{X_0}$-breaking and ${\rm SU}(3)_W \otimes {\rm U}(1)_{X_1}$-breaking VEVs, and can be diagonalized by the orthogonal transformation as follows
\beqs
\beqn
&& \Uc_L^{(0)} \Big(  \Mc_\Uc \Mc_\Uc^\dag  \Big)^{ (0)}  \Uc_L^{(0)\, T}  \approx {\rm diag}( 0\,, m_\cG^2\,, m_\UG^2\,, m_t^2 )\,, \\[1mm]
&& \left( \begin{array}{c} \hat c \\  \hat{\mathfrak{c}} \\ \hat{\mathfrak{U}} \\ \hat{t} \end{array} \right) = \Uc_L^{(0)} \cdot \left( \begin{array}{c} c \\  \cG \\  \UG  \\ t \end{array} \right)  \,, \quad  \Uc_L^{(0)} \approx \left(\begin{array}{cccc}
1&-\frac{ V_{\rep{3}} }{w_{\rep{4}} } &  0 & 0\\
\frac{V_{\rep{3}} }{\sqrt{2} w_{\rep{4}} }&\frac{1}{\sqrt{2}}&-\frac{1}{\sqrt{2}}&-\frac{Y_{\Tc \Cc }V_{\rep{3}}^\prime }{\sqrt{2} Y_\Cc w_{\rep{4}} }\\
\frac{V_{\rep{3}} }{\sqrt{2} w_{\rep{4}} }&\frac{1}{\sqrt{2}}&\frac{1}{\sqrt{2}}&-\frac{Y_{\Tc \Cc }V_{\rep{3}}^\prime }{\sqrt{2}Y_\Cc w_{\rep{4}} }\\
0&\frac{Y_{\Tc \Cc }V_{\rep{3}}^\prime }{Y_\Cc w_{\rep{4} } }&0&1
\end{array}\right) \,,
\eeqn
\eeqs
where we approximated the matrix to the order of $w_{ \rep{4}}^{-1}$.
By further including the terms from the $\Big(  \Mc_\Uc \Mc_\Uc^\dag  \Big)^{ (1)}$, which depend linearly on the EWSB VEVs of $(u_{\rep{2}}\,, u_{\rep{2}}^\prime \,, u_{\repb{2}})$, the mass eigenstates of the charm and the top quarks (denoted as $\hat{\hat{c}}$ and $\hat{\hat{t}}$) are related to the gauge eigenstates as
\beqn\label{eq:Uquark_leftMix}
\left(\begin{array}{c} \hat{\hat{c}}   \\ \hat{\hat{t}} \end{array} \right) &=& {\mathscr X}_\Uc \cdot  \left(\begin{array}{c} 
c \\ 
\cG  \\ 
t \end{array}\right) \,, \non
{\mathscr X}_\Uc &\approx& \left(\begin{array}{ccc}
1&-\frac{u_{\rep{2}} }{\sqrt{2} w_{\rep{4}}} -\frac{V_{\rep{3}} }{w_{\rep{4}} }-\frac{Y_{\Tc \Cc }^2 V_{\rep{3}}^\prime u_{\repb{2}}}{Y_\Tc Y_\Cc v_t w_{\rep{4}} }&-\frac{Y_{\Tc \Cc} u_{\repb{2}} }{Y_\Tc v_t}\\
\frac{Y_{\Tc \Cc }u_{\repb{2}} }{Y_\Tc v_t}&\frac{Y_{\Tc \Cc }V_{\rep{3}}^\prime }{Y_\Cc  w_{\rep{4}} }-\frac{Y_{\Tc \Cc }V_{\rep{3}} u_{\repb{2}} }{Y_\Tc v_t w_{\rep{4}} }-\frac{Y_{\Tc \Cc }u_{\rep{2}}^\prime}{\sqrt{2} Y_\Cc w_{\rep{4}} }  &  1
\end{array}\right) \,.
\eeqn


\para
The mass matrices for the down-type quarks with $Q_e=-\frac{1}{3}$ and charged leptons are correlated.
Here, we express their mass matrices in terms of the basis of $( \{ s\,, \sG\,, \SG \} \,, \{ b\,, \DG\,, \DG^\prime \} )$ and $( \{ \mu \,, \EG \,, \eG^{\prime\prime} \} \,, \{ \tau \,, \eG\,, \eG^{\prime} \} )$ as follows
\beqs
\beqn
\Mc_\Dc &=& \frac{1}{\sqrt{2} }  \left( \ba{cccccc}  
Y_\Sc u_{ \repb{2}\,, \dot 2 } &  Y_\Cc V_{ \rep{3}}  &  0  &    &    &    \\
 Y_\Sc u_{ \rep{2}\,, \dot 2 } &  Y_\Cc w_{ \rep{4}}  &  0    &  \multicolumn{3}{c}{ 0_{3\times 3}  }  \\ 
 0  &  Y_\Cc u_{ \rep{2}}   &  Y_\Sc V_{ \repb{3}\,, \dot {\rm I}}   &    &    &   \\ 
 0  &  - Y_{\Tc\Cc} V_{ \rep{3}}^\prime   & 0   &   Y_\Bc u_{ \repb{2}\,, 3 }   & 0  & 0    \\ 
 0  &  - Y_{\Tc\Cc} u_{ \rep{2}}^\prime   & 0   & 0  & Y_\Bc V_{ \repb{3}\,, {\rm I} }  & 0   \\ 
 0  &  - Y_{\Tc\Cc} u_{ \repb{2}}   & 0   &  0  & 0  & Y_\Bc  w_{ \repb{4}\,, {\rm II} }   \\    \ea \right) \,, \label{eq:Dquark_mass} \\[1mm]
\Mc_\Lc &=& \frac{1}{\sqrt{2} } \left( \ba{cccccc}
 Y_\Sc u_{ \repb{2}\,, \dot 2}  & Y_\Sc u_{ \rep{2}\,, \dot 2 }  & 0 & 0  & 0  & 0  \\
 Y_\Cc V_{\rep{3}}  & Y_\Cc w_{ \rep{4}} & -Y_\Cc u_{ \rep{2}}  & Y_{\Tc\Cc} V_{ \rep{3}}^\prime & Y_{\Tc\Cc} u_{ \repb{2}}  & Y_{\Tc\Cc} u_{ \rep{2}}^\prime   \\
 0  & 0   & Y_\Sc V_{ \repb{3}\,, \dot {\rm I}}  & 0  &  0 & 0  \\
  &  &  & Y_\Bc u_{ \repb{2}\,, 3}  & 0   &  0 \\
 \multicolumn{3}{c}{ 0_{3\times 3}  }  & 0 &  Y_\Bc w_{ \repb{4}\,, {\rm II}}  & 0 \\
  &  &  & 0 & 0 &  - Y_\Bc V_{ \repb{3}\,, {\rm I}}   \\  \ea \right)  \,, \label{eq:Lepton_mass}
\eeqn
\eeqs
according to Eqs.~\eqref{eq:masses_341_01}, \eqref{eq:masses_341_02}, \eqref{eq:masses_331_01}, \eqref{eq:masses_331_02}, \eqref{eq:masses_331_03}, \eqref{eq:masses_331_04}, \eqref{eq:masses_EW_01}, \eqref{eqs:masses_EW_02}, \eqref{eq:masses_EW_03}, and \eqref{eqs:masses_EW_04}.
Distinct from the mass matrix for the up-type quarks in Eq.~\eqref{eq:Uquark_mass}, they are both sparse matrices.
Both the second and the third generational $(s\,,\mu)$ and $(b\,,\tau)$ obtain their tree-level masses.
Given the patterns in Eqs.~\eqref{eq:Dquark_mass} and \eqref{eq:Lepton_mass}, one can naturally expect degenerate masses of $m_s = m_\mu$ and $m_b= m_\tau$ at the tree level.
Below, we focus on the down-type quark sector in order to address the electroweak mixing.
We find that
\beqn
&& \det \Big(  \Mc_\Dc \Mc_\Dc^\dag   \Big) = \frac{ Y_\Cc^2 Y_\Sc^4 Y_\Bc^6 }{64} w_{ \repb{4}\,, {\rm II}}^2 V_{  \repb{3}\,, {\rm I} }^2 V_{ \repb{3} \,, \dot {\rm I} }^2 u_{ \repb{2}\,, 3}^2 ( u_{ \repb{2}\,, \dot 2} w_{ \rep{4}} - u_{ \rep{2}\,, \dot 2} V_{ \rep{3} } )^2 = m_s^2 m_b^2 m_\sG^2 m_{ \SG}^2 m_\DG^2 m_{ \DG^\prime }^2 \,.
\eeqn

\para
One can expand the down-type quark mass matrix in terms of the VEV hierarchies as follows
%
%
\beqs\label{eqs:Dquark_diag}
\beqn
\Mc_\Dc \Mc_\Dc^\dag &\approx& \Big(  \Mc_\Dc \Mc_\Dc^\dag  \Big)^{ (0)} + \Big(  \Mc_\Dc \Mc_\Dc^\dag  \Big)^{ (1)}  \,, \\[1mm]
\Big(  \Mc_\Dc \Mc_\Dc^\dag  \Big)^{ (0)} &=& \hf  \left(\begin{array}{cccccc}
 Y_\Cc^2 V_{\rep{3}}^2 & Y_\Cc^2 V_{\rep{3}} w_{\rep{4}}  & 0 & -Y_\Cc Y_{\Tc \Cc }V_{\rep{3}} V_{\rep{3}}^\prime  & &\\
 Y_\Cc^2 V_{\rep{3}}  w_{\rep{4}} & Y_\Cc^2 w_{\rep{4}}^2 & 0 & - Y_\Cc Y_{\Tc \Cc } V_{\rep{3}}^\prime w_{\rep{4}} &&\\
0 & 0 & Y_\Sc^2V_{\repb{3}\,, \dot {\rm I} }^2 &0 &&\\
-Y_\Cc Y_{\Tc \Cc } V_{\rep{3}} V_{\rep{3}}^\prime & - Y_\Cc Y_{\Tc \Cc} V_{\rep{3}}^\prime w_{\rep{4}} & 0 & Y_{\Tc \Cc}^2 (V^\prime_3)^2 &&\\
&&&& Y_\Bc^2 V_{\repb{3}\,, {\rm I}}^2&\\
&&&&& Y_\Bc^2 w_{\repb{4}\,, {\rm II} }^2
\end{array}\right) \non
&&  \,.
\eeqn
\eeqs
The leading mass terms from the $\Big(  \Mc_\Dc \Mc_\Dc^\dag  \Big)^{ (0)}$ can be diagonalized by the orthogonal transformation as follows
\beqs
\beqn
&& \Dc_L^{(0)} \Big(  \Mc_\Dc \Mc_\Dc^\dag  \Big)^{ (0)}  \Dc_L^{(0)\, T}  \approx {\rm diag}( 0\,, m_\sG^2\,, m_\SG^2\,, 0 \,, m_\DG^2 \,, m_{ \DG^\prime }^2  )\,, \\[1mm]
&& \left( \begin{array}{c}  \hat{s}  \\ \hat{\sG}  \\  \hat{\SG} \\ \hat{ b} \\ \hat{\DG }  \\ \hat{\DG }^\prime \end{array} \right) = \Dc_L^{(0)} \cdot \left( \begin{array}{c}   s  \\  \sG  \\   \SG\\  b \\ \DG   \\  \DG^\prime   \end{array} \right)  \,, \quad
   \Dc_L^{(0)} \approx \left(\begin{array}{cccccc}
 c_{\alpha_1} & -s_{\alpha_1} & 0   & 0 &    &  \\
 s_{\alpha_1} c_{\alpha_2} & c_{\alpha_1} c_{\alpha_2} &  0  & -s_{\alpha_2}  &    &  \\
0 &  0 &  1  & 0 &    &  \\
s_{\alpha_1}  s_{\alpha_2}  & c_{\alpha_1} s_{\alpha_2}  & 0 & c_{\alpha_2}  &    &  \\
 &  &    &  &  1  &  \\
 &  &    &  &    & 1 \\
\end{array}\right) \,, \non
&&  t_{\alpha_1} = \frac{V_{ \rep{3}} }{ w_{ \rep{4}} } \,, \quad t_{\alpha_2} = \frac{ Y_{\Tc \Cc} V_{ \rep{3}}^\prime }{ Y_\Cc \sqrt{ w_{ \rep{4}}^2 + V_{ \rep{3}}^2 }  } \,.
\eeqn
\eeqs
By further including the $\Big(  \Mc_\Dc \Mc_\Dc^\dag  \Big)^{ (1)}$ mass terms, we find the strange and bottom quark masses of
\beqn\label{eq:sb_masses}
&& m_s \approx \frac{1}{ \sqrt{2} } Y_\Sc u_{ \repb{2}\,, \dot 2} \,, \quad m_b \approx \frac{1}{ \sqrt{2} } Y_\Bc u_{ \repb{2}\,,  3}
\eeqn
at the tree level.
The corresponding EWSB VEVs for their masses are generated from operators in Eqs.~\eqref{eq:Oc4A_VEVgen} and \eqref{eq:Oc4B1_VEVgen}, respectively.
In both operators of $\Oc_{\mathscr A}^{d=4}$ and $\Oc_{\mathscr B1}^{d=4}$, the minimal set of Higgs VEVs in Eqs.~\eqref{eqs:SU7_Higgs_VEVs_simple} are expected to be of the same order.
With the assumption of the natural Yukawa couplings of $Y_\Sc \sim Y_\Bc \sim \Oc(1)$, we have unrealistic mass relation of $m_s \approx m_b$ from Eq.~\eqref{eq:sb_masses}.
Their mass eigenstates (denoted as $\hat{\hat{s}}$ and $\hat{\hat{b}}$) are related to the gauge eigenstates as
\beqn\label{eq:Dquark_leftMix}
\left(\begin{array}{c} \hat{\hat{s}}  \\  \hat{\hat{b}}  \end{array} \right) &=& {\mathscr X}_\Dc \cdot  \left(\begin{array}{c} 
s \\ 
\sG  \\ 
b \end{array}\right) \,, \quad  {\mathscr X}_\Dc \approx \left(\begin{array}{ccc}
1 &   -\frac{V_{\rep{3}} }{ w_{\rep{4} } }  &  0   \\
0  &   \frac{ Y_{\Tc \Cc } V_{ \rep{3} }^\prime }{ Y_\Cc w_{ \rep{4} } }   &  1   \\
\end{array}\right) \,.
\eeqn
Together with the left-handed quark mixing matrix in Eq.~\eqref{eq:Dquark_leftMix}, we find the following approximation to the $2\times 2$ CKM matrix 
\beqn\label{eq:CKM22}
V_{\rm CKM}^{2 \times 2}&=&  {\mathscr X}_\Uc {\mathscr X}_\Dc^\dag \approx 
\left(\begin{array}{cc}
 1  &-\frac{Y_{\Tc \Cc } u_{\repb{2}}}{Y_\Tc  v_t}\\ \frac{Y_{\Tc \Cc }u_{\repb{2}} }{Y_\Tc v_t}&1\end{array}\right)  \,.
\eeqn
Indeed, the current $2\times 2$ CKM matrix resembles the observed feature of the realistic $3\times 3$ CKM matrix, namely, it is almost diagonal.
Notice that we have neglected all higher order correction terms suppressed by $w_{ \rep{4}}^{-1}$ in Eq.~\eqref{eq:CKM22}.
Given the mass matrices for the lepton sector in Eq.~\eqref{eq:Lepton_mass}, it is straightforward to infer the tree-level masses of
\beqn\label{eq:mutau_masses}
&& m_\mu \approx \frac{1}{ \sqrt{2} } Y_\Sc u_{ \repb{2}\,, \dot 2} \,, \quad m_\tau \approx \frac{1}{ \sqrt{2} } Y_\Bc u_{ \repb{2}\,,  3} \,.
\eeqn
This means the $b-\tau$ mass unification is extended to the $s-\mu$ mass unification at the tree level.
In the context of the Georgi-Glashow ${\rm SU}(5)$ model, the possible $b-\tau$ mass splitting was attributed to the renormalization group effects~\cite{Buras:1977yy}.
However, results therein cannot be naively applied to the $(b\,,\tau)$ and $(s\,, \mu)$ mass ratios in the {\it non-minimal} GUTs.
To fully evaluate their mass splitting, we expect two prerequisites of: (i) the evaluation of the intermediate symmetry breaking scales from an appropriate GUT group, and (ii) the identification of the SM fermion representations with the extended color and weak symmetries above the EW scale.
Both are distinctive features of the {\it non-minimal} GUTs, and we defer to analyze the details in the future work.

\subsection{The neutrino masses}

\para
We also summarize the neutrino masses in the ${\rm SU}(7)$ toy model.
According to our conventions in Eq.~\eqref{eq:SU7_fermion_names}, the neutral fermions include two EW active neutrinos of $(\nu_\mu \,, \nu_\tau)$, three vectorlike massive neutrinos of $(\nG\,, \nG^\prime\,, \nG^{\prime\prime})$ from the ${\rm SU}(2)_W$-doublets, nine left-handed sterile neutrinos, and one vectorlike massive sterile neutrinos of $\check \nG$.
From Eqs.~\eqref{eq:masses_341_01}, \eqref{eq:masses_331_01}, and \eqref{eq:masses_331_02}, we find the vectorlike neutrino masses of $m_{ \nG^\prime} = m_{ \check \nG} \sim \Oc( v_{341})$ and $m_{\nG} \sim m_{ \nG^{\prime\prime} } \sim \Oc( v_{331})$.
All other EW active neutrinos and sterile neutrinos are massless from the tree-level Yukawa couplings.
Neither can we find any tree-level Yukawa coupling that mixes the active neutrinos and the massive neutrinos in the current setup.
Meanwhile, it is known that the loop-level effects can be ubiquitous in the neutrino mass generation~\cite{Cai:2017jrq} and it is most appropriate to take the analysis in the three-generational {\it non-minimal} GUTs.

\subsection{The fermion gauge couplings with the extended weak symmetries and the flavor non-universality}

\para
We proceed to derive the fermion gauge couplings.
Since the color symmetry of ${\rm SU}(3)_c$ is always exact in the current context, we focus on the extended weak symmetries.
Some general features of the fermion gauge couplings in the ${\rm SU}(7)$ model can be outlined.
First, different SM fermion generations transform differently, as the current ${\rm SU}(7)$ model suggests through its fermion decompositions in Tabs.~\ref{tab:SU7_2gen_7barferm}, \ref{tab:SU7_2gen_21ferm}, and \ref{tab:SU7_2gen_35ferm}.
Consequently, we show manifestly that the flavor non-universality can be expected through the flavor-conserving neutral currents from the ${\rm SU}(4)_W \otimes {\rm U}(1)_{X_0}$ symmetry breaking.
Second, {\it non-minimal} GUTs such as ${\rm SU}(7)$ model and beyond contain vectorlike fermions in the spectrum.
This can be manifestly confirmed through the flavor-conserving neutral currents from the ${\rm SU}(3)_W \otimes {\rm U}(1)_{X_1}$ symmetry breaking in the current context.
Third, the tree-level currents include both flavor-changing charged currents (FCCC) mediated by $(W^\pm\,, W^{\prime\, \pm} \,, W^{\prime\prime\, \pm} )$, as well as flavor-changing neutral currents (FCNC) mediated by $(N_\mu\,, \bar N_\mu\,,  N_\mu^\prime\,, \bar N_\mu^\prime \,, N_\mu^{\prime\prime}\,, \bar N_\mu^{\prime\prime} )$.
The tree-level FCNCs in the current context never involve two different flavors of SM fermions.
Instead, they only involve one SM flavor with another heavy vectorlike fermion with the same electric charge, as will be explicitly given in Eqs.~\eqref{eq:FCNC_SU4W} and \eqref{eq:FCNC_SU3W}. 
Below, we organize the relevant couplings according to the symmetry breaking stages described in the current context.

\subsubsection{The ${\rm SU}(4)_W \otimes {\rm U}(1)_{X_0}$ gauge couplings}

\para
After the first-stage symmetry breaking, the FCCC and the FCNC are expressed as
\beqs
\beqn
\Lc_{ {\rm SU}(4)_W}^{\rm CC\,, \bcancel{\rm F}}&=&  \frac{ g_{4W} }{ \sqrt{2} } \Big(-  \overline{ \Ec_L^\Lambda } \gamma^\mu \check \Nc_L^{\Lambda^\prime }   + \overline{ \eG_R } \gamma^\mu \check \nG_R + \overline{ \tau_R } \gamma^\mu \nG_R^\prime + \overline{ \EG_R} \gamma^\mu \nG_R^{ \prime\prime }  \non
&+&  \overline{\DG_L^\prime } \gamma^\mu t_L  - \overline{ \sG_R} \gamma^\mu c_R - \overline{ s_L } \gamma^\mu \UG_L + \overline{ \SG_L } \gamma^\mu \cG_L  \Big) W_\mu^{\prime\prime\,-} + H.c. \,, \label{eq:FCCC_SU4W} \\[1mm]
\Lc_{ {\rm SU}(4)_W}^{\rm NC\,, \bcancel{\rm F}}&=&  \frac{ g_{4W} }{ \sqrt{2} } \Big(  \overline{ \check \Nc_L^{\Lambda^\prime } } \gamma^\mu \Nc_L^\Lambda - \overline{ \eG_R^\prime } \gamma^\mu \tau_R + \overline{ \check \nG_R } \gamma^\mu \nG_R  - \overline{ \eG_R^{ \prime\prime} } \gamma^\mu \EG_R  \non
&+& \overline{ b_L } \gamma^\mu \DG_L^\prime - \overline{ c_R } \gamma^\mu \cG_R + \overline{ \UG_L } \gamma^\mu c_L - \overline{ \sG_L } \gamma^\mu \SG_L  \Big) N_\mu^\prime \non
&+&  \frac{ g_{4W} }{ \sqrt{2} } \Big( - \overline{ \check \Nc_L^{\Lambda^\prime } } \gamma^\mu \check \Nc_L^\Lambda - \overline{ \nG_R^\prime } \gamma^\mu \nG_R - \overline{ \eG_R^\prime } \gamma^\mu \eG_R + \overline{ \mu_R } \gamma^\mu \EG_R   \non
&+& \overline{\DG_L } \gamma^\mu \DG_L^\prime - \overline{ c_R} \gamma^\mu \UG_R + \overline{ \sG_L } \gamma^\mu  s_L - \overline{ \cG_L } \gamma^\mu c_L    \Big) N_\mu^{\prime \prime} + H.c. \,. \label{eq:FCNC_SU4W}
\eeqn
\eeqs
%
%

\begin{table}[htp]
\begin{center}
\begin{tabular}{c|cccccc}
\hline \hline
  & $c$   &  $t$  &  $\cG$  &  $\UG$  & -  & -  \\ 
\hline
$g_f^{V\,\prime\prime }$  &  $\frac{1}{24} - \frac{1}{3} s_{\theta_G}^2 $  &  $\frac{1}{24} - \frac{1}{3} s_{\theta_G}^2 $  &  $\frac{1}{24} - \frac{1}{3} s_{\theta_G}^2 $  &  $ \frac{1}{24}  - \frac{1}{3} s_{\theta_G}^2 $   &  - & -  \\
$g_f^{A\,\prime\prime }$  &  $\frac{5}{24}  - \frac{1}{3} s_{\theta_G}^2 $  &  $ -\frac{1}{24} -\frac{1}{3} s_{\theta_G}^2 $  & $-\frac{1}{8}$   & $-\frac{1}{8}$   & -  & -  \\
\hline
  & $s$   &  $b$  &  $\sG$  &  $\SG$  &  $\DG$  &  $\DG^\prime$   \\ 
  \hline
$g_f^{V\,\prime\prime }$  &  $  - \frac{1}{12} + \frac{1}{6} s_{\theta_G}^2 $  & $\frac{1}{24}  + \frac{1}{6} s_{\theta_G}^2 $  &  $\frac{1}{24}  - \frac{1}{3} s_{\theta_G}^2 $  &  $- \frac{1}{12}  + \frac{1}{6} s_{\theta_G}^2 $  &  $\frac{1}{24}  + \frac{1}{6} s_{\theta_G}^2 $  & $ -\frac{1}{8} + \frac{1}{3} s_{\theta_G}^2 $  \\ 
$g_f^{A\,\prime\prime }$  &  $\frac{1}{12}  + \frac{1}{6} s_{\theta_G}^2 $  & $ - \frac{1}{24}  + \frac{1}{6} s_{\theta_G}^2$  &  $-\frac{1}{8}$  &    $\frac{1}{12}  + \frac{1}{6} s_{\theta_G}^2$ &  $- \frac{1}{24}  + \frac{1}{6} s_{\theta_G}^2$  & $\frac{1}{8}$   \\ 
\hline
  & $\mu$   &  $\tau$  &  $\eG$  &  $\eG^{\prime}$  &  $\eG^{\prime\prime }$  &  $\EG$   \\ 
  \hline
$g_f^{V\,\prime\prime }$  & $\frac{1}{2} s_{\theta_G}^2$   & $- \frac{1}{8}  + \frac{1}{2} s_{\theta_G}^2 $  &  $- \frac{1}{8}  + \frac{1}{2} s_{\theta_G}^2 $  &  $\frac{1}{24}  + \frac{1}{3} s_{\theta_G}^2 $ &  $\frac{1}{2} s_{\theta_G}^2 $  & $ - \frac{1}{8}  +  s_{\theta_G}^2$  \\ 
$g_f^{A\,\prime\prime }$  &  $\frac{1}{12} + \frac{1}{6} s_{\theta_G}^2$  & $- \frac{1}{24}  + \frac{1}{6}  s_{\theta_G}^2$  &  $- \frac{1}{24}  + \frac{1}{6} s_{\theta_G}^2$  &  $\frac{1}{8}$  &  $\frac{1}{12}  + \frac{1}{6} s_{\theta_G}^2$  &  $-\frac{1}{8}$  \\ 
\hline\hline
\end{tabular}
\end{center}
\caption{
The couplings of the flavor-conserving neutral currents mediated by $Z_\mu^{\prime\prime}$ in the $V-A$ basis, with $\theta_G$ being the ${\rm SU}(4)_W \otimes {\rm U}(1)_{X_0}$ mixing angle defined in Eq.~\eqref{eq:Glashow_angle}.
}
\label{tab:Zpp_VA}
\end{table}%

\para
The flavor-conserving neutral currents are expressed as follows in the $V-A$ basis
\beqn
\Lc_{ {\rm SU}(4)_W}^{\rm NC\,, F}&=& \frac{g_{X_1}}{ s_{\theta_G} c_{\theta_G} } ( g_f^{V\,\prime\prime } \overline f \gamma^\mu f + g_f^{A\, \prime\prime } \overline f \gamma^\mu \gamma_5 f ) Z_\mu^{\prime\prime } \,,
\eeqn
and we tabulate the vectorial and axial couplings of $( g_f^{V\,\prime\prime } \,, g_f^{A\,\prime\prime } )$ in Tab.~\ref{tab:Zpp_VA}. 
Manifestly, the gauge couplings of two-generational SM fermions with the flavor-conserving neutral boson of $Z_\mu^{\prime\prime}$ are distinctive, which is the source of flavor non-universality.
This is consistent from what we found from the fermion irreps in Tabs.~\ref{tab:SU7_2gen_7barferm}, \ref{tab:SU7_2gen_21ferm}, and \ref{tab:SU7_2gen_35ferm}.
The flavor non-universality is only possible with the extended weak symmetries of ${\rm SU}(4)_W \otimes {\rm U}(1)_{X_0}$, or in the ${\rm SU}(7)$ and beyond {\it non-minimal} GUTs.
As we have discussed previously in Sec.~\ref{section:model}, the non-trivial embedding of multiple SM generations requires at least rank-$3$ (or above) anti-symmetric irreps that are not self-conjugate.
One should also expect the flavor non-universality in a realistic {\it non-minimal} GUT from its extended weak and strong sectors.

\subsubsection{The ${\rm SU}(3)_W \otimes {\rm U}(1)_{X_1}$ gauge couplings}

\para
After the second-stage symmetry breaking, the FCCC and the FCNC are expressed as
\beqs
\beqn
\Lc_{ {\rm SU}(3)_W}^{\rm CC\,, \bcancel{\rm F}}&=&  \frac{ g_{3W} }{ \sqrt{2} } \Big( - \overline{ \Ec_L^\Lambda } \gamma^\mu \check \Nc_L^\Lambda + \overline{ \tau_R } \gamma^\mu \nG_R + \overline{ \mu_R } \gamma^\mu \nG_R^{ \prime \prime } - \overline{ \eG_R^\prime } \gamma^\mu  \check \nG_R \non
&+& \overline{ \DG_L } \gamma^\mu t_L  + \overline{ \SG_L } \gamma^\mu c_L - \overline{ \sG } \gamma^\mu \gamma_5 \UG \Big) W_\mu^{\prime \, -} + H.c. \,,   \label{eq:FCCC_SU3W} \\[1mm]
\Lc_{ {\rm SU}(3)_W}^{\rm NC\,, \bcancel{\rm F}}&=& \frac{ g_{3W} }{ \sqrt{2} } \Big( \overline{ \check \Nc_L^\Lambda } \gamma^\mu \Nc_L^\Lambda  - \overline{ \eG_R } \gamma^\mu  \tau_R - \overline{ \eG_R^{\prime \prime } } \gamma^\mu \mu_R - \overline{ \check \nG_R } \gamma^\mu  \nG_R^\prime  \non
&+& \overline{b_L} \gamma^\mu \DG_L + \overline{ s_L } \gamma^\mu \SG_L - \overline{ \UG } \gamma^\mu \cG  \Big) N_\mu  + H.c.   \,. \label{eq:FCNC_SU3W}
\eeqn
\eeqs

\begin{table}[htp]
\begin{center}
\begin{tabular}{c|ccccc}
\hline \hline
  & $c/t$   &     $\cG$  &  $\UG$  & -  & -  \\ 
\hline
$g_f^{V\,\prime }$  &  $\frac{1}{12} - \frac{5}{12} s_{\theta_S}^2  $     & $-\frac{1}{6}  -\frac{1}{6} s_{ \theta_S}^2  $   &  $\frac{1}{3} (1 -2 s_{ \theta_S}^2 ) $   &  - & -  \\
$g_f^{A\,\prime }$  &  $- \frac{1}{12} - \frac{1}{4} s_{\theta_S}^2$    &  $0$  & $0$  & -  & -  \\
\hline
  & $s/b$   &     $\sG$  &  $\SG$  &  $\DG$  &  $\DG^\prime$   \\ 
  \hline
$g_f^{V\,\prime }$  &  $\frac{1}{12} + \frac{1}{12} s_{\theta_S}^2$    & $-\frac{1}{6}  -\frac{1}{6} s_{ \theta_S}^2 $ & $-\frac{1}{6}  +\frac{1}{3} s_{ \theta_S}^2$ & $-\frac{1}{6}  +\frac{1}{3} s_{ \theta_S}^2$ & $\frac{1}{3} s_{ \theta_S}^2$   \\ 
$g_f^{A\,\prime }$  & $- \frac{1}{12} + \frac{1}{4} s_{\theta_S}^2$ &   $0$  & $\frac{1}{6}$ &  $\frac{1}{6}$  &  $0$   \\ 
\hline
  & $\mu/\tau$   &    $\eG$  &  $\eG^{\prime}$  &  $\eG^{\prime\prime }$  &  $\EG$   \\ 
  \hline
$g_f^{V\,\prime }$  & $- \frac{1}{4} + \frac{3}{4} s_{\theta_S}^2$ &   $\frac{1}{2} s_{\theta_S}^2 $   & $-\frac{1}{6} + \frac{1}{2} s_{ \theta_S}^2 $   & $\frac{1}{2} s_{\theta_S}^2 $  & $s_{ \theta_S}^2$   \\ 
$g_f^{A\,\prime }$  & $- \frac{1}{12} + \frac{1}{4} s_{\theta_S}^2$  & $\frac{1}{12}$  & $0$   & $\frac{1}{12}$ & $0$   \\ 
\hline\hline
\end{tabular}
\end{center}
\caption{
The couplings of the flavor-conserving neutral currents mediated by $Z_\mu^{\prime}$ in the $V-A$ basis, with $\theta_S$ being the ${\rm SU}(3)_W \otimes {\rm U}(1)_{X_1}$ mixing angle defined in Eq.~\eqref{eq:Salam_angle}.
}
\label{tab:Zp_VA}
\end{table}%

\para
The flavor-conserving neutral currents are expressed as follows in the $V-A$ basis
\beqn\label{eq:FconsNC_SU3W}
\Lc_{ {\rm SU}(3)_W}^{\rm NC\,, F}&=& \frac{g_{Y}}{ s_{\theta_S} c_{\theta_S} } ( g_f^{V\,\prime } \overline f \gamma^\mu f + g_f^{A\, \prime } \overline f \gamma^\mu \gamma_5 f ) Z_\mu^{\prime } \,,
\eeqn
and we tabulate the vectorial and axial couplings in Tab.~\ref{tab:Zp_VA}. 
The fermions of $(\cG\,, \UG\,, \sG\,, \DG^\prime\,, \eG^\prime\,, \EG)$ only exhibit vectorial gauge couplings with the $Z_\mu^\prime$.
This can be expected, since they already obtain vectorlike masses through the first-stage symmetry breaking, as one can find in Eqs.~\eqref{eq:masses_341_01} and \eqref{eq:masses_341_02}.
The SM fermions with the same electrical charges couple to the $Z_\mu^\prime$ universally.
This can also be expected, since the ${\rm SU}(6)$ GUT, which can unify the $\Gc_{331}$ gauge symmetries minimally, cannot have multiple fermions embedded non-trivially according to Georgi's counting rule and the third law.
Therefore, the flavor universality of the SM fermions should be expected through the flavor-conserving neutral currents of the effective $\Gc_{331}$ theory based on the {\it non-minimal} GUTs.
Consequently, the gauge couplings for the first generational SM fermions should be identical to the second and the third generational SM fermions as we have listed in Tab.~\ref{tab:Zp_VA}.
This is distinctive from the fermion contents in several previous 331 model studies~\cite{Pisano:1992bxx,Frampton:1992wt,Foot:1992rh,Montero:1992jk,Ng:1992st,Liu:1993gy,Pal:1994ba,Long:1995ctv,Ponce:2002sg,Dias:2003zt,Dias:2004dc,Ferreira:2011hm,Dong:2012bf,Buras:2012dp,Boucenna:2014ela,Boucenna:2014dia,Boucenna:2015zwa,Deppisch:2016jzl,Cao:2016uur,CarcamoHernandez:2021tlv,Hernandez:2021zje,Alves:2022hcp,Cherchiglia:2022zfy}, where the flavor non-universalities were assumed at the beginning.

\subsubsection{The electroweak gauge couplings}

\begin{table}[htp]
\begin{center}
\begin{tabular}{c|ccccc}
\hline \hline
  & $c/t$   &      $\cG$  &  $\UG$  & -  & -  \\ 
\hline
$g_f^{V }$  &  $\frac{1}{4} - \frac{2}{3} s_{ \theta_W}^2$    & $\frac{1}{2} - \frac{2}{3} s_{ \theta_W}^2$   & $ - \frac{2}{3} s_{ \theta_W}^2$  &  - & -  \\
$g_f^{A }$  & $-\frac{1}{4}$   &    $0$  &  $0$   & -  & -  \\
\hline
  & $s/b$   &   $\sG$  &  $\SG$  &  $\DG$  &  $\DG^\prime$   \\ 
  \hline
$g_f^{V }$  & $-\frac{1}{4} + \frac{1}{3} s_{ \theta_W}^2$  &     $-\frac{1}{2} + \frac{1}{3} s_{ \theta_W}^2$  &  $\frac{1}{3} s_{ \theta_W}^2$ &  $\frac{1}{3} s_{ \theta_W}^2$ &  $\frac{1}{3} s_{ \theta_W}^2$ \\ 
$g_f^{A }$  & $+\frac{1}{4}$   &    $0$  &  $0$ &  $0$  &  $0$  \\ 
\hline
  & $\mu/\tau$   &   $\eG$  &  $\eG^{\prime}$  &  $\eG^{\prime\prime }$  &  $\EG$   \\ 
  \hline
$g_f^{V }$  & $-\frac{1}{4} + s_{ \theta_W}^2 $  &   $-\frac{1}{4} +\frac{1}{2} s_{ \theta_W}^2 $  & $-\frac{1}{4} +\frac{1}{2} s_{ \theta_W}^2 $  & $-\frac{1}{4} +\frac{1}{2} s_{ \theta_W}^2 $ & $s_{ \theta_W}^2$   \\ 
$g_f^{A }$  & $+\frac{1}{4}$ &  $0$  & $0$  &  $0$  & $0$   \\ 
\hline\hline
\end{tabular}
\end{center}
\caption{
The couplings of the flavor-conserving neutral currents mediated by $Z_\mu$ in the $V-A$ basis, with $\theta_W$ being the Weinberg angle.
}
\label{tab:Z_VA}
\end{table}%

\para
At the stage of the EWSB, the flavor-conserving neutral currents are expressed as follows in the $V-A$ basis
\beqn
\Lc_{ {\rm SU}(2)_W}^{\rm NC\,, F}&=& \frac{e }{ s_{\theta_W} c_{\theta_W} } ( g_f^{V } \overline f \gamma^\mu f + g_f^{A } \overline f \gamma^\mu \gamma_5 f ) Z_\mu\,,
\eeqn
and we tabulate the vectorial and axial couplings in Tab.~\ref{tab:Z_VA}. 
Consistent SM fermion gauge couplings are obtained.
Besides of the SM fermion couplings, all heavy partner fermions in the spectrum only have vectorial gauge couplings to the $Z_\mu$ boson.

\subsection{The search for the $Z_\mu^\prime$ gauge boson}

\para
The {\it non-minimal} GUTs such as the two-generational ${\rm SU}(7)$ model predict an extended $\Gc_{331}$ effective theory above the EWSB scale according to the symmetry breaking pattern in Eq.~\eqref{eq:SU7_pattern}.
All massive gauge bosons and vectorlike fermions are expected to have masses of $\sim v_{331}$ according to the previous analyses.
According to Eqs.~\eqref{eq:FCCC_SU3W}, \eqref{eq:FCNC_SU3W} and \eqref{eq:FconsNC_SU3W}, only the flavor-conserving neutral $Z_\mu^\prime$ can couple to SM fermions and anti-fermions, while all other massive gauge bosons always mediate between a SM fermion and a heavy vectorlike fermion, or between heavy vectorlike fermions.
Accordingly, we can mostly expect the current and/or future terrestrial collider searches for the $Z_\mu^\prime$.
Based on the related gauge couplings in Tab.~\ref{tab:Zp_VA}, we estimate the leptonic decay branching ratios of ${\rm Br}[Z_\mu^\prime \to \ell^+ \ell^-]_{ 331}=0.2$ by assuming that the $Z_\mu^\prime$ cannot decay into the vectorlike fermions of $( \SG\,, \DG\,, \eG\,, \eG^{\prime\prime} )$.
The current LHC searched for the $Z_\mu^\prime$ via the di-lepton final states and assumed the sequential SM (SSM) scenario~\cite{ATLAS:2019erb} was assumed.
The corresponding leptonic decay branching ratio reads ${\rm Br}[Z_\mu^\prime \to \ell^+ \ell^-]_{\rm SSM}=0.06$ according to the gauge couplings in Tab.~\ref{tab:Z_VA}.
A rescaling of the couplings leads to the signal strength of
\beqn
\mu&\equiv& \frac{ \sigma[ p p \to Z_\mu^\prime ]_{331} \times {\rm Br}[Z_\mu^\prime \to \ell^+ \ell^-]_{331} }{   \sigma[ p p \to Z_\mu^\prime ]_{\rm SSM} \times  {\rm Br}[Z_\mu^\prime \to \ell^+ \ell^-]_{\rm SSM} } \,.
\eeqn
The ratio of the production cross sections is found to be
\beqn
&& \frac{ \sigma[ p p \to Z_\mu^\prime ]_{331}  }{\sigma[ p p \to Z_\mu^\prime ]_{\rm SSM}  } = 0.13\, \frac{ \kappa_{d/u} + 1.14 }{ \kappa_{d/u} + 0.78} \,,
\eeqn
where we estimate the ratio of~\cite{Mao:2017hpp}
\beqn
\kappa_{d/u}&=& \frac{ \int dx_1 dx_2 f_d(x_1)  f_{\bar d}(x_2) \delta( x_1 x_2 - m_{Z^\prime}^2/s) }{  \int dx_1 dx_2 f_u(x_1)  f_{\bar u}(x_2) \delta( x_1 x_2 - m_{Z^\prime}^2/s)  } \,,
\eeqn
by using the MSTW2008PDF~\cite{Martin:2009iq}.
Altogether, we find that $\mu \simeq 0.57$, which sets a limit of $m_{Z^\prime} \gtrsim 4.8\,{\rm TeV}$ to our current model setup according to the LHC searches for the SSM $Z_\mu^\prime$~\cite{ATLAS:2019erb}.
By further using the gauge boson mass in Eq.~\eqref{eq:331_GBmass02} and the mixing angle relation in Eq.~\eqref{eq:SW_relation}, we estimate that the current LHC searches have set a limit of $v_{331} \gtrsim 12\,{\rm TeV}$ to the {\it non-minimal} GUTs.

\section{Discussions}
\label{section:conclusion}

\para
In this work, we initiate a study of the {\it non-minimal} GUTs with multiple generational SM fermions that transform differently in the UV theory.
A two-generational ${\rm SU}(7)$ unified theory is a reasonable step towards more realistic model building.
Remarkably, the anomaly-free fermion contents in the UV theory do not display any generational structure, as one can find in Eq.~\eqref{eq:SU7_2gen_fermions}.
It will be straightforward to find that the first {\it non-minimal} GUT with the minimal fermion contents that can lead to three generations at the electroweak scale has a unified group of ${\rm SU}(8)$~\cite{Barr:1979xt,Ma:1981pr,Barr:2008pn}, which is also composed by both the rank-$2$ and the rank-$3$ anti-symmetric fermions, plus the anti-fundamental fermions.
Regarding this, some of the results in the two-generational ${\rm SU}(7)$ model will be expected to be relevant in a more realistic model construction.
Below, we summarize some major results of the current work.

\para
Firstly, the {\it non-minimal} GUTs are built based on the conjectured third law of the flavor unification.
Naturally, the unified gauge symmetry undergoes multiple intermediate stages of symmetry breaking below the GUT scale.
In the current context, we focus on the symmetry breaking pattern where the weak symmetries are extended. 
It can also be expected that the realistic symmetry breaking pattern of a three-generational theory includes both extended strong and weak sectors beyond the SM gauge symmetries.
At each stage of symmetry breaking, the vectorlike fermions that acquire masses are predictable through the anomaly-free conditions.
Similar to the ${\rm SU}(6)$ model, we found that the Higgs field of $\rep{35_H}$ can only develop VEV for the EWSB and give the top quark mass in the given symmetry breaking pattern.
These results suggest that the fermion masses are acquired through the natural Yukawa couplings of $Y_f \sim \Oc(1)$ in the {\it non-minimal} GUTs.

\para
Secondly, the emergent global DRS symmetries are generally expected with our conjectured third law on the anomaly-free fermion content. 
We determine the Higgs fields according to the minimal set of Yukawa couplings with the corresponding DRS symmetries.
With the minimal set of Higgs VEVs in Eqs.~\eqref{eqs:SU7_Higgs_VEVs_simple}, there can be additional Higgs VEVs generated through the Higgs mixing terms.
They are responsible to give masses to other electrically charged SM fermions other than the top quark.
In other words, the previously observed VEV generations through a $d=3$ operator in the ${\rm SU}(6)$ model~\cite{Chen:2021zwn} are generalized in the current context. 
In the context of the ${\rm SU}(7)$ model, the additional Higgs VEVs are either generated from $d=4$ renormalizable operators, or from $d\geq 5$ non-renormalizable operators.
To have a complete VEV generation chain as we have shown in Fig.~\ref{fig:VEV_chain} in the current model, the non-renormalizable operators are inevitably DRS-charged, and they can only be allowed with the gravitational effect that violates the global symmetries in general.

\para 
Thirdly, the observed fermion mass spectrum can partially explain the observed mass hierarchies of the SM fermion.
In particularly, the charm quark mass can be reasonably suppressed from the top quark.
Notice that Yukawa texture that leads to vanishing tree-level charm quark mass in Eq.~\eqref{eq:masses_EW_03} is purely due to the symmetries in the current setup.
Besides, we have successfully displayed the observed CKM mixing pattern with two generations.
These results may be viewed as positive hints for the future construction of the three-generational model.
However, the current model predicts the unrealistic mass relations of $m_s \approx m_b$ and $m_\mu \approx m_\tau$.
As we have mentioned in the second point, such VEV generation chains may be completely due to the non-renormalizable operators with the gravitational effects.
Besides, the hierarchical SM fermion masses among three generations were long conjectured due to the radiative mechanism with extended gauge bosons and/or Higgs fields~\cite{Barr:1976bk,Barr:1978rv,Weinberg:2020zba}, including in the class of the ${\rm SU}(N)$ GUTs~\cite{Barr:1979xt}.
Altogether with the distinct gauge symmetries and the emergent DRS symmetries in the three-generational {\it non-minimal} GUTs, it is premature to conclude such degenerate fermion mass pattern among different generations. 
Of course, the usual $b-\tau$ mass unification is further extended to the second generational fermions of $(s\,,\mu)$ at the tree level.
Thus, it will be reasonable to investigate the renormalization group effects~\cite{Buras:1977yy} in a realistic {\it non-minimal} GUT, with two prerequisites of: (i) the identification of the intermediate symmetry-breaking scales, and (ii) the identification of the SM fermion representations with the extended color and weak symmetries.


\para
Lastly, we wish to point out the early constructions of three-generational ${\rm SU}(7)$ model by Frampton~\cite{Frampton:1979cw,Frampton:1979tj} cannot be realistic models.
Our previous argument was that the partition of the fermions into irreducible anomaly-free sets in Eqs.~\eqref{eqs:Frampton_SU7_partition} violates our conjectured third law.
According to the current analyses, the ${\rm SU}(7)$ can only undergo two more intermediate symmetry breaking stages above the EWSB scale.
One can naturally expect that the first generational fermion masses, in particular the down quark and the electron, will be degenerate with SM fermions that carry the same electric charges.

\para
In addition to the study of the SM fermion masses and their weak mixings, several new physics ingredients have automatically emerged in the current context, which include
\begin{itemize}

\item Emergence of the DRS symmetries that may give rise to high-quality axion~\cite{Georgi:1981pu,Barr:1992qq,Kamionkowski:1992mf,Holman:1992us,Ghigna:1992iv,Dobrescu:1996jp} for the strong CP problem~\cite{Peccei:1977hh}.

\item The vectorlike mirror quark doublets~\cite{Maalampi:1988va} and the sterile neutrinos from the minimal fermion contents.

\item Flavor-changing neutral currents and flavor non-universal gauge couplings~\cite{Crivellin:2021sff} from the extended weak symmetries.
In particular, the flavor non-universality originates from the non-trivial embedding of multiple generations into the unified theory.

\end{itemize}
These ingredients are expected to be general in {\it non-minimal} GUTs with three generational SM fermions, given their anomaly-free fermion content according to our conjectured third law of flavor unification, as well as the realistic symmetry breaking patterns from the group theoretical considerations.


\section*{Acknowledgements}
%
%
\para
We would like to thank Haipeng An, Chee Sheng Fong, Yuan Sun, Xun-Jie Xu, Wenbin Yan, Chang-Yuan Yao, Shuang-Yong Zhou and Ye-Ling Zhou for very enlightening discussions at different stages of this work. 
N.C. would like to thank Peking University and Tsinghua University for hospitality when preparing this work.
N.C. is partially supported by the National Natural Science Foundation of China (under Grants No. 12035008 and No. 12275140), and Nankai University.
Y.N.M. is partially supported by the National Natural Science Foundation of China (under Grant No. 12205227), the Fundamental Research Funds for the Central Universities (WUT: 2022IVA052), and Wuhan University of Technology.

\appendix

\section{Conventions, rules of decompositions, and charge quantizations}
\label{section:Br}

\para
In this section, we list the decomposition rules of the fermions and Higgs fields that are relevant in the $\gSU(7)$ breaking patterns.
After the GUT symmetry breaking, we define the ${\rm U}(1)_{X_0}$ charges as follows
\beqn
\hat X_0( \rep{7} ) &\equiv& {\rm diag} ( \underbrace{ - \frac{1}{3}  \,, - \frac{1}{3}\,, - \frac{1}{3} }_{ \rep{3_c} }\,, \underbrace{ +\frac{1}{4} \,, +\frac{1}{4}\,, +\frac{1}{4}\,, +\frac{1}{4} }_{ \rep{4_W} } )\,.\label{eq:X0charge}
\eeqn
Sequentially, the ${\rm U}(1)_{X_1}$ and ${\rm U}(1)_{Y}$ charges are defined according to the ${\rm SU}(4)_W$ fundamental representation as follows
\beqs
\beqn
\hat X_1(\rep{4_W}) &\equiv&\frac{1}{ \sqrt{6} } T_{ {\rm SU}(4)}^{15} + \Xc_0 \,\mathbb{I}_4 = {\rm diag} ( \frac{1}{12}+\Xc_0 \,, \frac{1}{12}+ \Xc_0 \,, \frac{1}{12} + \Xc_0\,, -\frac{1}{4}+ \Xc_0 ) \,,\label{eq:X1charge}\\[1mm]
\hat Y ( \rep{4_W} )&\equiv&  \frac{1}{ \sqrt{3} } T_{ {\rm SU}(4)}^{8} + \Xc_1\,\mathbb{I}_4 = {\rm diag} ( \frac{1}{6} + \Xc_1\,, \frac{1}{6} + \Xc_1\,, - \frac{1}{3} + \Xc_1 \,, \Xc_1 ) \non
&=& {\rm diag} ( \frac{1}{4}+\Xc_0 \,,   \frac{1}{4}+ \Xc_0  \,, -\frac{1}{4} + \Xc_0  \,, -\frac{1}{4} + \Xc_0  )  \,. \label{eq:Ycharge}
\eeqn
\eeqs
Explicitly, the ${\rm SU}(4)$ Cartan generators are listed as follows
\beqs\label{eqs:SU4_Cartan}
\beqn
T_{ {\rm SU}(4)}^{3} &=& \frac{1}{2 }\,{\rm diag}(1\,, -1\,, 0\,,0) \,,\\[1mm]
T_{ {\rm SU}(4)}^{8} &=& \frac{1}{2 \sqrt{3} }\,{\rm diag}(1\,, 1\,, -2\,,0) \,,\\[1mm]
T_{ {\rm SU}(4)}^{15} &=& \frac{1}{ 2\sqrt{6} } \,{\rm diag}(1\,, 1\,, 1\,, -3) \,.
\eeqn
\eeqs 
Correspondingly, the electric charge quantization is given by
\beqn\label{eq:Qcharge_4fund}
\hat Q_e( \rep{4_W} )&\equiv& T_{ {\rm SU}(4) }^3 + \Yc \,\mathbb{I}_4 = {\rm diag} ( \frac{3}{4}+ \Xc_0 \,,   - \frac{1}{4}+ \Xc_0  \,, -\frac{1}{4} + \Xc_0  \,, -\frac{1}{4} + \Xc_0  )   \,.
\eeqn
For the ${\rm SU}(4)_W$ adjoint, the $4\times 4$ electric charge matrix becomes
\beqn\label{eq:Qcharge_4adj}
\hat Q_e( \rep{15_W} )&\equiv&{\rm diag} ( \frac{3}{4} \,,   - \frac{1}{4}  \,, -\frac{1}{4}  \,, -\frac{1}{4}  ) \,,
\eeqn
with $\Xc_0=0$.
We also define our convention of indices for different gauge groups in Tab.~\ref{tab:notations}.
The fundamental and anti-fundamental representations will be denoted by superscripts and subscripts, respectively.

\begin{table}[htp]
\begin{center}
\begin{tabular}{c|ccc}
\hline \hline
Indices   &  group  & irrep &  range    \\
\hline
$a\,,b\,,c$   &  ${\rm SU}(3)_c$   &  fundamental  & $1\,,2\,,3$   \\
   &      & anti-fundamental  &   \\
 $A\,,B\,,C$   &     ${\rm SU}(3)_c$  & adjoint  & $1\,,...\,,8$  \\ \hline
 $\bar i\,,\bar j\,, \bar k$   &  ${\rm SU}(4)_W$   &  fundamental  & $1\,,2\,,3\,,4$   \\
   &      & anti-fundamental  &   \\
  $\bar I\,,\bar J\,, \bar K$   &     ${\rm SU}(4)_W$  & adjoint  & $1\,,...\,,15$  \\ \hline
 $\tilde i \,, \tilde j \,, \tilde k $   &  ${\rm SU}(3)_W$   &  fundamental  & $1\,,2\,,3$   \\
   &      & anti-fundamental  &   \\
  $\tilde I \,, \tilde J \,, \tilde K $   &     ${\rm SU}(3)_W$  & adjoint  & $1\,,...\,,8$  \\ \hline
 $ i\,, j\,, k$   &  ${\rm SU}(2)_W$   &  fundamental  & $1\,,2$   \\
   &      & anti-fundamental  &   \\
  $ I\,, J\,, K$   &     ${\rm SU}(2)_W$  & adjoint  & $1\,,2\,,3$  \\
\hline\hline
\end{tabular}  
\end{center}
\caption{
Definition of indices for various gauge groups. 
}
\label{tab:notations}
\end{table}%

\section{Name scheme of the ${\rm SU}(7)$ fermions}
\label{section:name}
%
%

\para
Here, we list the names for fermions in Tabs.~\ref{tab:SU7_2gen_7barferm}, \ref{tab:SU7_2gen_21ferm}, and \ref{tab:SU7_2gen_35ferm} according to the symmetry breaking patten analyzed in Sec.~\ref{section:breaking}.
Names of all heavy partner fermions are expressed in $\mathfrak{Gothic}$ fonts.
According to Tabs.~\ref{tab:SU7_2gen_7barferm}, \ref{tab:SU7_2gen_21ferm}, and \ref{tab:SU7_2gen_35ferm}, we name fermions according to their electric charges as follows
\beqn\label{eq:SU7_fermion_names}
Q_e=+\frac{2}{3}~&:&~ c\,, ~ t\,, ~ \cG\,, ~ \UG \,, \non
Q_e=-\frac{1}{3}~&:&~ s \,, ~ b\,, ~ \sG\,,~ \DG\,,~ \DG^\prime \,, ~ \SG \,, \non
Q_e= \pm 1~&:&~ \mu\,, ~ \tau \,,~ \eG\,, ~ \eG^{\prime}\,, ~ \eG^{\prime\prime} \,, ~ \EG\,, \non
Q_e=0~&:&~ \nu_\mu \,, ~ \nu_\tau \,, ~ \nG\,, ~\nG^\prime \,,~ \nG^{\prime\prime} \non 
&& \check \Nc_L^{ \Lambda} ~( \Lambda = {\rm I}\,, 3\,, \dot {\rm I}\,, \dot 2) \,,~ \check \Nc_L^{ \Lambda^\prime } ~( \Lambda = {\rm I}\,, {\rm II} \,, 3\,, \dot {\rm I}\,, \dot 2 ) \,,~ \check \nG \,.
\eeqn
Neutrinos marked with $\check{}$ are sterile neutrinos, since they are SM singlets.

\section{The Higgs mixing operators in the ${\rm SU}(7)$ Higgs potential}
\label{section:SU7_HiggsOp}

\para
In Sec.~\ref{section:Higgs}, we argue that the VEV terms in the ${\rm SU}(7)$ Higgs potential should be determined through gauge-invariant operators listed in Tabs.~\ref{tab:SU7_HiggsMix_renOp} and \ref{tab:SU7_HiggsMix_nonrenOp}.
In this section, we show explicitly whether each term contributes to a VEV term or not according to the Higgs decompositions in Eqs.~\eqref{eqs:SU7_Higgs_Br}.
This can be made by checking whether a specific operator composed by the Higgs components framed with boxes is gauge-invariant.

\subsection{General rules}

\para
We list some general rules to obtain the VEV term contributions to the Higgs potential.
They read as follows
\begin{enumerate}

\item One should always choose the Higgs VEV components, and look for gauge-invariant operators at each stage of symmetry breaking.
In the current context, the VEV components are framed with boxes in Eqs.~\eqref{eqs:SU7_Higgs_Br}.

\item For VEV components from DRS-transforming Higgs fields that develop VEVs, their VEVs at one specific symmetry-breaking stage should not appear more than once in the Higgs mixing operators.
Otherwise, the VEVs at the specific stage will be vanishing due to the DRS-invariant $\epsilon$-tensors in the Higgs mixing operators.
In the current context, the Higgs fields of $( \repb{7_H}_{\,,\lambda} \,, \repb{21_H}_{\,,\dot \lambda})$ are DRS-transforming fields, while others are DRS singlets, as can be seen in Tab.~\ref{tab:SU7_setup}.

\end{enumerate}

\subsection{The $d=3$ operators}

\para
For two $d=3$ operators of $\Oc_{ \mathscr A}^{d=3}$ and $\Oc_{ \mathscr B}^{d=3}$, we decompose them as follows
\beqs\label{eqs:Oc3_decomp}
\beqn
\Oc_{ \mathscr A}^{d=3}&=&( \rep{21_H})^2 \cdot \rep{35_H} \supset  ( \rep{1}\,, \rep{6} \,, +\frac{1}{2} )_{\mathbf{H}} \otimes ( \rep{1}\,, \rep{6} \,, +\frac{1}{2} )_{\mathbf{H}} \otimes ( \rep{1}\,, \repb{4} \,, +\frac{3}{4} )_{\mathbf{H}} \non
&\Rightarrow&  \textrm{no gauge-invariant term} \,,\\[1mm]
\Oc_{ \mathscr B}^{d=3} &=& \rep{7_H} \cdot  ( \rep{35_H} )^2  \supset ( \rep{1}\,, \rep{4} \,, +\frac{1}{4} )_{\mathbf{H}} \otimes ( \rep{1}\,, \repb{4} \,, +\frac{3}{4} )_{\mathbf{H}} \otimes ( \rep{1}\,, \repb{4} \,, +\frac{3}{4} )_{\mathbf{H}} \non
&\Rightarrow&  \textrm{no gauge-invariant term}  \,.
\eeqn
\eeqs
Obviously, none of the VEV components in Eqs.~\eqref{eqs:Oc3_decomp} are gauge-invariant after the decomposition.

\subsection{The $d=4$ operators}

\para
For the operator of $\Oc_{\mathscr A}^{d=4}$, we find
\beqn\label{eq:Oc4A_decomp}
&& \epsilon^{ \lambda \kappa \delta }\, \repb{7_H}_{\,,\lambda }  \repb{7_H}_{\,,\kappa} \repb{7_H}_{\,,\delta} \rep{35_H} \non
&\supset& \epsilon^{ \lambda \kappa \delta}  \, ( \rep{1}\,, \repb{4} \,, -\frac{1}{4} )_{\mathbf{H}\,,  \lambda } \otimes ( \rep{1}\,, \repb{4} \,, -\frac{1}{4} )_{\mathbf{H}\,, \kappa } \otimes ( \rep{1}\,, \repb{4} \,, -\frac{1}{4} )_{\mathbf{H}\,, \delta }  \otimes ( \rep{1}\,, \repb{4} \,, +\frac{3}{4} )_{\mathbf{H} } \non
&\supset& w_{ \repb{4}\,, {\rm II}} \epsilon^{ \lambda \kappa } \,   ( \rep{1}\,, \repb{3} \,, -\frac{1}{3} )_{\mathbf{H}\,,  \lambda } \otimes ( \rep{1}\,, \repb{3} \,, -\frac{1}{3} )_{\mathbf{H}\,,  \kappa }   \otimes ( \rep{1}\,, \repb{3} \,, +\frac{2}{3} )_{\mathbf{H}}^\prime  \,.
\eeqn
We took $\delta={\rm II}$ according to the fermion-Higgs matching pattern at the second step.
Thus the remaining flavor indices are $(\lambda\,, \kappa)=( {\rm I}\,,3)$.
With the VEV assignment of $\langle ( \rep{1}\,, \repb{3} \,, -\frac{1}{3} )_{\mathbf{H}\,,  {\rm I} }  \rangle= (0\,, 0\,, V_{ \repb{3}\,,{\rm I}})^T$ and $ \langle ( \rep{1}\,, \repb{3} \,, +\frac{2}{3} )_{\mathbf{H} } \rangle=(v_t \,, 0\,, 0)^T$, it is unavoidable to have $( \rep{1}\,, \repb{2} \,, -\frac{1}{2} )_{\mathbf{H}\,,  3 } \subset ( \rep{1}\,, \repb{3} \,, -\frac{1}{3} )_{\mathbf{H}\,,  3}$ develop the VEV of $\langle  ( \rep{1}\,, \repb{2} \,, -\frac{1}{2} )_{\mathbf{H}\,,  3 } \rangle =(0\,,u_{ \repb{2}\,,3})^T$.
Explicitly, this VEV term reads
\beqn\label{eq:Oc4A_VEVgen}
\Oc_{\mathscr A}^{d=4}&\sim& g_{ 4\mathscr A}  w_{ \repb{4}\,,  {\rm II}}  V_{ \repb{3}\,, {\rm I}} u_{\repb{2}\,,3 } v_t   \Rightarrow \boxed{ u_{\repb{2}\,,3} } \,.
\eeqn
According to the minimal VEV assignments for the fermion-Higgs matching pattern in Eqs.~\eqref{eqs:SU7_Higgs_VEVs_simple}, this operator leads to a VEV of $u_{\repb{2}\,,3}$ to avoid the tadpole term.

\para
Obviously, the gauge-invariant VEV term in Eq.~\eqref{eq:Oc4A_decomp} plays the similar role as the $\nu$-term in the ${\rm SU}(6)$ model~\cite{Chen:2021zwn} and can lead to a tadpole term.
By denoting the coefficient of this operator as $g_{ 4\mathscr A}$, we have $\nu_{ 4\mathscr A}=g_{ 4\mathscr A} w_{ \repb{4}\,, {\rm II}}$.
According to the study of the third-generational ${\rm SU}(6)$ model, one expects $\nu\sim \Oc(100)\,{\rm GeV}$.
Similar to the ${\rm SU}(6)$ model, a fine-tuning is also observed.
Thus, the $\nu$-problem persists in the current context and will be solved in the realistic models.

\para
For the operator of $\Oc_{\mathscr B}^{d=4}$, we find
\beqs\label{eqs:Oc4B_decomp}
\beqn
&& \epsilon^{ \dot \lambda  \dot \kappa} \repb{21_H}_{\,, \dot \lambda }  \repb{21_H}_{\,,\dot \kappa} \cdot \rep{7_H} \cdot \rep{35_H} \non
&\supset& \epsilon^{ \dot \lambda  \dot \kappa } \, ( \rep{1}\,, \rep{6} \,, -\frac{1}{2} )_{\mathbf{H}\,, \dot \lambda } \otimes ( \rep{1}\,, \rep{6} \,, -\frac{1}{2} )_{\mathbf{H}\,, \dot \kappa }  \otimes ( \rep{1}\,, \rep{4} \,, +\frac{1}{4} )_{\mathbf{H} }  \otimes  ( \rep{1}\,, \repb{4} \,, +\frac{3}{4} )_{\mathbf{H}} \non
&\supset& \Oc_{\mathscr B1}^{d=4} + \Oc_{\mathscr B2}^{d=4} + \Oc_{\mathscr B3}^{d=4}    \,,\\[1mm]
\Oc_{\mathscr B1}^{d=4}&=& w_{ \rep{4}}  \epsilon^{ \dot \lambda  \dot \kappa }\, ( \rep{1}\,, \repb{3} \,, -\frac{1}{3} )_{\mathbf{H}\,, \dot \lambda } \otimes ( \rep{1}\,, \repb{3} \,, -\frac{1}{3} )_{\mathbf{H}\,, \dot \kappa}  \otimes  ( \rep{1}\,, \repb{3} \,, +\frac{2}{3} )_{\mathbf{H}}^\prime \,, \label{eq:Oc4B_decomp01}  \\[1mm]
\Oc_{\mathscr B2}^{d=4}&=& \epsilon^{ \dot \lambda  \dot \kappa } \, ( \rep{1}\,, \rep{3} \,, -\frac{2}{3} )_{\mathbf{H}\,, \dot \lambda} \otimes ( \rep{1}\,, \repb{3} \,, -\frac{1}{3} )_{\mathbf{H}\,, \dot \kappa}  \otimes  ( \rep{1}\,, \rep{3} \,, +\frac{1}{3} )_{\mathbf{H}} \otimes  ( \rep{1}\,, \repb{3} \,, +\frac{2}{3} )_{\mathbf{H}}^\prime  \,, \label{eq:Oc4B_decomp02}  \\[1mm]
\Oc_{\mathscr B3}^{d=4}&=& \epsilon^{ \dot \lambda  \dot \kappa } \, ( \rep{1}\,, \rep{2} \,, -\frac{1}{2} )_{\mathbf{H}\,, \dot \lambda} \otimes ( \rep{1}\,, \repb{2} \,, -\frac{1}{2} )_{\mathbf{H}\,, \dot \kappa}  \otimes  ( \rep{1}\,, \rep{2 } \,, +\frac{1}{2} )_{\mathbf{H}} \otimes  ( \rep{1}\,, \repb{2} \,, +\frac{1 }{2} )_{\mathbf{H}}^\prime \,. \label{eq:Oc4B_decomp03}  
\eeqn
\eeqs
Clearly, the operator of $\Oc_{\mathscr B1}^{d=4}$ plays the similar role as the operator of $\Oc_{\mathscr A}^{d=4}$. 
One can think of this as a $\nu$-term for the second generational fermions, since two $( \rep{1}\,, \repb{3} \,, -\frac{1}{3} )_{\mathbf{H}\,, \dot \lambda }$ carry the dotted flavor indices.
Explicitly, the VEV term from Eq.~\eqref{eq:Oc4B_decomp01} read
\beqs
\beqn
\Oc_{\mathscr B1}^{d=4}&\sim& g_{4 \mathscr B} w_{ \rep{4}}  V_{\repb{3}\,, \dot {\rm I}} u_{\repb{2}\,, \dot 2} v_t \Rightarrow \boxed{ u_{\repb{2}\,, \dot 2} } \,, \label{eq:Oc4B1_VEVgen}\\[1mm]
 \Oc_{\mathscr B2}^{d=4} &\sim& g_{4 \mathscr B} V_{ \rep{3}}  V_{\repb{3}\,, \dot {\rm I}} u_{\rep{2}\,, \dot 2} v_t \Rightarrow  \boxed{V_{ \rep{3}}} \,, \boxed{u_{\rep{2}\,, \dot 2} }\,, \\[1mm]
 \Oc_{\mathscr B3}^{d=4} &\sim& g_{4 \mathscr B} u_{\rep{2}\,, [ \dot 2} u_{\repb{2}\,, \dot {\rm I} ] } u_{\rep{2}} v_t  \Rightarrow  \boxed{ u_{\rep{2}\,, \dot 2}}\,, \boxed{u_{\repb{2}\,, \dot {\rm I}}}\,, \boxed{u_{\rep{2}} } ~{\rm or }~ \boxed{ u_{\rep{2}\,, \dot {\rm I}}} \,, \boxed{ u_{\repb{2}\,, \dot 2}} \,, \boxed{ u_{\rep{2}} } \,.  
\eeqn
\eeqs

\para
For the operator of $\Oc_{\mathscr C}^{d=4}$, we find
\beqs\label{eqs:Oc4C_decomp}
\beqn
&& \epsilon^{ \dot \lambda  \dot \kappa } \, \repb{21_H}_{\,, \dot \lambda }  \repb{21_H}_{\,,\dot \kappa }  \cdot ( \rep{21_H} )^2
 \supset  \Oc_{\mathscr C1}^{d=4} + \Oc_{\mathscr C2}^{d=4} + \Oc_{\mathscr C3}^{d=4} + \Oc_{\mathscr C4}^{d=4}  \,,\\[1mm]
\Oc_{\mathscr C1}^{d=4}  
&=& \epsilon^{ \dot \lambda  \dot \kappa } \,  ( \rep{1}\,, \rep{3} \,, -\frac{2}{3} )_{\mathbf{H}\,, \dot \lambda }^{ [ \tilde r} \otimes ( \rep{1}\,, \rep{3} \,, -\frac{2}{3} )_{\mathbf{H}\,, \dot \kappa}^{\tilde s ] } \otimes ( \rep{1}\,, \repb{3} \,, +\frac{2}{3} )_{\mathbf{H}\,, \tilde r} \otimes ( \rep{1}\,, \repb{3} \,, +\frac{2}{3} )_{\mathbf{H}\,, \tilde s}  \non
&\Rightarrow & 0 \,, \label{eq:Oc4C_decomp01}  \\[1mm]
\Oc_{\mathscr C2}^{d=4}  &=& \epsilon^{ \dot \lambda  \dot \kappa }  \, ( \rep{1}\,, \rep{3} \,, -\frac{2}{3} )_{\mathbf{H}\,, \dot \lambda } \otimes  ( \rep{1}\,, \repb{3} \,, -\frac{1}{3} )_{\mathbf{H}\,, \dot \kappa} \otimes ( \rep{1}\,, \repb{3} \,, +\frac{2}{3} )_{\mathbf{H}} \otimes ( \rep{1}\,, \rep{3} \,, +\frac{1}{3} )_{\mathbf{H}}^\prime  \,, \label{eq:Oc4C_decomp02}  \\[1mm]
\Oc_{\mathscr C3}^{d=4}  &=& \epsilon^{ \dot \lambda  \dot \kappa }  \, ( \rep{1}\,, \rep{2} \,, -\frac{1 }{2} )_{\mathbf{H}\,, \dot \lambda } \otimes  ( \rep{1}\,, \repb{2 } \,, -\frac{1}{2 } )_{\mathbf{H}\,, \dot \kappa} \otimes ( \rep{1}\,, \repb{2} \,, +\frac{1 }{2} )_{\mathbf{H}} \otimes ( \rep{1}\,, \rep{2} \,, +\frac{1}{2 } )_{\mathbf{H}}^\prime   \,, \label{eq:Oc4C_decomp03}  \\[1mm]
\Oc_{\mathscr C4}^{d=4}  &=&  \epsilon^{ \dot \lambda  \dot \kappa } \, ( \rep{1}\,, \repb{3} \,, -\frac{1}{3} )_{\mathbf{H}\,, \dot \lambda}  \otimes ( \rep{1}\,, \repb{3} \,, -\frac{1}{3} )_{\mathbf{H}\,, \dot \kappa}  \otimes ( \rep{1}\,, \rep{3} \,, +\frac{1}{3} )_{\mathbf{H}}^{\prime}  \otimes ( \rep{1}\,, \rep{3} \,, +\frac{1}{3} )_{\mathbf{H}}^{ \prime}   \,. \label{eq:Oc4C_decomp04}
\eeqn
\eeqs
In Eq.~\eqref{eq:Oc4C_decomp01}, the possible EWSB VEVs from the $( \rep{1}\,, \rep{3} \,, -\frac{2}{3} )_{\mathbf{H}\,, \dot \lambda }$ can only be developed with $\tilde r = \tilde s=1$.
The anti-symmetrizing of two ${\rm SU}(3)_W$ indices clearly makes this term vanishing.
Explicitly, the VEV terms from Eqs.~\eqref{eq:Oc4C_decomp02} and \eqref{eq:Oc4C_decomp03} read
\beqs
\beqn
  \Oc_{\mathscr C2}^{d=4} &\sim& g_{4 \mathscr C}  V_{ \rep{3}}^\prime V_{ \repb{3}\,,  \dot {\rm I}} u_{ \rep{2}\,, \dot 2} u_{\repb{2}}  \Rightarrow  \boxed{ u_{\rep{2}\,, \dot 2} } \,, \boxed{ u_{\repb{2}}}  \,,\\[1mm]
  \Oc_{\mathscr C3}^{d=4} &\sim& g_{4 \mathscr C}  u_{ \rep{2}\,, \dot 2} u_{\repb{2}\,,  \dot {\rm I}} u_{\repb{2}} u_{\rep{2}}^\prime  \Rightarrow  \boxed{  u_{ \rep{2}\,, \dot 2}} \,, \boxed{  u_{\repb{2}\,,  \dot {\rm I}}}  \,, \boxed{  u_{\repb{2}}} \,,\boxed{  u_{\rep{2}}^\prime } \,,\\[1mm]
\Oc_{\mathscr C4}^{d=4}  &\sim&  g_{4 \mathscr C} V_{ \rep{3}}^\prime V_{ \repb{3}\,,  \dot {\rm I}} u_{ \repb{2}\,, \dot 2} u_{\rep{2}}^\prime  \Rightarrow \boxed{u_{ \repb{2}\,, \dot 2} } \,, \boxed{ u_{\rep{2}}^\prime }  \,.
\eeqn
\eeqs

\para
For the operator of $\Oc_{\mathscr D}^{d=4}$, we find
\beqn\label{eq:Oc4D_decomp}
\Oc_{ \mathscr D}^{d=4}  &=& \rep{7_H} \cdot (\rep{21_H} )^3 \supset  ( \rep{1}\,, \rep{4} \,, +\frac{1}{4} )_{\mathbf{H}} \otimes ( \rep{1}\,, \rep{6} \,, +\frac{1}{2} )_{\mathbf{H}}  \otimes ( \rep{1}\,, \rep{6} \,, +\frac{1}{2} )_{\mathbf{H}}  \otimes ( \rep{1}\,, \rep{6} \,, +\frac{1}{2} )_{\mathbf{H}}  \non
 &\Rightarrow& \textrm{no gauge-invariant term} \,.
\eeqn
Hence, this term does not contribute to a VEV term in the Higgs potential.

\subsection{The $d=5$ operators}

\para
For the operator of $\Oc_{\mathscr A}^{d=5}$, we find
\beqn\label{eq:Oc5A_decomp}
&& \epsilon^{\lambda \kappa \gamma } \repb{7_H}_{\,,\lambda }  \repb{7_H}_{\,,\kappa } \repb{7_H}_{\,,\gamma}  \cdot  \epsilon^{ \dot \rho  \dot \delta } \repb{21_H}_{\,, \dot \rho}  \repb{21_H}_{\,,\dot \delta} \non
&\supset&  \epsilon^{\lambda \kappa \gamma } \epsilon^{ \dot \rho  \dot \delta }\, ( \rep{1}\,, \repb{4} \,, -\frac{1}{4} )_{\mathbf{H}\,, \lambda} \otimes ( \rep{1}\,, \repb{4} \,, -\frac{1}{4} )_{\mathbf{H}\,, \kappa} \otimes ( \rep{1}\,, \repb{4} \,, -\frac{1}{4} )_{\mathbf{H}\,, \gamma} \non
& \otimes&  ( \rep{1}\,, \rep{6} \,, -\frac{1}{2} )_{\mathbf{H}\,, \dot \rho} \otimes ( \rep{1}\,, \rep{6} \,, -\frac{1}{2} )_{\mathbf{H}\,, \dot \delta}  \Rightarrow \textrm{no gauge-invariant term} \,.
\eeqn
Hence, this term does not contribute to a VEV term in the Higgs potential.

\para
For the operator of $\Oc_{\mathscr B}^{d=5}$, we find
\beqn\label{eq:Oc5B_decomp}
&& (\rep{7_H} )^4  \rep{35_H}  \non
&\supset&  ( \rep{1}\,, \rep{4} \,, +\frac{1}{4} )_{\mathbf{H}} \otimes ( \rep{1}\,, \rep{4} \,, +\frac{1}{4} )_{\mathbf{H}} \otimes ( \rep{1}\,, \rep{4} \,, +\frac{1}{4} )_{\mathbf{H}} \otimes ( \rep{1}\,, \rep{4} \,, +\frac{1}{4} )_{\mathbf{H}} \otimes  ( \rep{1}\,, \repb{4} \,, +\frac{3}{4} )_{\mathbf{H}} \non
  &\Rightarrow& \textrm{no gauge-invariant term} \,.
\eeqn
Hence, this term does not contribute to a VEV term in the Higgs potential.

\para
For the operator of $\Oc_{\mathscr C}^{d=5}$, we find
\beqn\label{eq:Oc5C_decomp}
&&  ( \epsilon^{\lambda \kappa \delta }\, \repb{7_H}_{\,,\lambda }  \repb{7_H}_{\,,\kappa } \repb{7_H}_{\,,\delta}  ) \cdot \rep{7_H} \cdot \rep{21_H} \non
&\supset& \epsilon^{\lambda \kappa \delta }  \, ( \rep{1}\,, \repb{4} \,, -\frac{1}{4} )_{\mathbf{H}\,, \lambda} \otimes ( \rep{1}\,, \repb{4} \,, -\frac{1}{4} )_{\mathbf{H}\,, \kappa } \otimes ( \rep{1}\,, \repb{4} \,, -\frac{1}{4} )_{\mathbf{H}\,,\delta }  \otimes ( \rep{1}\,, \rep{4} \,, +\frac{1}{4} )_{\mathbf{H}} \otimes ( \rep{1}\,, \rep{6} \,, +\frac{1}{2} )_{\mathbf{H}} \non
&\supset&  \Oc_{\mathscr C1}^{d=5} + \Oc_{\mathscr C2}^{d=5} + \Oc_{\mathscr C3}^{d=5} \,.
\eeqn
Explicitly, the VEV term from Eq.~\eqref{eq:Oc5C_decomp} reads
\beqn
\Oc_{\mathscr C1}^{d=5}&\sim& g_{5 \mathscr C} w_{ \repb{4}\,, {\rm II} } V_{ \repb{3}\,, {\rm I}} u_{ \repb{2}\,,3}  V_{ \rep{3}}^\prime u_{\rep{2}}  \Rightarrow \boxed{ u_{ \repb{2}\,,3} } \,, \boxed{ u_{\rep{2}}} \,, \non
\Oc_{\mathscr C2}^{d=5}&\sim& g_{5 \mathscr C} w_{ \repb{4}\,, {\rm II} } V_{ \repb{3}\,, {\rm I}} u_{ \repb{2}\,,3}  V_{ \rep{3}} u_{ \rep{2}}^\prime \Rightarrow \boxed{ u_{ \repb{2}\,,3} } \,,  \boxed{ V_{\rep{3}} } \,, \boxed{ u_{ \rep{2}}^\prime}  \non
\Oc_{\mathscr C3}^{d=5}&\sim& g_{5 \mathscr C} w_{ \repb{4}\,, {\rm II} } V_{ \repb{3}\,, {\rm I}} u_{ \repb{2}\,,3}  w_{ \rep{4}} u_{ \repb{2}} \Rightarrow \boxed{ u_{ \repb{2}\,,3} } \,, \boxed{ u_{ \repb{2}} } \,.
\eeqn

\para
For the operator of $\Oc_{\mathscr D}^{d=5}$, we find
\beqn\label{eq:Oc5D_decomp}
&&  ( \epsilon^{\dot \lambda \dot \kappa }\, \repb{21_H}_{\,, \dot \lambda }  \repb{21_H}_{\,,\dot \kappa }   )^2 \cdot \rep{7_H}  \non
&\supset& \epsilon^{\dot \lambda_1 \dot \kappa_1 } \epsilon^{\dot \lambda_2 \dot \kappa_2 }\, ( \rep{1}\,, \rep{6} \,, -\frac{1}{2} )_{\mathbf{H}\,,\dot \lambda_1 } \otimes ( \rep{1}\,, \rep{6} \,, -\frac{1}{2} )_{\mathbf{H}\,,\dot \kappa_1 } \otimes ( \rep{1}\,, \rep{6} \,, -\frac{1}{2} )_{\mathbf{H}\,,\dot \lambda_2 } \otimes ( \rep{1}\,, \rep{6} \,, -\frac{1}{2} )_{\mathbf{H}\,,\dot \kappa_2 } \non
& \otimes&  ( \rep{1}\,, \rep{4} \,, +\frac{1}{4} )_{\mathbf{H} } \Rightarrow \textrm{no gauge-invariant term} \,.
\eeqn
Hence, this term does not contribute to a VEV term in the Higgs potential.

\subsection{The $d=6$ operators}

\para
For the operator of $\Oc_{\mathscr A}^{d=6}$, we find
\beqn\label{eq:Oc6A_decomp} 
&& (\epsilon^{ \lambda \kappa \delta }\,  \repb{7_H}_{\,,\lambda }  \repb{7_H}_{\,,\kappa} \repb{7_H}_{\,,\delta} )\cdot  ( \rep{7_H} )^3 \non
&\supset& \epsilon^{ \lambda \kappa \delta } \, ( \rep{1}\,, \repb{4} \,, -\frac{1}{4} )_{\mathbf{H}\,,  \lambda } \otimes ( \rep{1}\,, \repb{4} \,, -\frac{1}{4} )_{\mathbf{H}\,, \kappa} \otimes ( \rep{1}\,, \repb{4} \,, -\frac{1}{4} )_{\mathbf{H}\,, \delta} \non
&\otimes & ( \rep{1}\,, \rep{4} \,, +\frac{1}{4} )_{\mathbf{H}}  \otimes ( \rep{1}\,, \rep{4} \,, +\frac{1}{4} )_{\mathbf{H}} \otimes ( \rep{1}\,, \rep{4} \,, +\frac{1}{4} )_{\mathbf{H}}   \non
&\supset&   w_{\repb{4}\,,{\rm II} }  w_{\rep{4}} \epsilon^{  \kappa \delta } \,  ( \rep{1}\,, \repb{3} \,, -\frac{1}{3} )_{\mathbf{H}\,, \kappa } \otimes  ( \rep{1}\,, \repb{3} \,, -\frac{1}{3} )_{\mathbf{H}\,, \delta } \otimes ( \rep{1}\,, \rep{3} \,, +\frac{1}{3} )_{\mathbf{H} } \otimes ( \rep{1}\,, \rep{3} \,, +\frac{1}{3} )_{\mathbf{H} }  \,.
\eeqn
According to the minimal fermion-Higgs matching pattern, both the $( \rep{1}\,, \repb{4} \,, -\frac{1}{4} )_{\mathbf{H}\,,  \lambda={\rm II} }$ and the $( \rep{1}\,, \rep{4} \,, +\frac{1}{4} )_{\mathbf{H}}$ develop the $\Gc_{341}$-breaking VEVs.
Explicitly, the VEV term from Eq.~\eqref{eq:Oc6A_decomp} reads
\beqn
\Oc_{\mathscr A}^{d=6}&\sim& g_{ 6 \mathscr A} w_{\repb{4}\,,{\rm II} }  w_{\rep{4}} V_{ \repb{3}\,, {\rm I}} u_{ \repb{2}\,, 3} V_{ \rep{3}} u_{\rep{2}} \Rightarrow  \boxed{ u_{ \repb{2}\,,3} } \,,  \boxed{ V_{\rep{3}} } \,,\boxed{ u_{ \rep{2}} }\,.
\eeqn

\para
For the operator of $\Oc_{\mathscr B}^{d=6}$, we find
\beqs
\beqn
\Oc_{\mathscr B}^{d=6}&=& \epsilon^{ \dot \lambda  \dot \kappa } \, \repb{21_H}_{\,, \dot \lambda }  \repb{21_H}_{\,,\dot \kappa } \cdot ( \rep{7_H})^4 \non
&\supset& \epsilon^{ \dot \lambda  \dot \kappa }  \, ( \rep{1}\,, \rep{6} \,, -\frac{1}{2} )_{\mathbf{H}\,, \dot \lambda } \otimes ( \rep{1}\,, \rep{6} \,, -\frac{1}{2} )_{\mathbf{H}\,,  \dot \kappa } \non
&\otimes&  ( \rep{1}\,, \rep{4} \,, +\frac{1}{4} )_{\mathbf{H}} \otimes  ( \rep{1}\,, \rep{4} \,, +\frac{1}{4} )_{\mathbf{H}} \otimes  ( \rep{1}\,, \rep{4} \,, +\frac{1}{4} )_{\mathbf{H}} \otimes ( \rep{1}\,, \rep{4} \,, +\frac{1}{4} )_{\mathbf{H}} \,.
\eeqn
\eeqs
%
%
Obviously, the possible VEV term vanishes due to the ${\rm SU}(4)_W$-invariant $\epsilon$-tensor.

\para
For the operator of $\Oc_{\mathscr C}^{d=6}$, we find
\beqn
&& ( \rep{21_H}  )^4 \cdot  ( \rep{35_H}  )^2 \non
&\supset& ( \rep{1}\,, \rep{6} \,, +\frac{1}{2} )_{\mathbf{H}}  \otimes ( \rep{1}\,, \rep{6} \,, +\frac{1}{2} )_{\mathbf{H}}  \otimes ( \rep{1}\,, \rep{6} \,, +\frac{1}{2} )_{\mathbf{H}}  \otimes ( \rep{1}\,, \rep{6} \,, +\frac{1}{2} )_{\mathbf{H}} \non
&\otimes& ( \rep{1}\,, \repb{4} \,, +\frac{3}{4} )_{\mathbf{H}}  \otimes ( \rep{1}\,, \repb{4} \,, +\frac{3}{4} )_{\mathbf{H}}  \Rightarrow \textrm{no gauge-invariant term}  \,.
\eeqn
Hence, this operator does not contribute to a VEV term in the Higgs potential.

\para
For the operator of $\Oc_{\mathscr D}^{d=6}$, we find
\beqn
&& \rep{7_H} \cdot ( \rep{21_H}  )^2 \cdot  ( \rep{35_H}  )^3 \non
&\supset& ( \rep{1}\,, \rep{4} \,, +\frac{1}{4} )_{\mathbf{H}} \otimes ( \rep{1}\,, \rep{6} \,, +\frac{1}{2} )_{\mathbf{H}} \otimes ( \rep{1}\,, \rep{6} \,, +\frac{1}{2} )_{\mathbf{H}}  \non
&\otimes& ( \rep{1}\,, \repb{4} \,, +\frac{3}{4} )_{\mathbf{H}} \otimes ( \rep{1}\,, \repb{4} \,, +\frac{3}{4} )_{\mathbf{H}} \otimes ( \rep{1}\,, \repb{4} \,, +\frac{3}{4} )_{\mathbf{H}}  \Rightarrow \textrm{no gauge-invariant term}  \,.
\eeqn
Hence, this operator does not contribute to a VEV term in the Higgs potential.

\subsection{The $d=8$ operators}

\para
For the operator of $\Oc^{d=8}$, we find
\beqn
&&   ( \epsilon^{ \dot \lambda_1   \dot \lambda_2 }  \repb{21_H}_{\,,\dot \lambda_1}  \repb{21_H}_{\,,\dot \lambda_2 } ) \cdot ( \epsilon^{ \dot \kappa_1   \dot \kappa_2 }  \repb{21_H}_{\,,\dot \kappa_1}  \repb{21_H}_{\,,\dot \kappa_2 }  ) \cdot (  \epsilon^{ \dot \delta_1   \dot \delta_2 }  \repb{21_H}_{\,,\dot \delta_1}  \repb{21_H}_{\,,\dot \delta_2 }) \cdot \rep{21_H}\cdot \rep{35_H} \non
&\supset&   \Big[   \epsilon^{\dot \rho \dot \delta}  ( \rep{1}\,, \rep{6} \,, -\frac{1}{2} )_{\mathbf{H}\,, \dot \rho} \otimes  ( \rep{1}\,, \rep{6} \,, -\frac{1}{2} )_{\mathbf{H}\,, \dot \delta} \Big]^3  \otimes   ( \rep{1}\,, \rep{6} \,, +\frac{1}{2} )_{\mathbf{H}} \otimes  ( \rep{1}\,, \repb{4} \,, +\frac{3}{4} )_{\mathbf{H}} \non
&\Rightarrow& \textrm{no gauge-invariant term} \,.
\eeqn
Hence, this operator does not contribute to a VEV term in the Higgs potential.

\subsection{The $d=9$ operators}

\para
For the operator of $\Oc_{\mathscr A}^{d=9}$, we find
\beqn\label{eq:Oc9A_decomp}
&&   ( \epsilon^{\lambda \kappa \delta } \repb{7_H}_{\,,\lambda}  \repb{7_H}_{\,,\kappa }  \repb{7_H}_{\,,\delta }  )^2 \cdot (\rep{21_H})^3 \non
&\supset&   \Big[  \epsilon^{\lambda_1 \lambda_2  \lambda_3} ( \rep{1}\,, \repb{4} \,, -\frac{1}{4} )_{\mathbf{H}\,, \lambda_1} \otimes ( \rep{1}\,, \repb{4} \,, -\frac{1}{4} )_{\mathbf{H}\,, \lambda_2 } \otimes ( \rep{1}\,, \repb{4} \,, -\frac{1}{4} )_{\mathbf{H}\,, \lambda_3} \Big]^2 \non
&\otimes& ( \rep{1}\,, \rep{6} \,, +\frac{1}{2} )_{\mathbf{H}} \otimes ( \rep{1}\,, \rep{6} \,, +\frac{1}{2} )_{\mathbf{H}}  \otimes ( \rep{1}\,, \rep{6} \,, +\frac{1}{2} )_{\mathbf{H}}  \non
&\supset& ( w_{\repb{4}\,, {\rm II}})^2 \epsilon^{\lambda_1 \lambda_2  } \epsilon^{\kappa_1 \kappa_2  } \, ( \rep{1}\,, \repb{3} \,, -\frac{1}{3} )_{\mathbf{H}\,, \lambda_1} \otimes ( \rep{1}\,, \repb{3} \,, -\frac{1}{3} )_{\mathbf{H}\,, \lambda_2 } \otimes ( \rep{1}\,, \repb{3} \,, -\frac{1}{3} )_{\mathbf{H}\,, \kappa_1} \otimes ( \rep{1}\,, \repb{3} \,, -\frac{1}{3} )_{\mathbf{H}\,, \kappa_2 }  \non
&\otimes& ( \rep{1}\,, \repb{3} \,, +\frac{2}{3} )_{\mathbf{H} } \otimes  ( \rep{1}\,, \rep{3} \,, +\frac{1}{3} )_{\mathbf{H} }^\prime \otimes  ( \rep{1}\,, \rep{3} \,, +\frac{1}{3} )_{\mathbf{H} }^\prime  \,.
\eeqn
Explicitly, the VEV term from Eq.~\eqref{eq:Oc9A_decomp} reads
\beqn
\Oc_{\mathscr A}^{d=9}&\sim& g_{ 9 \mathscr A} ( w_{\repb{4}\,,{\rm II} }   V_{ \repb{3}\,, {\rm I}} u_{ \repb{2}\,, 3} )^2 V_{ \rep{3}}^\prime u_{\rep{2}}^\prime u_{\repb{2}} \Rightarrow  \boxed{ u_{ \repb{2}\,,3} } \,,  \boxed{ u_{\rep{2}}^\prime } \,,  \boxed{ u_{\repb{2}} }\,.
\eeqn

\para
For the operator of $\Oc_{\mathscr B}^{d=9}$, we find
\beqn
&&     \Big[ \epsilon\cdot ( \repb{21_H}_{\,,\dot \lambda})^2 \Big]^3\cdot \rep{7_H} \cdot (\rep{21_H})^2 \non
&\supset&  \Big[   \epsilon^{\dot \rho \dot \delta}  ( \rep{1}\,, \rep{6} \,, -\frac{1}{2} )_{\mathbf{H}\,, \dot \rho} \otimes  ( \rep{1}\,, \rep{6} \,, -\frac{1}{2} )_{\mathbf{H}\,, \dot \delta} \Big]^3 \otimes  ( \rep{1}\,, \rep{4} \,, +\frac{1}{4} )_{\mathbf{H}}   \otimes ( \rep{1}\,, \rep{6} \,, +\frac{1}{2} )_{\mathbf{H}} \otimes ( \rep{1}\,, \rep{6} \,, +\frac{1}{2} )_{\mathbf{H}} \non
&\Rightarrow& \textrm{no gauge-invariant term} \,.
\eeqn

\para
For the operator of $\Oc_{\mathscr C}^{d=9}$, we find
\beqn
&&  ( \epsilon^{\dot \rho \dot \delta} \repb{21_H}_{\,,\dot \rho}  \repb{21_H}_{\,,\dot \delta} )^4\cdot \rep{21_H} \non
&\supset& \Big[   \epsilon^{\dot \rho \dot \delta}  ( \rep{1}\,, \rep{6} \,, -\frac{1}{2} )_{\mathbf{H}\,, \dot \rho} \otimes  ( \rep{1}\,, \rep{6} \,, -\frac{1}{2} )_{\mathbf{H}\,, \dot \delta} \Big]^4 \otimes ( \rep{1}\,, \rep{6} \,, +\frac{1}{2} )_{\mathbf{H}} \non
&\Rightarrow& \textrm{no gauge-invariant term} \,.
\eeqn

\subsection{Other higher-dimensional operators}

\begin{table}[htp]
\begin{center}
\begin{tabular}{c|cc|c}
\hline \hline
  $\Oc_{\rm mix}$  &   ${\rm U}(1)_1$  &  ${\rm U}(1)_2$  & VEV terms \\
\hline
$\Oc^{d=10} \equiv \Big[ \epsilon\cdot (\repb{7_H}_{\,,\lambda})^3  \Big]^3 \cdot \rep{21_H}$  &  $-(9 p_1 + 10 q_1)\,,(0)$  &  $-q_2\,,(\frac{p_2}{2})$ &  \xmark  \\[1mm]
\hline
$\Oc^{d=12} \equiv  \Big[ \epsilon\cdot ( \repb{21_H}_{\,,\dot \lambda})^2 \Big]^5\cdot (\rep{35_H} )^2$  & $-4 q_1\,,(0)$   & $-10 ( p_2 + q_2)\,,(-5p_2)$  & \xmark  \\[1mm]
\hline
$\Oc_{\mathscr A}^{d=13} \equiv  \Big[ \epsilon\cdot ( \repb{21_H}_{\,,\dot \lambda})^2 \Big]^5 \cdot (\rep{21_H} )^3$  & $-3 q_1\,,(0)$   & $-( 10 p_2 + 13 q_2)\,,(-\frac{7}{2}p_2)$  & \xmark  \\
$\Oc_{\mathscr B}^{d=13} \equiv  \Big[ \epsilon\cdot ( \repb{21_H}_{\,,\dot \lambda})^2 \Big]^6 \cdot \rep{35_H}$  & $-2q_1\,,(0)$   & $-12 (p_2 + q_2)\,,(-6p_2)$  & \xmark   \\[1mm]
\hline
$\Oc_{\mathscr A}^{d=14} \equiv \Big[ \epsilon\cdot (\repb{7_H}_{\,,\lambda})^3  \Big]^4 \cdot \rep{21_H} \cdot \rep{35_H}$  &  $-3(4 p_1 + 5 q_1)\,,(0)$  &  $-q_2\,,(\frac{p_2}{2})$ &  \xmark  \\
  $\Oc_{\mathscr B}^{d=14} \equiv \Big[ \epsilon\cdot ( \repb{21_H}_{\,,\dot \lambda})^2 \Big]^7 $  & $0$   & $-14(p_2 + q_2 )\,,(-7p_2)$  & \xmark   \\[1mm]
  \hline
 $\Oc^{d=16} \equiv  \Big[ \epsilon\cdot (\repb{7_H}_{\,,\lambda})^3 \Big]^5 \cdot \rep{7_H} $  &  $-15( p_1 + q_1)\,,(0)$  &  $-2 q_2\,,(p_2)$ & \xmark   \\[1mm]
  \hline
 $\Oc_{\mathscr A}^{d=20} \equiv  \Big[ \epsilon\cdot (\repb{7_H}_{\,,\lambda})^3 \Big]^6 \cdot (\rep{21_H} )^2$  &  $-2(9 p_1 + 10 q_1)\,,(0)$  &  $-2 q_2\,,(p_2)$ &  \xmark  \\[1mm]
 $\Oc_{\mathscr B}^{d=20} \equiv \Big[ \epsilon\cdot (\repb{7_H}_{\,,\lambda})^3 \Big]^6 \cdot \rep{7_H} \cdot \rep{35_H} $  &  $-2(9 p_1 + 10 q_1)\,,(0)$  &  $-2 q_2\,,(p_2)$ & \xmark \\[1mm]
\hline
  $\Oc^{d=21} \equiv \Big[  \epsilon\cdot (\repb{7_H}_{\,,\lambda})^3 \Big]^7 $  &  $-21 (p_1 + q_1)\,,(0)$  &  $0$ &  \xmark   \\[1mm]
\hline\hline
\end{tabular}
\end{center}
\caption{
The non-renormalizable ${\rm SU}(7)$ Higgs mixing operators with $d\geq 10$ and their ${\rm U}(1)_{1\,,2}$ charges.
The \cmark and \xmark represent whether the specific operator can contribute to VEV terms in the Higgs potential or not.
The charge assignments with $p_1=q_1=0$ and $q_2=- \hf p_2 \neq 0$ are marked in parentheses.
}
\label{tab:SU7_HiggsMix_higherOp}
\end{table}%

\para
For the operator of $\Oc^{d=10}$, we find
\beqn
&&  ( \epsilon^{\lambda \kappa \delta } \repb{7_H}_{\,, \lambda }  \repb{7_H}_{\,, \kappa}  \repb{7_H}_{\,, \delta} )^3 \cdot \rep{21_H} \non
&\supset& \Big[  \epsilon^{\lambda \kappa \delta } ( \rep{1}\,, \repb{4} \,, -\frac{1}{4} )_{\mathbf{H}\,, \lambda} \otimes ( \rep{1}\,, \repb{4} \,, -\frac{1}{4} )_{\mathbf{H}\,, \kappa} \otimes ( \rep{1}\,, \repb{4} \,, -\frac{1}{4} )_{\mathbf{H}\,, \delta} \Big]^3 \otimes ( \rep{1}\,, \rep{6} \,, +\frac{1}{2} )_{\mathbf{H}}   \non
&\Rightarrow& \textrm{no gauge-invariant term}  \,.
\eeqn


\para
For the operator of $\Oc^{d=12}$, we find
\beqn
&& ( \epsilon^{\dot \rho \dot \delta} \repb{21_H}_{\,,\dot \rho}  \repb{21_H}_{\,,\dot \delta}  )^5\cdot (\rep{35_H} )^2\non
&\supset& \Big[   \epsilon^{\dot \rho \dot \delta}  ( \rep{1}\,, \rep{6} \,, -\frac{1}{2} )_{\mathbf{H}\,, \dot \rho} \otimes  ( \rep{1}\,, \rep{6} \,, -\frac{1}{2} )_{\mathbf{H}\,, \dot \delta}  \Big]^5  \otimes  ( \rep{1}\,, \repb{4} \,, +\frac{3}{4} )_{\mathbf{H}}  \otimes  ( \rep{1}\,, \repb{4} \,, +\frac{3}{4} )_{\mathbf{H}} \non
&\Rightarrow& \textrm{no gauge-invariant term}   \,.
\eeqn

\para
For the operator of $\Oc_{\mathscr A}^{d=13}$, we find
\beqn
&& ( \epsilon^{\dot \rho \dot \delta} \repb{21_H}_{\,,\dot \rho}  \repb{21_H}_{\,,\dot \delta} )^5 \cdot (\rep{21_H} )^3\non
&\supset& \Big[   \epsilon^{\dot \rho \dot \delta}  ( \rep{1}\,, \rep{6} \,, -\frac{1}{2} )_{\mathbf{H}\,, \dot \rho} \otimes  ( \rep{1}\,, \rep{6} \,, -\frac{1}{2} )_{\mathbf{H}\,, \dot \delta}  \Big]^5  \otimes ( \rep{1}\,, \rep{6} \,, +\frac{1}{2} )_{\mathbf{H}}  \otimes ( \rep{1}\,, \rep{6} \,, +\frac{1}{2} )_{\mathbf{H}}  \otimes ( \rep{1}\,, \rep{6} \,, +\frac{1}{2} )_{\mathbf{H}}   \non
&\Rightarrow& \textrm{no gauge-invariant term}   \,.
\eeqn

\para
For the operator of $\Oc_{\mathscr B}^{d=13}$, we find
\beqn
&& ( \epsilon^{\dot \rho \dot \delta} \repb{21_H}_{\,,\dot \rho}  \repb{21_H}_{\,,\dot \delta}  )^6 \cdot \rep{21_H} \non
&\supset& \Big[   \epsilon^{\dot \rho \dot \delta}  ( \rep{1}\,, \rep{6} \,, -\frac{1}{2} )_{\mathbf{H}\,, \dot \rho} \otimes  ( \rep{1}\,, \rep{6} \,, -\frac{1}{2} )_{\mathbf{H}\,, \dot \delta}  \Big]^6  \otimes ( \rep{1}\,, \rep{6} \,, +\frac{1}{2} )_{\mathbf{H}}  \non
&\Rightarrow& \textrm{no gauge-invariant term}   \,.
\eeqn
%
%


\para
For the operator of $\Oc_{\mathscr A}^{d=14}$, we find
\beqn\label{eq:Oc14A_decomp}
&& (  \epsilon^{\lambda \kappa \delta } \repb{7_H}_{\,, \lambda }  \repb{7_H}_{\,, \kappa}  \repb{7_H}_{\,, \delta} )^4 \cdot \rep{21_H} \cdot \rep{35_H}  \non
&\supset& \Big[  \epsilon^{\lambda \kappa \delta } ( \rep{1}\,, \repb{4} \,, -\frac{1}{4} )_{\mathbf{H}\,, \lambda} \otimes ( \rep{1}\,, \repb{4} \,, -\frac{1}{4} )_{\mathbf{H}\,, \kappa} \otimes ( \rep{1}\,, \repb{4} \,, -\frac{1}{4} )_{\mathbf{H}\,, \delta} \Big]^4  \otimes ( \rep{1}\,, \rep{6} \,, +\frac{1}{2} )_{\mathbf{H}}  \otimes ( \rep{1}\,, \repb{4} \,, +\frac{3}{4} )_{\mathbf{H}}   \non
&\Rightarrow& \textrm{no gauge-invariant term}   \,.
\eeqn

\para
For the operator of $\Oc_{\mathscr B}^{d=14}$, we find
\beqn\label{eq:Oc14B_decomp}
&& ( \epsilon^{\dot \rho \dot \delta} \repb{21_H}_{\,,\dot \rho}  \repb{21_H}_{\,,\dot \delta} )^7  \supset \Big[ \epsilon^{\dot \rho \dot \delta}  ( \rep{1}\,, \rep{6} \,, -\frac{1}{2} )_{\mathbf{H}\,, \dot \rho} \otimes  ( \rep{1}\,, \rep{6} \,, -\frac{1}{2} )_{\mathbf{H}\,, \dot \delta}  \Big]^7 \non
  &\Rightarrow& \textrm{no gauge-invariant term}    \,.
\eeqn


\para
For the operator of $\Oc^{d=16}$, we find
\beqn
&& ( \epsilon^{\lambda \kappa \delta } \repb{7_H}_{\,, \lambda }  \repb{7_H}_{\,, \kappa}  \repb{7_H}_{\,, \delta}  )^5 \cdot \rep{7_H} \non
&\supset&  \Big[ \epsilon^{\lambda \kappa \delta } ( \rep{1}\,, \repb{4} \,, -\frac{1}{4} )_{\mathbf{H}\,, \lambda} \otimes ( \rep{1}\,, \repb{4} \,, -\frac{1}{4} )_{\mathbf{H}\,, \kappa} \otimes ( \rep{1}\,, \repb{4} \,, -\frac{1}{4} )_{\mathbf{H}\,, \delta}  \Big]^5  \otimes ( \rep{1}\,, \rep{4} \,, +\frac{1}{4} )_{\mathbf{H}} \non
  &\Rightarrow& \textrm{no gauge-invariant term}  \,.
\eeqn
%
%


\para
For the operator of $\Oc_{\mathscr A}^{d=20}$, we find
\beqn
&& ( \epsilon^{\lambda \kappa \delta } \repb{7_H}_{\,, \lambda }  \repb{7_H}_{\,, \kappa}  \repb{7_H}_{\,, \delta}  )^6 \cdot (\rep{21_H} )^2 \non
&\supset&  \Big[ \epsilon^{\lambda \kappa \delta } ( \rep{1}\,, \repb{4} \,, -\frac{1}{4} )_{\mathbf{H}\,, \lambda} \otimes ( \rep{1}\,, \repb{4} \,, -\frac{1}{4} )_{\mathbf{H}\,, \kappa} \otimes ( \rep{1}\,, \repb{4} \,, -\frac{1}{4} )_{\mathbf{H}\,, \delta}  \Big]^6  \otimes  ( \rep{1}\,, \rep{6} \,, +\frac{1}{2} )_{\mathbf{H}} \otimes  ( \rep{1}\,, \rep{6} \,, +\frac{1}{2} )_{\mathbf{H}} \non
  &\Rightarrow& \textrm{no gauge-invariant term}  \,.
\eeqn

\para
For the operator of $\Oc_{\mathscr B}^{d=20}$, we find
\beqn\label{eq:Oc20B_decomp}
&& ( \epsilon^{\lambda \kappa \delta } \repb{7_H}_{\,, \lambda }  \repb{7_H}_{\,, \kappa}  \repb{7_H}_{\,, \delta} )^6 \cdot \rep{7_H} \cdot \rep{35_H} \non
&\supset& \Big[ \epsilon^{\lambda \kappa \delta } ( \rep{1}\,, \repb{4} \,, -\frac{1}{4} )_{\mathbf{H}\,, \lambda} \otimes ( \rep{1}\,, \repb{4} \,, -\frac{1}{4} )_{\mathbf{H}\,, \kappa} \otimes ( \rep{1}\,, \repb{4} \,, -\frac{1}{4} )_{\mathbf{H}\,, \delta}  \Big]^6  \otimes  ( \rep{1}\,, \rep{4} \,, +\frac{1}{4} )_{\mathbf{H}} \otimes ( \rep{1}\,, \repb{4} \,, +\frac{3}{4} )_{\mathbf{H}} \non
  &\Rightarrow& \textrm{no gauge-invariant term}    \,.
\eeqn

\para
For the operator of $\Oc^{d=21}$, we find
\beqn\label{eq:Oc21_decomp}
\Oc^{d=21}&=& ( \epsilon^{\lambda \kappa \delta } \repb{7_H}_{\,, \lambda }  \repb{7_H}_{\,, \kappa}  \repb{7_H}_{\,, \delta} )^7 \non
&\supset& \Big[ \epsilon^{\lambda \kappa \delta } ( \rep{1}\,, \repb{4} \,, -\frac{1}{4} )_{\mathbf{H}\,, \lambda} \otimes ( \rep{1}\,, \repb{4} \,, -\frac{1}{4} )_{\mathbf{H}\,, \kappa} \otimes ( \rep{1}\,, \repb{4} \,, -\frac{1}{4} )_{\mathbf{H}\,, \delta}  \Big]^7 \non
  &\Rightarrow& \textrm{no gauge-invariant term}  \,.
\eeqn
%
%


\providecommand{\href}[2]{#2}\begingroup\raggedright\endgroup

\end{document}